%% file: excited_states.tex
\documentstyle [floats,epsf,aps,eqsecnum]{revtex}
\draft
\global\firstfigfalse

\newcommand{\eqnref}[1]{Eq.~(\ref{#1})}
\newcommand{\eqnlessref}[1]{(\ref{#1})}

\def\Re{{\rm Re}}
\def\Im{{\rm Im}}
\def\Det{{\rm Det}}

\def\csch{{\rm csch}}

\def\Tr{{\rm Tr}}
\def\scrD{{\cal D}}
\def\lsim{\mathrel{\lower0.3em\hbox{$\stackrel{\textstyle <}{\sim}$}}}
\def\gsim{\mathrel{\lower0.3em\hbox{$\stackrel{\textstyle >}{\sim}$}}}
\def\scrK{{\cal K}}
\def\negspace{\kern -0.4em}

\def\ehat{{\bf \hat e}}

\def\det{{\rm det}}
\def\tr{{\rm tr}}
\def\dvec{\raise 0.3 em \hbox{$^\leftrightarrow$} \kern -0.77 em}

\def\omegahat{\hat%
	{\setbox0=\hbox{$\omega$}%
		\kern-.025em\copy0\kern-\wd0
		\kern.05em\copy0\kern-\wd0
		\kern-.025em\raise.0433em\box0}}

\begin{document}

\title{Semiclassical Quantization of Effective String Theory and Regge Trajectories}
\author{M. Baker and R. Steinke}

\address{University of Washington, P.O. Box 351560, Seattle, WA 98195-1560, USA\\
        E-mail: baker@phys.washington.edu, rsteinke@w-link.net}
\preprint{UW/PT 02-01}

\maketitle

\begin{abstract}
We begin with an effective string theory for long distance QCD, and evaluate
the semiclassical expansion of this theory about a classical rotating
string solution, taking into account the the dynamics of the boundary of
the string. We show that, after renormalization, the zero point energy
of the string fluctuations remains finite when the masses of the quarks on
the ends of the string approach zero. The theory is then conformally invariant
in any spacetime dimension $D$. For $D=26$ the energy spectrum of the rotating
string formally coincides with that of the open string in classical Bosonic
string theory. However, its physical origin is different. It is a semiclassical
spectrum of an effective string theory valid only for large values of the
angular momentum. For $D=4$, the first semiclassical correction adds the constant
$1/12$ to the classical Regge formula.
\end{abstract}

\newpage

\input{intro}
\input{string_calc}
\input{results}

% bibliography

%\nocite{*}
\bibliographystyle{plain}
\bibliography{excited_states}

% appendices

\appendix

\input{string_quant}
\input{Geval}
\input{boundary_partition}

\end{document}

%% file: intro.tex
\section{Introduction}

String models provide a simple picture of quark confinement, and
have been used to understand the hadronic spectrum since well before
QCD was established as the theory of strong interactions.
A straight rotating string gives rise to linear Regge
trajectories relating the angular momenta of mesons
composed of light quarks to the squares of their masses.
A fixed straight string gives a linear potential between
heavy quarks, and the zero point energy of the long
wavelength fluctuations of this string gives rise to
a universal correction to the linear potential~\cite{Luscher1}.
Excited states of a fluctuating string with fixed ends give potentials
of hybrid mesons~\cite{Isgur+Paton}. In this paper we
calculate the effect of string fluctuations on the
Regge trajectories of mesons.

In a previous paper~\cite{Baker+Steinke2}, we derived an
effective string theory of vortices, beginning with a field theory
containing classical vortex solutions (dual superconducting
vortices)~\cite{Abrikosov,Baker+Ball+Brambilla+Prosperi+Zachariasen,%
Baker+Ball+Zachariasen:1990,Baker+Ball+Zachariasen:1995,%
Mandelstam,Nambu,Nielsen+Olesen,tHooft}. The field theory itself
was an effective field theory of long distance QCD, describing
phenomena at distances greater than the radius of the
flux tube whose center is the location of the vortex.
The resulting effective string theory was obtained as a development
of earlier work by many authors~\cite{ACPZ,Allen+Olsson+Veseli:1998,%
Dubin+Kaidalov+Simonov2,Dubin+Kaidalov+Simonov,%
Forster,Gervais+Sakita,Kikkawa+Kotani+Sato,LaCourse+Olsson,Luscher2,%
Luscher1,Pol+Strom,Polyakov:book}.
We then used this effective string theory to calculate the zero
point energy of the string fluctuations around a straight,
rotating string with quarks on its ends. The classical
equations of motion determined the distance between the
quarks in terms of their angular velocity $\omega$,
and the fluctuations of the ends of the string
were not taken into account. The calculated zero point energy
gave a correction to the classical formula for the leading Regge
trajectory. For static quarks separated by a fixed distance $R$,
the expression for the zero point energy
reduced to $-\pi/12R$, the result of L\"uscher for the
contribution of string fluctuations to the static
quark--antiquark potential. However, for rotating quarks
the zero point energy diverged logarithmically as the quark
mass $m$ approached zero, and we were not able to calculate
Regge trajectories for zero mass quarks.

In this paper, we show how to take the mass zero limit. We also treat the
quark motion quantum mechanically,
so that the boundaries of the string become dynamical variables
which couple to the interior degrees of freedom of the string.
We evaluate the contribution of string fluctuations to Regge
trajectories in the limit of massless quarks, and in the
limit where one quark is massless and the other is heavy.
Finally, we generalize our expressions for Regge trajectories
to $D$ spacetime dimensions, and compare with the spectrum
of the classical bosonic string.

\section{Outline}

In section~\ref{previous work} we review the results obtained
in~\cite{Baker+Steinke2}, giving the expression for the
functional integral representation of the effective string
theory. We give the expression for the contribution of
string fluctuations to the Wilson loop of the effective
string theory, calculated in the classical background of a worldsheet
with rotating quarks on its ends. This expression exhibits
a logarithmic divergence as the quark mass goes to
zero. In section~\ref{geodesic curvature section},
we show how to remove the logarithmic
divergence by renormalization, and take the zero mass limit.

In section~\ref{end motion section}, we take into account the quantum fluctuations of
the positions of the quark and antiquark at the ends of the
string. We obtain an effective Lagrangian for the rotating string
from which the meson energy levels can be determined.
This effective Lagrangian $L_{{\rm eff}}(\omega)$ is determined
by a functional integral of the effective string theory evaluated
in the steepest descent approximation about a classical rotating
string solution. The action determining this path integral
is the Nambu--Goto action, added to the action of the
point particles on the ends of the string.

In sections~\ref{quadratic section}~through~\ref{decoupling section},
we expand the action to quadratic order in small fluctuations about
the classical rotating string solution. These fluctuations separate
into two classes, interior degrees of freedom determining the positions
of the interior points of the string, and boundary degrees of freedom
determining the fluctuations of the positions of the quark--antiquark
pair at the ends of the string. The fluctuations of the ends of the
string excite the interior points, which in turn react back on the
ends, producing an effective quark--antiquark interaction. The
remaining interior string fluctuations are decoupled from the
fluctuations of the positions of the quarks. In
section~\ref{interior integral section},
we calculate the zero point energy of these interior fluctuations,
generalizing the results of~\cite{Baker+Steinke2} to the case where
the quark masses are not equal.

In section~\ref{boundary integral section}, we find the propagator for the boundary
fluctuations from the effective boundary action. We show that
for zero mass quarks, the poles in the propagator are at
integral multiples of $\omega$. These frequencies are the same as the
frequencies of the harmonic oscillators determining the
interior Lagrangian. That is, for massless quarks, we find that
taking the fluctuations of the boundary into account does not
change the spectrum of the excited states of the rotating string.
We also evaluate the position of the poles in the propagator when
one quark is heavy and the other is massless. We find that the
spectrum in this case is shifted when boundary effects are taken into
account.

In section~\ref{quark mass limit section}, we evaluate the results
of sections \ref{interior integral section} and \ref{boundary integral section}
in certain physical limits. Our results are valid when the masses
of the quarks are either zero or very large, and when the string
length is large compared to its thickness.
In section~\ref{meson spectrum section}, we calculate the Regge trajectories of mesons
containing zero mass quarks. The ground state of the rotating string
gives the leading Regge trajectory, and the excited states of the
rotating string give rise to daughter Regge trajectories determining
the spectrum of hybrid mesons. We also calculate the Regge trajectories
for mesons composed of one heavy quark and one light quark.

In section~\ref{D dimen section}, we extend the calculations of this paper to $D$
spacetime dimensions, and compare with the spectrum of classical
Bosonic string theory.

\section{Previous Work}
\label{previous work}

\subsection{The Effective String Theory}

In reference~\cite{Baker+Steinke2}, we began with a quantum field
theory having classical vortex solutions. The dual Abelian Higgs model
is an example of such a field theory. The surface $\tilde x^\mu(\xi^1,\xi^2)$
of zeros of the complex dual Higgs field is the location of the
vortex sheet, and electric flux is confined to tubes of radius $a$,
where $a^{-1} = M$, the mass of the vector particle in the theory.

The path integral, which defines the Wilson loop $W[\Gamma]$ of the field
theory, goes over all field configurations containing a vortex sheet
bounded by the loop $\Gamma$ formed by the worldlines of the
trajectories of the quark and antiquark on the ends of the vortex.
The action $S_{{\rm eff}}[\tilde x^\mu]$ of the effective string
theory is obtained by first integrating only over field configurations
containing a vortex on a particular surface $\tilde x^\mu$. The
remaining integral over the surfaces $\tilde x^\mu$ then gives
$W[\Gamma]$ the form of an effective string theory of
vortices.

The action $S_{{\rm eff}}[\tilde x^\mu]$ is invariant under
reparameterizations $\xi^a \to {\xi'}^a(\xi)$, $a = 1\,,2$
of the worldsheet $\tilde x^\mu(\xi)$ of the vortex. We
choose a particular parameterization of $\tilde x^\mu$
in terms of the amplitudes $f^a(\xi)$, $a = 1\,,2$ of
the two transverse fluctuations of the vortex,
\begin{equation}
\tilde x^\mu =  x^\mu\left(f^1(\xi), f^2(\xi), \xi^1, \xi^2\right) \,.
\end{equation}
This gives $W[\Gamma]$ the form
\begin{equation}
W[\Gamma] = \int \scrD f^1 \scrD f^2 \Delta_{FP} e^{i S_{{\rm eff}}[\tilde x^\mu]} \,,
\label{Wilson loop def}
\end{equation}
where
\begin{equation}
\Delta_{FP} = \Det\left[\frac{\epsilon^{\mu\nu\alpha\beta}}{\sqrt{-g}}
\frac{\partial x^\mu}{\partial f^1}
\frac{\partial x^\nu}{\partial f^2}
\frac{\partial \tilde x^\alpha}{\partial \xi^1}
\frac{\partial \tilde x^\beta}{\partial \xi^2}\right]
\end{equation}
is the Faddeev--Popov determinant produced by gauge fixing the
reparameterization symmetry, and where $\sqrt{-g}$ is the
square root of the determinant of the induced metric $g_{ab}$,
\begin{equation}
g_{ab} = \frac{\partial \tilde x^\mu}{\partial \xi^a}
\frac{\partial \tilde x_\mu}{\partial \xi^b} \,.
\end{equation}
The path integral \eqnlessref{Wilson loop def} goes over string
fluctuations with wavelengths greater than the radius $1/M$ of the flux tube.
The measure of the path integral \eqnlessref{Wilson loop def} is universal
and parameterization invariant. The factor $\Delta_{FP}$ came from rewriting the
original field theory path integral as a ratio of path integrals of two string
theories~\cite{ACPZ,Baker+Steinke2}.

The action $S_{{\rm eff}}[\tilde x^\mu]$ can be expanded in powers of
the extrinsic curvature tensor $\scrK^A_{ab}$ of the worldsheet
$\tilde x^\mu$,
\begin{equation}
S_{{\rm eff}}[\tilde x^\mu] = -\sigma \int d^2\xi \sqrt{-g}
- \beta \int d^2\xi \sqrt{-g} \left(\scrK^A_{ab}\right)^2 + ...
\label{expanded action}
\end{equation}
The extrinsic curvature tensor is
\begin{equation}
\scrK^A_{ab} = n_\mu^A(\xi) \frac{\partial^2 \tilde x^\mu}
{\partial \xi^a \partial \xi^b} \,,
\end{equation}
where $n_\mu^A(\xi)$, $A = 1,2$ are vectors normal to the worldsheet
at the point $\tilde x^\mu(\xi)$. The string tension $\sigma$ and the
rigidity $\beta$ are determined by the parameters of the underlying
effective field theory.

The extrinsic curvature $\scrK^A_{ab}$ is of the order of magnitude of the
angular velocity $\omega$, and the expansion parameter in the semiclassical
approximation is $\omega^2/\sigma \sim 1/J$, where $J$ is the angular momentum
of the rotating string. Therefore, in the region of large $J$ where the
effective theory is applicable, the action \eqnlessref{expanded action}
can be replaced by the Nambu--Goto action $S_{{\rm NG}}$,
\begin{equation}
S_{{\rm eff}}[\tilde x^\mu] = S_{{\rm NG}}
= -\sigma \int d^2\xi \sqrt{-g} \,.
\label{Nambu Goto action}
\end{equation}

\subsection{The Semiclassical Calculation in the Background of a Rotating String}

\label{old L string review}

\begin {figure}[Ht]
    \begin{center}
	\null \hfill \epsfbox{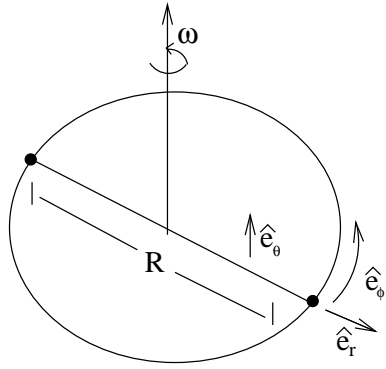} \hfill \null
    \end{center}
    \caption{The string coordinate system}
    \label{rot string fig}
\end {figure}

Using \eqnlessref{Wilson loop def} and \eqnlessref{Nambu Goto action},
we calculated $W[\Gamma]$ in the leading semiclassical approximation
in the background of a worldsheet generated by a straight string
attached to quarks rotating with uniform angular velocity $\omega$
(See Fig.~\ref{rot string fig}).
The quarks have masses $m_1$ and $m_2$,
move with velocities $v_1 = \omega R_1$ and $v_2 = \omega R_2$,
and are separated by a fixed distance $R = R_1 + R_2$. The
parameters $\xi = (t, r)$ are the time $t$ and the coordinate $r$,
which runs along the string from $-R_1$ to $R_2$, so that the
transverse velocity of the straight string is zero when $r=0$.

The amplitudes $f(\xi)$ of the transverse fluctuations are
the spherical coordinates $\theta(r,t)$ and $\phi(r,t)$
of a point on the string. These angles are defined in
an unconventional manner so that $\theta(r,t) = \phi(r,t) = 0$
is a straight string rotating in the $xy$ plane. The ends
of the string are fixed to their classical trajectories,
\begin{equation}
\theta(-R_1, t) = \theta(R_2, t) = \phi(-R_1, t) = \phi(R_2, t) = 0 \,.
\label{classical boundary angles}
\end{equation}
The fluctuating worldsheet $\tilde x^\mu$ then has the parameterization
$\tilde x^\mu(r,t)$ given by
\begin{eqnarray}
\tilde x^\mu(r,t) &=&  x^\mu(\theta(r, t), \phi(r, t), r, t)
\nonumber \\
&=& t \ehat_0^\mu + r \Big[ \cos\theta(r,t) \cos\left(\phi(r,t) + \omega t\right)
\ehat_1^\mu + \cos\theta(r,t) \sin\left(\phi(r,t) + \omega t\right) \ehat_2^\mu
\nonumber \\
& & - \sin\theta(r,t) \ehat_3^\mu \Big] \,,
\label{x mu param}
\end{eqnarray}
where $\ehat_\alpha^\mu$, $\alpha = 0\,,...\,3$ are unit vectors along the
four fixed spacetime axes, $\ehat_\alpha^\mu = \delta_\alpha^\mu$.

The classical rotating straight string $\bar x^\mu(r,t)$ has the
parameterization \eqnlessref{x mu param} with
\newline
$\theta = \phi = 0$,
\begin{eqnarray}
\bar x^\mu(r,t) &=& x^\mu\left( \theta(r,t) = 0, \phi(r,t) = 0, r, t \right)
\nonumber \\
&=& t \ehat_0^\mu + r \left[ \cos\omega t \ehat_1^\mu + \sin\omega t \ehat_2^\mu
\right] \,.
\end{eqnarray}
The corresponding metric $\bar g_{ab} = g_{ab}[\bar x^\mu]$
and the classical action $S_{{\rm NG}}[\bar x^\mu]$ are independent of
the time $t$, so that $W[\Gamma]$ has the form
\begin{equation}
W[\Gamma] = e^{iTL^{{\rm string}}(R_1, R_2, \omega)} \,,
\label{Wilson loop integrated}
\end{equation}
where $T$ is the elapsed time. For massless quarks, the ends of the
string move with the velocity of light, and singularities
appear in $L^{{\rm string}}$. To regulate these singularities,
we retain the quark mass as a cutoff and take the massless limit
at the end when evaluating physical quantities for massless quarks.

The Lagrangian $L^{{\rm string}}$ is the sum of a classical part
$L^{{\rm string}}_{{\rm cl}}$ and a fluctuating part $L^{{\rm string}}_{{\rm fluc}}$,
\begin{equation}
L^{{\rm string}} = L^{{\rm string}}_{{\rm cl}} + L^{{\rm string}}_{{\rm fluc}} \,,
\label{L string separated}
\end{equation}
where
\begin{equation}
L^{{\rm string}}_{{\rm cl}} = -\frac{\sigma}{T} \int d^2\xi \sqrt{-\bar g}
= -\sigma \int_{-R_1}^{R_2} dr \sqrt{1 - r^2 \omega^2} \,.
\end{equation}
The effective Lagrangian for the quark--antiquark pair is obtained by adding
quark mass terms to $L^{{\rm string}}$:
\begin{equation}
L_{{\rm eff}}(R_1, R_2, \omega) = - \sum_{i=1}^2 m_i \sqrt{1-(\omega R_i)^2}
+ L^{{\rm string}}(R_1, R_2, \omega) \,.
\end{equation}
The effective Lagrangian is the sum of a classical part and a fluctuating part,
\begin{equation}
L_{{\rm eff}}(R_1, R_2, \omega) = L_{{\rm cl}}(R_1, R_2, \omega)
+ L^{{\rm string}}_{{\rm fluc}}(R_1, R_2, \omega) \,,
\end{equation}
where
\begin{eqnarray}
L_{{\rm cl}} &=& - \sum_{i=1}^2 m_i \sqrt{1-(\omega R_i)^2}
 - \sigma \int_{-R_1}^{R_2} dr \sqrt{1 - r^2 \omega^2}
\nonumber \\
&=& -\sum_{i=1}^2 \left[ \sigma \frac{R_i}{2} \left( \frac{\arcsin(v_i)}{v_i}
+ \gamma_i^{-1} \right) + m_i \gamma_i^{-1} \right] \,,
\label{L_cl def formal}
\end{eqnarray}
with
\begin{equation}
\gamma_i = \frac{1}{\sqrt{1-v_i^2}} \,, \kern 1 in v_i = \omega R_i \,.
\end{equation}

The expression for $L^{{\rm string}}_{{\rm fluc}}$ is obtained
from \eqnlessref{Wilson loop integrated} and the semiclassical
calculation of $W[\Gamma]$. It contains terms which are quadratically,
linearly, and logarithmically divergent in the cutoff $M$.
The quadratically divergent term is a renormalization of the
string tension, the linearly divergent term is a renormalization
of the quark mass, and the logarithmically divergent term is
proportional to the integral of the scalar curvature
over the whole worldsheet~\cite{Luscher1}. After absorbing the
quadratically and linearly divergent terms into renormalizations,
we obtained an expression for
$L^{{\rm string}}_{{\rm fluc}}$~\cite{Baker+Steinke2}.
The following is a generalization of that expression to
the case of unequal quark masses
(see Section~\ref{interior integral section}):
\begin{equation}
L^{{\rm string}}_{{\rm fluc}}(R_1, R_2, \omega)
= \frac{\pi}{12 R_p} - \sum_{i=1}^2 \frac{\omega v_i \gamma_i}{\pi}
\left[ \ln\left(\frac{M R_i}{\gamma_i^2 - 1}\right) + 1 \right]
+ \frac{1}{2} \omega + \omega f(v_1, v_2) \,,
\label{L fluc eval}
\end{equation}
where $R_p$ is the proper length of the string,
\begin{equation}
R_p = \frac{1}{\omega} \left( \arcsin v_1 + \arcsin v_2 \right) \,,
\label{proper length}
\end{equation}
and
\begin{equation}
f(v_1, v_2) = -\frac{1}{\pi} \int_0^\infty ds \ln\left[
\frac{s^2 + (v_1\gamma_1 + v_2\gamma_2) s \coth(s R_p \omega)
+ v_1 \gamma_1 v_2 \gamma_2}{(s + v_1\gamma_1)(s + v_2\gamma_2)} \right] \,.
\label{f v_1 v_2 def}
\end{equation}
The function $f(v_1,v_2)$ vanishes when $v_1$ and $v_2$ approach
unity, so that the last term in \eqnlessref{L fluc eval} is
small for relativistic quarks.

In the limit $\omega \to 0$, $R_p \to R_1 + R_2 = R$, and
$L^{{\rm string}}_{{\rm fluc}}$ reduces to the result of L\"uscher
for the correction to the static quark--antiquark potential
due to string fluctuations,
\begin{equation}
V_{\hbox{\scriptsize L\"uscher}} = - L_{{\rm fluc}}^{{\rm string}}
(R_1, R_2, \omega = 0) = - \frac{\pi}{12R} \,.
\end{equation}
For $\omega \ne 0$,
$L^{{\rm string}}_{{\rm fluc}}$ contains a logarithmically divergent
part. We simplify this term using the classical equation of
motion,
\begin{equation}
\frac{\partial L_{{\rm cl}}}{\partial R_i} \Big|_{R_i = \bar R_i} = 0 \,,
\label{canonical classical eqn}
\end{equation}
to express $\bar R_i$ in terms of $\omega$. \eqnref{canonical classical eqn}
gives the relation
\begin{equation}
\sigma \bar R_i = m_i \left(\bar \gamma_i^2 - 1\right) \,,
\label{classical eqn}
\end{equation}
where $\bar\gamma_i$ is equal to $\gamma_i$ evaluated at $R_i = \bar R_i$.
The solution of \eqnlessref{classical eqn} for $\bar R_i$ as a function of $\omega$ is
\begin{equation}
\bar R_i = \frac{1}{\omega} \left(\sqrt{\left(\frac{m_i\omega}{2\sigma}\right)^2
+ 1} - \frac{m_i\omega}{2\sigma}\right) \,.
\label{solve for R_i}
\end{equation}
Using the relation \eqnlessref{classical eqn} in \eqnlessref{L fluc eval}
gives
\begin{equation}
L^{{\rm string}}_{{\rm fluc}} = \frac{\pi}{12 R_p} - \sum_{i=1}^2
\frac{\omega v_i \bar\gamma_i}{\pi} \left[ \ln\left(\frac{Mm_i}{\sigma}\right) + 1 \right]
+ \frac{1}{2} \omega + \omega f(v_1, v_2) \,.
\label{L fluc pre renorm}
\end{equation}
The logarithmically divergent quantity in the square brackets is
independent of the dynamical parameter $\omega$. This is important,
because the quantity $\omega v_i \bar \gamma_i$
diverges when $m_i \to 0$.

In the next section, we will show that the term containing the logarithmic
divergence can be absorbed by renormalization of a
contribution to the string action called the geodesic curvature.
When this divergence is removed, the theory will be finite
in the $m_i \to 0$ limit. This renormalization was not done in
\cite{Baker+Steinke2}, and is important because it will produce a
finite limit of the theory for massless quarks.

\section{Renormalization of the Geodesic Curvature}

\label{geodesic curvature section}

We now define the geodesic curvature~\cite{Alvarez:1983}, and
renormalize \eqnlessref{L fluc pre renorm}.
In the same way that the action for
the string \eqnlessref{expanded action} can be expanded in
powers of the extrinsic curvature, the action for the boundary
can be expanded in powers of the geodesic curvature.
Using the notation $x^\mu_i$ for the positions of the ends of the string,
\begin{equation}
x^\mu_i(t) \equiv \tilde x^\mu((-1)^i R_i(t), t) \,,
\label{x mu i def}
\end{equation}
the boundary part $S_b$ of the action is
\begin{equation}
S_b = -\sum_{i=1}^2 m_i \int dt \sqrt{-\hbox{$\dot x^\mu_i$}^2}
- \sum_{i=1}^2 \kappa_i \int dt \frac{\dot x^\mu_i}{\sqrt{-\hbox{$\dot x^\mu_i$}^2}}
\left((-1)^i t_{\mu\nu}\right) \Big|_{r = (-1)^i R_i(t)}
\frac{d}{dt} \frac{\dot x^\nu_i}{\sqrt{-\hbox{$\dot x^\mu_i$}^2}} \,,
\label{boundary action}
\end{equation}
where $t_{\mu\nu}$ is the antisymmetric string worldsheet orientation tensor,
\begin{equation}
t^{\mu\nu} \equiv \frac{\epsilon^{ab}}{\sqrt{-g}}
\frac{\partial\tilde x^\mu}{\partial\xi^a}
\frac{\partial\tilde x^\nu}{\partial\xi^b} \,.
\end{equation}
The first term in \eqnlessref{boundary action} is the quark mass term.
The second is the contribution of the geodesic curvature of the boundary
(the extrinsic curvature of the boundary in the plane of the string
worldsheet). The factor of $(-1)^i$ multiplying $t_{\mu\nu}$ is present
so that $\dot x^\mu_i (-1)^i t_{\mu\nu}$ is always an outward pointing
radial vector.

For a straight string rotating
with angular velocity $\omega$, the geodesic curvature is equal
to $\omega v_i\gamma_i^2$. Inserting this in \eqnlessref{boundary action}
and dropping the integral over time gives the boundary Lagrangian,
\begin{eqnarray}
L^{{\rm boundary}} &=& -\sum_{i=1}^2 \gamma_i^{-1} \left[ m_i
+ \kappa_i \omega v_i \gamma_i^2 + ... \right] \,,
\label{L boundary expand}
\end{eqnarray}
where $\kappa_i$ is the coefficient of the first order term in this
expansion. The logarithmic divergence
in \eqnlessref{L fluc pre renorm} can then be regarded as a renormalization
of $\kappa_i$. In the limit where the quark is massless,
we must take the $m_i \to 0$ limit before we take
the cutoff $M$ to infinity, since we have an effective theory.
In the $m_i \to 0$ limit, the requirement that the action
\eqnlessref{L boundary expand} is finite then forces the
renormalized value of $\kappa_i$ to be zero (note that this does not
prevent $\kappa_i$ from being zero for nonzero $m_i$).
The logarithmic divergence in $M$ may therefore
be absorbed into a renormalization of the geodesic curvature in
the case where either $m_i = 0$ or $v_i << 1$.

Removing the terms in \eqnlessref{L fluc pre renorm} proportional
to $\omega v_i \gamma_i$ gives an expression for $L^{{\rm string}}_{{\rm fluc}}$
which is applicable in the massless quark limit,
\begin{equation}
L_{{\rm fluc}}^{{\rm string}} = \frac{\pi}{12R_p} + \frac{1}{2} \omega
+ \omega f(v_1, v_2) \,.
\label{L fluc renorm}
\end{equation}
In the case of two light mesons with $m_1 = m_2 = 0$
($\bar\gamma_1, \bar\gamma_2 \to \infty$), Eqs.~\eqnlessref{proper length}
and \eqnlessref{f v_1 v_2 def} give $R_p = \pi/\omega$ and
$f(v_1, v_2) = 0$, so that \eqnlessref{L fluc renorm} becomes
\begin{equation}
L^{{\rm string}}_{{\rm fluc}}(\omega) \Big|_{m_1 = m_2 = 0}
= \frac{\omega}{12} + \frac{\omega}{2} = \frac{7}{12} \omega \,.
\label{L fluc ll}
\end{equation}
In the case of one heavy and one light quark, $m_1 \to \infty$ ($v_1 \to 0$)
and $m_2 = 0$ ($\bar\gamma_2 \to \infty$), $R_p = \pi/2\omega$ and
\begin{equation}
f(v_1 = 0, v_2 \to 1) = -\frac{1}{\pi} \int_0^\infty \ln\coth\left(\frac{\pi}{2} s\right)
= -\frac{\omega}{4} \,,
\end{equation}
so
\begin{equation}
L^{{\rm string}}_{{\rm fluc}}(\omega) \Big|_{{m_1 \to \infty} \atop {m_2 = 0}}
= \frac{\omega}{6} + \frac{\omega}{2} - \frac{\omega}{4} = \frac{5}{12} \omega \,.
\label{L fluc hl}
\end{equation}

\section{Fluctuations in the Motion of the Quarks at the Ends of the String}

\label{end motion section}

In the previous discussion, the quark--antiquark pair moved in
a fixed classical trajectory in the $xy$ plane (See Fig.~\ref{rot string fig}
and Eq's \eqnlessref{classical boundary angles} and \eqnlessref{solve for R_i}).
We now take into account the fluctuations of the positions ${\bf \vec x}_1(t)$
and ${\bf \vec x}_2(t)$ of the quarks at the ends of the rotating string,
so that these coordinates are no longer fixed by
\eqnlessref{classical boundary angles} and \eqnlessref{solve for R_i}.
The radial coordinates $R_1(t)$ and $R_2(t)$, along with the
angular coordinates $\theta(-R_1(t), t)$, $\phi(-R_1(t), t)$,
$\theta(R_2(t), t)$ and $\phi(R_2(t), t)$, parameterize the
endpoints ${\bf \vec x}_1(t)$ and ${\bf \vec x}_2(t)$ of the
string in a reference frame rotating with angular velocity $\omega$
in the $xy$ plane,
\begin{eqnarray}
\vec x_1(t) &=& -R_1(t) \Big(\cos\theta(-R_1(t),t) \cos(\phi(-R_1(t),t)
+ \omega t) \ehat_1
\nonumber \\
& & + \cos\theta(-R_1(t),t) \sin(\phi(-R_1(t),t) + \omega t) \ehat_2
- \sin\theta(-R_1(t),t) \ehat_3 \Big) \,,
\nonumber \\
\vec x_2(t) &=& R_2(t) \Big(\cos\theta(R_2(t),t) \cos(\phi(R_2(t),t)
+ \omega t) \ehat_1
\nonumber \\
& & + \cos\theta(R_2(t),t) \sin(\phi(R_2(t),t) + \omega t) \ehat_2
- \sin\theta(R_2(t),t) \ehat_3 \Big) \,.
\label{boundary param}
\end{eqnarray}
The values of the coordinates $r$ and $t$ at the ends of the string
are determined by the equations $r = -R_1(t)$ and $r = R_2(t)$,
so the string has the representation \eqnlessref{x mu param}
with
\begin{equation}
-R_1(t) \le r \le R_2(t) \,.
\end{equation}

We extend the functional integral \eqnlessref{Wilson loop def} to
include a path integral over $\vec x_1(t)$ and
${\bf\vec x}_2(t)$, and add the action of the quarks to the
string action \eqnlessref{Nambu Goto action}. This
extension replaces $W[\Gamma]$ by the ``partition function'' $Z$,
\begin{equation}
Z = \frac{1}{Z_b} \int \scrD f^1(\xi) \scrD f^2(\xi) \scrD \vec x_1(t)
\scrD {\bf\vec x}_2(t) \Delta_{FP} e^{-i\sigma \int d^2\xi \sqrt{-g}
- i \sum_{i=1}^2 m_i \int_{-T/2}^{T/2} dt \sqrt{1 - \dot{\vec x}_i^2(t)}} \,,
\label{string partition}
\end{equation}
where $Z_b$ is the partition function of two free (scalar) quarks,
\begin{equation}
Z_b = \int \scrD {\bf\vec x}_1(t)\scrD {\bf\vec x}_2(t)
e^{-i \sum_{i=1}^2 m_i \int_{-T/2}^{T/2} dt \left(\sqrt{1 - \dot{\vec x}_i^2(t)}
\right)} \,.
\end{equation}
Dividing by $Z_b$ removes the vacuum energy of the quarks.

The partition function $Z$ sums over all string states. In choosing
the parameterization \eqnlessref{x mu param} for $\tilde x^\mu(\xi)$,
we have replaced $Z$ with a partition function which contains
a sum over those string states with a particular value of the
average angular velocity $\omega$. We denote this partition
function by $Z(\omega)$.

Under the parameterization \eqnlessref{x mu param},
the integration measure $\scrD f^1 \scrD f^2 \Delta_{FP}$
for the interior of the string becomes
\begin{equation}
\scrD f^1(\xi) \scrD f^2(\xi) \Delta_{FP} = \scrD\left(\sin\theta(r,t)\right)
\scrD\phi(r,t) \Det\left[ \frac{r^2}{\sqrt{-g}} \right] \,.
\end{equation}
The integration measure for the endpoints is
\begin{equation}
\scrD {\bf\vec x}_1 \scrD {\bf\vec x}_2 =
\scrD(\sin\theta)\Big|_{r = -R_1(t)} \scrD\phi\Big|_{r = -R_1(t)} \scrD R_1
\scrD(\sin\theta)\Big|_{r = R_2(t)} \scrD\phi\Big|_{r = R_2(t)} \scrD R_2
\Det[R_1^2] \Det[R_2^2] \,.
\end{equation}
The path integral \eqnlessref{string partition}, with the
parameterizations \eqnlessref{x mu param} and \eqnlessref{boundary param},
is then
\begin{equation}
Z(\omega) = \frac{1}{Z_b} \int \scrD(\sin\theta) \scrD\phi
\scrD R_1 \scrD R_2 \Det\left[\frac{r^2}{\sqrt{-g}}\right]
\Det[R_1^2] \Det[R_2^2] e^{i \int_{-T/2}^{T/2} dt L[\tilde x^\mu]} \,,
\label{partition param}
\end{equation}
where the Lagrangian $L[\tilde x^\mu]$ is
\begin{equation}
L[\tilde x^\mu] = -\sigma \int_{-R_1(t)}^{R_2(t)} dr \sqrt{-g}
- \sum_{i=1}^2 m_i \sqrt{ - \hbox{$\dot x^\mu_i$}^2} \,,
\label{L full def}
\end{equation}
where $\dot x^\mu_i$ is the time derivative of $x^\mu_i$,
defined in \eqnlessref{x mu i def}.

The functional integral \eqnlessref{partition param} evaluated
in the steepest descent approximation about the classical
solution $\theta(r,t) = \phi(r,t) = 0$, $R_i(t) = \bar R_i$
determines the effective Lagrangian for the rotating quark--antiquark
pair,
\begin{equation}
Z(\omega) = e^{iT L_{{\rm eff}}(\omega)} \,.
\label{partition effective Lagrangian}
\end{equation}
\eqnref{partition effective Lagrangian} is the extension
of \eqnlessref{Wilson loop integrated} to include fluctuations
of the boundary. The effective Lagrangian is the sum of
a classical part and a fluctuating part,
\begin{equation}
L_{{\rm eff}}(\omega) = L_{{\rm cl}}(\omega) + L_{{\rm fluc}}(\omega) \,,
\label{L eff separation}
\end{equation}
where $L_{{\rm cl}}(\omega)$ is given by \eqnlessref{L_cl def formal}.
The fluctuation part $L_{{\rm fluc}}$ contains contributions
both from fluctuations of the interior of the string
and from fluctuations of the boundary of the string.

In Appendix~\ref{DHN appendix}, using the methods of Dashen,
Hasslacher, and Neveu~\cite{DHN}, we obtain from the full partition
function $Z$ a quantization condition on the angular momentum,
and show how to find the energies of the physical meson states.
We summarize these results here. The angular momentum $J$ is
\begin{equation}
J = \frac{d L_{{\rm eff}}(\omega)}{d\omega} \,,
\end{equation}
where $L_{{\rm eff}}(\omega)$ is determined by
\eqnlessref{partition effective Lagrangian} in terms of the particular
partition function \eqnlessref{partition param}.
The angular momentum is fixed by
the WKB quantization condition,
\begin{equation}
J = l + \frac{1}{2} \kern 1 in l = 0, 1, 2,... \,.
\label{J quant}
\end{equation}
The energy $E(\omega)$ is given by the corresponding Hamiltonian,
\begin{equation}
E(\omega) = \omega \frac{d L_{{\rm eff}}(\omega)}{d\omega}
- L_{{\rm eff}}(\omega) \,.
\end{equation}
The energy is equal to the classical energy $E_{{\rm cl}}(\omega)$,
plus a correction due to fluctuations. To first order in the
perturbation $L_{{\rm fluc}}(\omega)$, the correction to the energy
is minus the correction to the Lagrangian~\cite{Landau+Lifshitz:Mechanics},
\begin{equation}
E(\bar\omega) = E_{{\rm cl}}(\bar\omega) - L_{{\rm fluc}}(\bar\omega) \,,
\end{equation}
where
\begin{equation}
E_{{\rm cl}}(\bar\omega) = \bar\omega \frac{d L_{{\rm cl}}(\bar\omega)}{d\bar\omega}
- L_{{\rm cl}}(\bar\omega) \,,
\label{E class}
\end{equation}
and where $\bar\omega$ is given
as a function of $J$ by the classical relation
\begin{equation}
J = \frac{d L_{{\rm cl}}(\bar\omega)}{d\bar\omega} \,.
\label{J class}
\end{equation}
The zero point energy of the fluctuations is then
\begin{equation}
E_{{\rm fluc}}(J) = - L_{{\rm fluc}}(\bar\omega(J)) \,.
\label{E of J}
\end{equation}
Eq's \eqnlessref{J class} and \eqnlessref{E of J} give the
leading semiclassical correction to the energies of mesons
on the leading Regge trajectory.

%% file: string_calc.tex
\section{Quadratic Expansion of the Action}

\label{quadratic section}

To evaluate the effective Lagrangian $L_{{\rm eff}}(\omega)$ from
\eqnlessref{partition effective Lagrangian}, we must first expand
the Lagrangian $L[\tilde x^\mu]$ to quadratic order in the
small fluctuations about the classical solution. We call
$r_1(t)$ and $r_2(t)$ the fluctuations of $R_1(t)$ and $R_2(t)$
about the classical values $\bar R_1$ and $\bar R_2$
\eqnlessref{solve for R_i},
\begin{equation}
R_1(t) = \bar R_1 + r_1(t) \,, \kern 1in R_2(t) = \bar R_2 + r_2(t) \,.
\label{R_i expand}
\end{equation}
To quadratic order in small fluctuations, the angular coordinates
of the ends of the string $\theta((-1)^i R_i(t), t)$ and
$\phi((-1)^i R_i(t), t)$ can be evaluated at the classical
values of the $R_i(t)$.

The degrees of freedom in the Lagrangian can be viewed as a
combination of string degrees of freedom ($\phi(r,t)$ and $\theta(r,t)$
for $-\bar R_1 < r < \bar R_2$) and quark degrees of freedom
\begin{equation}
\theta(-\bar R_1,t) \,, \kern 0.25in
\theta(\bar R_2,t) \,, \kern 0.25in
\phi(-\bar R_1,t) \,, \kern 0.25in
\phi(\bar R_2,t) \,, \kern 0.25in
r_1(t) \,, \kern 0.25in
r_2(t) \,.
\label{boundary degrees}
\end{equation}
The quark degrees of freedom depend only upon $t$ (and not $r$), and
we refer to them as ``boundary'' degrees of freedom. The string
degrees of freedom depend on both $t$ and $r$, and we refer to
them as the ``interior'' degrees of freedom. Eq's
\eqnlessref{partition effective Lagrangian} and \eqnlessref{L eff separation}
for $L_{{\rm eff}}(\omega)$ reduce to Eq's \eqnlessref{Wilson loop integrated}
and \eqnlessref{L string separated} when the boundary degrees
of freedom are set equal to zero, and become boundary conditions on
the interior degrees of freedom. Inclusion of these
boundary degrees of freedom gives an additional contribution
to $L_{{\rm eff}}(\omega)$.

We now expand the Lagrangian \eqnlessref{L full def} and
the corresponding action $\int_{-T/2}^{T/2} dt L[\tilde x^\mu]$
to quadratic order in the small fluctuations $\theta(r,t)$,
$\phi(r,t)$ and $r_i(t)$ about the classical
solution $\theta = \phi = 0$, $r_i = 0$. We first
evaluate the string tension term. Using the
parameterization \eqnlessref{x mu param} of the
worldsheet, we obtain the tangent vectors to $\tilde x^\mu$,
\begin{eqnarray}
\dot x^\mu &=& \ehat_0^\mu - r \dot\theta \left( \sin\theta
\cos(\phi + \omega t) \ehat_1^\mu
+ \sin\theta \sin(\phi + \omega t) \ehat_2^\mu
+ \cos\theta \ehat_3^\mu \right)
\nonumber \\
& & + r \left( \dot\phi + \omega \right) \cos\theta
\left( -\sin(\phi + \omega t) \ehat_1^\mu
+ \cos(\phi + \omega t) \ehat_2^\mu \right) \,,
\nonumber \\
x'^\mu &\equiv& \frac{\partial x^\mu}{\partial r}
= \left( \cos\theta \cos(\phi + \omega t) \ehat_1^\mu
+ \cos\theta \sin(\phi + \omega t) \ehat_2^\mu
- \sin\theta \ehat_3^\mu \right)
\nonumber \\
& & - r \theta' \left( \sin\theta
\cos(\phi + \omega t) \ehat_1^\mu
+ \sin\theta \sin(\phi + \omega t) \ehat_2^\mu
+ \cos\theta \ehat_3^\mu \right)
\nonumber \\
& & + r \phi' \cos\theta
\left( -\sin(\phi + \omega t) \ehat_1^\mu
+ \cos(\phi + \omega t) \ehat_2^\mu \right) \,.
\label{x mu tangents}
\end{eqnarray}
The components of the metric are
\begin{eqnarray}
g_{tt} = \dot x^\mu \dot x_\mu &=& -1 + r^2 \dot\theta^2
+ r^2 \left( \dot\phi + \omega \right)^2 \cos^2\theta \,,
\nonumber \\
g_{rt} = \dot x^\mu x'_\mu &=& r^2 \dot\theta \theta'
+ r^2 \left( \dot\phi + \omega \right) \phi' \cos^2\theta \,,
\nonumber \\
g_{rr} = x'^\mu x'_\mu &=& 1 + r^2 \theta'^2 + r^2 \phi'^2 \cos^2\theta \,.
\end{eqnarray}
To quadratic order in $\theta$ and $\phi$, the square root of
the determinant of the metric is
\begin{equation}
\sqrt{-g} = \gamma^{-1} - r^2 \omega \gamma \dot\phi - \frac{1}{2} r^2\gamma
\left(\dot\theta^2 - \omega^2 \theta^2 - \gamma^{-2} \theta'^2\right)
- \frac{1}{2} r^2\gamma^3 \left(\dot\phi^2 - \gamma^{-2} \phi'^2\right) \,,
\end{equation}
where $\gamma^{-1} = \sqrt{1 - r^2 \omega^2}$.

To evaluate the string tension term to quadratic order, we must also
expand the limits of integration $-R_1(t)$ and $R_2(t)$ about
$-\bar R_1$ and $\bar R_2$ respectively. Using \eqnlessref{R_i expand}
and \eqnlessref{x mu tangents}, we obtain
\begin{eqnarray}
\int_{-R_1(t)}^{R_2(t)} dr \sqrt{-g}
&=& \int_{-\bar R_1}^{\bar R_2} dr \Bigg[
\gamma^{-1} - r^2 \omega \gamma \dot\phi
\nonumber \\
& & - \frac{1}{2} r^2 \gamma \left( \dot\theta^2 - \omega^2 \theta^2
- \gamma^{-2} \theta'^2 \right) - \frac{1}{2} r^2 \gamma^3
\left( \dot\phi^2 - \gamma^{-2} \phi'^2 \right) \Bigg]
\nonumber \\
& & + \sum_i \left[ r_i \left( \gamma^{-1} - r^2 \omega \gamma
\dot\phi \right) \Big|_{r = (-1)^i \bar R_i} + \frac{1}{2} (-1)^i r_i^2
\frac{d \gamma^{-1}}{dr} \Big|_{r = (-1)^i \bar R_i} \right] \,.
\label{tension term expand}
\end{eqnarray}
The second term in \eqnlessref{tension term expand}, which is linear in $\phi$,
is a perfect time derivative, and contributes a term in the action
given by
\begin{equation}
-\int_{-T/2}^{T/2} dt \int_{-\bar R_1}^{\bar R_2} dr r^2 \gamma \omega
\dot\phi = - \int_{-\bar R_1}^{\bar R_2} dr r^2 \gamma \omega
\left[\phi\left(\frac{T}{2},r\right) - \phi\left(-\frac{T}{2},r\right)\right]
= 0 \,.
\label{phi linear term}
\end{equation}
The quantity in square brackets is zero, since
the angular velocity $\omega$ is defined to be the angle traversed by the
string in time $T$, divided by the time $T$. The constraint
\eqnlessref{phi linear term} is the condition that the fluctuation
$\dot\phi$ does not contribute to first order to the angular
momentum of the string, and hence contributes only to
vibrational modes.

Next we expand the quark mass term in \eqnlessref{L full def},
\begin{eqnarray}
\sqrt{-\left(\dot{\tilde x}^\mu + \dot r \tilde x'^\mu\right)^2}
\Bigg|_{r = (-1)^i R_i(t)} &=&
\bar\gamma_i^{-1} - R_i(t)^2 \omega \gamma \Big|_{r = (-1)^i R_i(t)}
\frac{d\phi((-1)^i R_i(t),t)}{dt}
\nonumber \\
& & - \frac{1}{2} \bar R_i^2 \bar\gamma_i \left( \dot\theta^2
- \omega^2 \theta^2 \right) \Big|_{r = (-1)^i \bar R_i}
- \frac{1}{2} \bar R_i^2 \bar\gamma_i^3 \dot\phi^2 \Big|_{r =
(-1)^i \bar R_i}
\nonumber \\
& & - \frac{1}{2} \bar\gamma_i \dot R_i(t)^2
\nonumber \\
&=& \bar\gamma_i^{-1} + (-1)^i r_i \frac{d(\gamma^{-1})}{dr}
\Bigg|_{r = (-1)^i \bar R_i} + \frac{1}{2} r_i^2
\frac{d^2(\gamma^{-1})}{dr^2} \Bigg|_{r = (-1)^i \bar R_i}
\nonumber \\
& & - \bar R_i^2 \bar\gamma_i \omega
\frac{d\phi(R_i(t),t)}{dt} - (-1)^i r_i \dot\phi \omega
\frac{d(r^2\gamma)}{dr} \Bigg|_{r = (-1)^i \bar R_i}
\nonumber \\
& & - \frac{1}{2} \bar R_i^2 \bar\gamma_i \left( \dot\theta^2
- \omega^2 \theta^2 \right) \Big|_{r = (-1)^i \bar R_i}
- \frac{1}{2} \bar R_i^2 \bar\gamma_i^3 \dot\phi^2 \Big|_{r =
(-1)^i \bar R_i}
\nonumber \\
& & - \frac{1}{2} \bar\gamma_i \dot r_i(t)^2 \,.
\label{mass term expand}
\end{eqnarray}

Inserting \eqnlessref{tension term expand} and \eqnlessref{mass term expand}
into \eqnlessref{L full def} and applying the constraint
\eqnlessref{phi linear term} gives
\begin{eqnarray}
L[\tilde x^\mu] &=& L_{{\rm cl}}(\omega)
+ \frac{1}{2} \sigma \int_{-\bar R_1}^{\bar R_2} dr
\left[ r^2 \gamma \left( \dot\theta^2 - \omega^2 \theta^2
- \gamma^{-2} \theta'^2 \right) + r^2 \gamma^3 \left( \dot\phi^2
- \gamma^{-2} \phi'^2 \right) \right]
\nonumber \\
& & + \sum_{i=1}^2 \Bigg[ m_i \bar R_i^2 \left( \dot\theta^2 - \omega^2 \theta^2 \right)
\Big|_{r = (-1)^i \bar R_i} + m_i \bar R_i^2 \dot\phi^2 \Big|_{r = (-1)^i \bar R_i}
\nonumber \\
& & + \left( -\sigma\bar\gamma_i^{-1} + m_i \bar\gamma_i \bar R_i \omega^2 \right) r_i
+ \frac{1}{2} m_i \bar\gamma_i \dot r_i^2
+ \frac{1}{2} \left( \sigma \bar\gamma_i \bar R_i \omega^2
+ m_i \bar\gamma_i^3 \omega^2 \right) r_i^2
\nonumber \\
& & + \left( \sigma \bar\gamma_i \bar R_i^2 \omega
+ m_i \bar\gamma_i \left( \bar\gamma_i^2 + 1 \right)
\bar R_i \omega \right) r_i \dot\phi \Big|_{r = (-1)^i \bar R_i}
\Bigg] \,.
\label{L before cancel}
\end{eqnarray}
The term in \eqnlessref{L before cancel} which is linear in $r_i$ vanishes,
 because $\bar R_i$ satisfies the classical equation of motion
\eqnlessref{solve for R_i}.
Replacing $\sigma$ using \eqnlessref{solve for R_i} in the terms
in \eqnlessref{L before cancel} which contain $r_i$ gives
\begin{eqnarray}
L[\tilde x^\mu] &=& L_{{\rm cl}}(\omega)
+ \frac{1}{2} \sigma \int_{-\bar R_1}^{\bar R_2} dr
\left[ r^2 \gamma \left(  \dot\theta^2 - \omega^2 \theta^2
- \gamma^{-2} \theta'^2 \right)
+ r^2 \gamma^3 \left( \dot\phi^2 - \gamma^{-2} \phi'^2 \right) \right]
\nonumber \\
& & + \sum_i m_i \bar\gamma_i \left[
\frac{1}{2} \bar R_i^2 \Big( \dot\theta^2 - \omega^2 \theta^2 \right)
\Big|_{r = (-1)^i \bar R_i} + \frac{1}{2} \bar R_i^2 \bar\gamma_i^2
\dot\phi^2 \Big|_{r = (-1)^i \bar R_i}
\nonumber \\
& & + \frac{1}{2} \dot r_i^2
+ \frac{1}{2} \left( 2\bar\gamma_i^2 - 1 \right) \omega^2 r_i^2
+ 2 \bar\gamma_i^2 \bar R_i \omega r_i \dot\phi
\Big|_{r = (-1)^i \bar R_i} \Big] \,.
\label{L quadratic}
\end{eqnarray}
\eqnref{L quadratic} is the complete quadratic expansion of the Lagrangian,
and it includes both interior and boundary degrees of freedom. The
interior--interior interactions are all contained in the integral over $r$,
and the boundary--boundary interactions in the last term are proportional
to the quark masses. Interior--boundary interactions occur in the terms
in the integral containing $\theta'$ and $\phi'$, which couple interior
and boundary parts of $\theta$ and $\phi$ through the spatial derivative.

\section{Coupling of Quarks to External Sources}

\label{sources section}

The effective Lagrangian \eqnlessref{partition effective Lagrangian}
determines the energy \eqnlessref{E of J} of the ground state of
a rotating quark--antiquark pair having angular momentum $J$.
We will also calculate the energies of the excited states of the mesons
(hybrid mesons lying on daughter Regge trajectories) by examining the
poles in the Green's function which describes the coupling of the
endpoints of the string. To obtain this Green's function, we add
to $L[\tilde x^\mu]$ a term $L_{{\rm source}}$ coupling the
positions ${\bf\vec x}_1(t)$ and ${\bf\vec x}_2(t)$ of the
quarks to external forces $\vec\rho^{\,\,i}(t)$, $i=1\,,2$,
\begin{equation}
L_{{\rm source}} = \sum_{i=1}^2 \vec\rho^{\,\,i}(t) \cdot {\bf\vec x}_i(t) \,,
\end{equation}
where
\begin{equation}
\vec\rho^{\,\,i} = \rho_\phi^i \sin\delta_i \ehat_1 + \rho_\phi^i \cos\delta_i \ehat_2
+ \rho_\theta^i \ehat_3 \,.
\end{equation}
The sources $\rho_\theta^i$ couple to the fluctuations $\theta$
transverse to the plane of rotation, while $\rho_\phi^i$ and $\delta_i$
are the polar coordinates of the forces coupling to the fluctuations
$\phi$, $r_1$, and $r_2$ lying in this plane. The Lagrangian
\eqnlessref{L quadratic} couples
the interior and boundary degrees of freedom, so that sources acting
on the boundary will also couple to the interior degrees of
freedom, and are capable of generating the excited states of the
string.

Inserting the expression \eqnlessref{boundary param} for the
quark coordinate ${\bf\vec x}_i$ into $L_{{\rm source}}$,
and keeping the leading term in small fluctuations gives
\begin{equation}
L_{{\rm source}} = L_{{\rm source}}^\theta + L_{{\rm source}}^\phi \,,
\end{equation}
where
\begin{equation}
L_{{\rm source}}^\theta = \sum_{i=1}^2 (-1)^i \rho_\theta^i(t) \bar R_i
\theta((-1)^i \bar R_i, t) \,,
\end{equation}
and
\begin{equation}
L_{{\rm source}}^\phi = \sum_{i=1}^2 (-1)^i \rho_\phi^i(t) \left[
r_i(t) \sin(\omega t + \delta_i(t)) + \bar R_i \phi((-1)^i \bar R_i, t)
\cos(\omega t + \delta_i(t)) \right] \,.
\end{equation}
The Lagrangian $L_{{\rm source}}^\theta$ gives the coupling of the
source to the transverse fluctuations $\theta((-1)^i \bar R_i, t)$,
while $L_{{\rm source}}^\phi$ gives the coupling of the sources to the
in plane degrees of freedom $r_i(t)$ and $\phi((-1)^i \bar R_i, t)$
(See Fig.~\ref{rot string fig}).
The phases $\delta_i$ give the direction of the external force in the plane
of rotation in the space--fixed system, and the angles $\omega t + \delta_i$
give the angle between this force and the instantaneous position of
the rotating string.

The Lagrangian \eqnlessref{L quadratic} does not couple the traverse
degrees of freedom $\theta(r, t)$ to the in-plane degrees of
freedom $\phi(r, t)$ and $r_i(t)$, so we can treat them independently.
We can write
\begin{equation}
L + L_{{\rm source}} = L_{{\rm cl}} + L_\theta + L_\phi \,,
\label{L division}
\end{equation}
where
\begin{eqnarray}
L_{\theta} &=& \frac{1}{2} \sigma \int_{-\bar R_1}^{\bar R_2} dr \gamma r^2
\left(\dot\theta^2 - \omega^2 \theta^2 - \gamma^{-2} {\theta'}^2 \right)
+ \hat L_\theta \,,
\nonumber \\
L_\phi &=& \frac{1}{2} \sigma \int_{-\bar R_1}^{\bar R_2} dr \gamma^3 r^2
\left( \dot\phi^2 - \gamma^{-2} {\phi'}^2 \right)
+ \hat L_\phi \,,
\label{L both in text}
\end{eqnarray}
and
\begin{eqnarray}
\hat L_\phi &=& \sum_i m_i \bar\gamma_i \left[ \frac{1}{2} \dot r_i^2
+ \frac{1}{2} \left( 2\bar\gamma_i^2 - 1 \right) \omega^2 r_i^2 
+ 2 \bar R_i \omega \bar\gamma_i^2 r_i \dot\phi
+ \frac{1}{2} \bar R_i^2 \bar\gamma_i^2 \dot\phi^2 \right]
\Bigg|_{r = (-1)^i \bar R_i}
\nonumber \\
& & + \sum_i (-1)^i \rho_\phi^i \left[ r_i \sin\left(\omega t
+ \delta_i\right)
+ \bar R_i \phi \Big|_{r = (-1)^i \bar R_i} \cos\left(\omega t + \delta_i\right)
\right] \,,
\nonumber \\
\hat L_\theta &=& \sum_{i=1}^2 \left[ \frac{1}{2} m_i \bar R_i^2 \bar\gamma_i
\left( \dot\theta^2 - \omega^2 \theta^2 \right)
+ (-1)^i \rho_\theta^i \bar R_i \theta \right] \Bigg|_{r = (-1)^i \bar R_i} \,.
\label{L hat in text}
\end{eqnarray}

The quantities $\hat L_\theta$ and $\hat L_\phi$ contain the quark mass terms
and the source terms, and depend only on the boundary values
\eqnlessref{boundary degrees}. The remaining terms in
\eqnlessref{L both in text} are the contributions of the string
Lagrangian to $L_\theta$ and $L_\phi$, and they depend upon both the
interior and boundary degrees of freedom. In the next section we will
decouple the interior and boundary degrees of freedom, and will obtain
an expression for the action as a sum of an interior contribution and a boundary
contribution. We will do this separately in each ``sector'' ($\theta$
and $\phi$) using a common procedure.

\section{Decoupling the Interior from the Boundary}

\label{decoupling section}

We write the two equations \eqnlessref{L both in text} for $L_\theta$
and $L_\phi$ as specializations of an equation for $L_\psi$ ($\psi = \theta,\phi$):
\begin{eqnarray}
L_\psi &=& \frac{1}{2} \sigma \int_{-\bar R_1}^{\bar R_2} dr \Sigma(r)
\left( \dot\psi^2(r,t) - C \psi^2(r,t) - \gamma^{-2} {\psi'}^2(r,t) \right)
+ \hat L_\psi \,.
\label{L sector in text}
\end{eqnarray}
The constant $C$ is $\omega^2$ in the $\theta$ sector,
and zero in the $\phi$ sector. The function $\Sigma(r)$ is
\begin{equation}
\Sigma(r) \Big|_{L_\theta} = \gamma r^2 \,,
\kern 1 in
\Sigma(r) \Big|_{L_\phi} = \gamma^3 r^2 \,.
\label{Sigma def}
\end{equation}
The action $S_\psi$ for each sector can be expressed in terms of the Fourier
transform of $\psi$ with respect to time,
\begin{equation}
\tilde\psi(r, \nu) \equiv \int dt e^{-i\nu t} \psi(r, t) \,.
\end{equation}
\begin{eqnarray}
S_\psi = \int dt L_\psi
&=& -\frac{\sigma}{2} \int \frac{d\nu}{2\pi} \int_{-\bar R_1}^{\bar R_2}
dr \tilde\psi^* \left[-\frac{\partial}{\partial r} \Sigma(r) \gamma^{-2}
\frac{\partial}{\partial r} - (\nu^2 - C) \Sigma(r)\right] \tilde\psi
\nonumber \\
& & - \frac{\sigma}{2} \sum_{i=1}^2 (-1)^i \Sigma((-1)^i \bar R_i) \bar\gamma_i^{-2}
\tilde\psi_{b,i}^* \left(\frac{\partial\tilde\psi}{\partial r}\right)
\Bigg|_{r = (-1)^i \bar R_i} + \int dt \hat L_\psi \,,
\label{L psi diffeq}
\end{eqnarray}
where $\tilde\psi_{b,i}$ are the values of $\tilde\psi$ evaluated
at the ends $(-1)^i \bar R_i$ of the string,
\begin{equation}
\tilde\psi_{b,i}(\nu) \equiv \tilde\psi((-1)^i \bar R_i, \nu) \,.
\end{equation}

We next define the ``boundary part'' $\tilde\psi_B(r, \nu)$ of $\tilde\psi(r, \nu)$
to be the solution to the differential equation
\begin{equation}
\left[-\frac{\partial}{\partial r} \Sigma(r) \gamma^{-2}
\frac{\partial}{\partial r} - (\nu^2 - C) \Sigma(r)\right]
\tilde\psi_B(r, \nu) = 0 \,,
\label{psi B eqn}
\end{equation}
satisfying the boundary conditions
\begin{equation}
\tilde\psi_B((-1)^i \bar R_i, \nu) = \tilde\psi_{b,i}(\nu) \,.
\label{psi B bc}
\end{equation}
We define the ``interior part'' $\tilde\psi_I(r, \nu)$ of $\tilde\psi$ as
\begin{equation}
\tilde\psi_I(r,\nu) = \tilde\psi(r,\nu) - \tilde\psi_B(r,\nu) \,.
\end{equation}
Due to the boundary condition \eqnlessref{psi B bc}, $\tilde\psi_B$
is entirely determined by the $\tilde\psi_{b,i}$. \eqnref{psi B eqn}
guarantees that $\tilde\psi_B$ does not couple to $\tilde\psi_I$.
Notice that, by definition,
\begin{equation}
\tilde\psi_I((-1)^i \bar R_i, \nu) = 0 \,,
\end{equation}
so that the fields $\tilde\psi_I$ do not involve the boundary
fluctuations.

Replacing $\tilde\psi$ in \eqnlessref{L psi diffeq} with
$\tilde\psi_B$ and $\tilde\psi_I$ and integrating
by parts yields
\begin{equation}
S_\psi = S_{I,\psi} + S_{B,\psi} \,,
\end{equation}
with
\begin{equation}
S_{I,\psi} = -\frac{\sigma}{2} \int \frac{d\nu}{2\pi} \int_{-\bar R_1}^{\bar R_2}
dr \tilde\psi_I^* \left[-\frac{\partial}{\partial r} \Sigma(r) \gamma^{-2}
\frac{\partial}{\partial r} - (\nu^2 - C) \Sigma(r)\right] \tilde\psi_I \,,
\label{L_I def}
\end{equation}
and
\begin{equation}
S_{B,\psi} = - \frac{\sigma}{2} \sum_{i=1}^2 (-1)^i \Sigma((-1)^i \bar R_i)
\bar\gamma_i^{-2} \int \frac{d\nu}{2\pi}
\tilde\psi_{b,i}^* \left(\frac{\partial\tilde\psi_B}{\partial r}\right)
\Bigg|_{r = (-1)^i \bar R_i} + \int dt \hat L_\psi \,.
\label{L_B def}
\end{equation}
The interior action depends only on the interior degrees of freedom
(the values of $\tilde\psi(r, \nu)$ for $-\bar R_1 < r < \bar R_2$),
while the boundary action only depends of $\tilde\psi_{b,i}(\nu)$.

To express $\tilde\psi_B$ in terms of the $\tilde\psi_{b,i}$,
we use the Green's function $G(r,r',\nu)$
satisfying the equation
\begin{equation}
\left[-\frac{\partial}{\partial r} \Sigma(r) \gamma^{-2}
\frac{\partial}{\partial r} - (\nu^2 - C) \Sigma(r)\right]
G(r,r',\nu) = \delta(r - r') \,,
\end{equation}
for $-\bar R_1 < r < \bar R_2$, and the boundary conditions
\begin{equation}
G((-1)^i \bar R_i, r', \nu) = 0 \,.
\end{equation}
The solution of \eqnlessref{psi B eqn} with boundary conditions
\eqnlessref{psi B bc} is
\begin{equation}
\tilde\psi_B(r,\nu) = \sum_{i=1}^2 (-1)^{i+1} \Sigma((-1)^i \bar R_i)
\bar\gamma_i^{-2} \tilde\psi_{b,i}(\nu) \frac{\partial}{\partial r'} G(r,r', \nu)
\Bigg|_{r' = (-1)^i \bar R_i} \,.
\label{expand psi B}
\end{equation}

Inserting the expression \eqnlessref{expand psi B} for
$\tilde\psi_B$ into the definition \eqnlessref{L_B def}
of $S_{B,\psi}$, we find
\begin{equation}
S_{B,\psi} = \int dt \hat L_\psi  - \frac{\sigma}{2}
\sum_{i,j=1}^2 \int \frac{d\nu}{2\pi}
\tilde\psi_{b,i}^* G^{ij}_\psi(\nu) \tilde\psi_{b,j} \,,
\label{S_B Greens}
\end{equation}
where
\begin{equation}
G^{ij}_\psi(\nu) \equiv (-1)^{i+j} \Sigma((-1)^i \bar R_i)
\Sigma((-1)^j \bar R_j) \bar\gamma_i^{-2} \bar\gamma_j^{-2}
\frac{\partial^2}{\partial r \partial r'} G(r,r',\nu)
\Bigg|_{{r' = (-1)^i \bar R_i} \atop {r = (-1)^j \bar R_j}} \,.
\label{G^ij def}
\end{equation}
\eqnref{S_B Greens} gives the boundary action in terms of
$\tilde\psi_{b,i}(\nu)$ and the functions $G_\psi^{ij}(\nu)$
which are evaluated in Appendix~\ref{G^ij appendix}
(Eqs.~\eqnlessref{G^ij_theta} and \eqnlessref{G^ij_phi}).
The term involving $G^{ij}_\psi$ in \eqnlessref{S_B Greens}
represents the ``back reaction'' of the interior degrees of
freedom to the boundary variables.

Inserting \eqnlessref{L hat in text} into \eqnlessref{S_B Greens}
gives the boundary actions $S_{B,\theta}$ and $S_{B,\phi}$,
\begin{eqnarray}
S_{B,\theta} &=& \int \frac{d\nu}{2\pi} \Bigg\{ \sum_{i=1}^2
\frac{1}{2} m_i \bar R_i^2 \bar\gamma_i
\left( \nu^2 - \omega^2 \right) \left|\tilde\theta_{b,i}(\nu)\right|^2
- \frac{\sigma}{2} \sum_{i,j=1}^2 \tilde\theta^*_{b,i}(\nu) G_\theta^{ij}(\nu)
\tilde\theta_{b,j}(\nu)
\nonumber \\
& & + \sum_{i=1}^2 (-1)^i \hbox{$\tilde\rho^i_\theta$}^{\negthinspace*}(\nu)
\bar R_i \tilde\theta_{b,i}(\nu) \Bigg\} \,,
\label{theta boundary action}
\end{eqnarray}
and
\begin{eqnarray}
S_{B,\phi} &=& \int \frac{d\nu}{2\pi} \Bigg\{ \sum_{i=1}^2 m_i \bar\gamma_i
\Bigg[ \frac{1}{2} \left(\nu^2 + \left( 2\bar\gamma_i^2 - 1 \right)
\omega^2 \right) \left|\tilde r_i(\nu)\right|^2
- 2 \bar R_i \omega \bar\gamma_i^2 \nu
\Im\left(\tilde r_i^*(\nu) \tilde\phi_{b,i}(\nu)\right)
\nonumber \\
& & + \frac{1}{2} \bar R_i^2 \bar\gamma_i^2 \nu^2
\left|\tilde\phi_{b,i}(\nu)\right|^2 \Bigg]
+ \sum_{i=1}^2 \frac{(-1)^i}{2} \Bigg[
\hbox{$\tilde\rho^i_\phi$}^{\negthinspace*}(\nu + \omega)
\left( \bar R_i \tilde\phi_{b,i}(\nu) - i \tilde r_i(\nu) \right)
\nonumber \\
& & + \hbox{$\tilde\rho^i_\phi$}^{\negthinspace*}(\nu - \omega)
\left( \bar R_i \tilde\phi_{b,i}(\nu) + i \tilde r_i(\nu) \right) \Bigg]
- \frac{\sigma}{2} \sum_{i,j=1}^2 \tilde\phi^*_{b,i}(\nu) G_\phi^{ij}(\nu)
\tilde\phi_{b,j}(\nu) \Bigg\} \,.
\label{phi boundary action}
\end{eqnarray}
We have introduced the Fourier transforms of $r_i$ and the
sources $\rho_\theta^i$ and $\rho_\phi^i$,
\begin{equation}
\tilde r_i(\nu) \equiv \int dt e^{-i\nu t} r_i(t) \,,
\kern 0.35 in
\tilde\rho_\theta^i(\nu) \equiv \int dt e^{-i\nu t} \rho_\theta^i(t) \,,
\kern 0.35 in
\tilde\rho_\phi^i(\nu) \equiv \int dt e^{-i\nu t - i\delta_i(t)} \rho_\phi^i(t) \,.
\end{equation}
The function $\tilde\rho_\phi^i(\nu)$ incorporates two degrees of freedom,
$\rho_\phi^i(t)$ and $\delta_i(t)$, so it is an arbitrary complex function.
The other two Fourier transforms satisfy the reality conditions
\begin{equation}
\tilde r_i(-\nu) = \tilde r_i^*(\nu) \,, \kern 1 in
\tilde\rho_\theta^i(-\nu) = \hbox{$\tilde\rho^i_\theta$}^{\negthinspace*}(\nu) \,.
\end{equation}

Using the definition \eqnlessref{Sigma def} of $\Sigma(r)$ and $C$
in \eqnlessref{L_I def} gives the interior actions
\begin{equation}
S_{I,\theta} = -\frac{\sigma}{2} \int \frac{d\nu}{2\pi}
\int_{\bar R_1}^{\bar R_2} dr \tilde\theta_I^* \left[
-\frac{\partial}{\partial r} \gamma^{-1} r^2 \frac{\partial}{\partial r}
- (\nu^2 - \omega^2) \gamma r^2 \right] \tilde\theta_I \,,
\end{equation}
and
\begin{equation}
S_{I,\phi} = -\frac{\sigma}{2} \int \frac{d\nu}{2\pi}
\int_{\bar R_1}^{\bar R_2} dr \tilde\phi_I^* \left[
-\frac{\partial}{\partial r} \gamma r^2 \frac{\partial}{\partial r}
- \nu^2 \gamma^3 r^2 \right] \tilde\phi_I \,.
\end{equation}
The total action is the sum of the independent contributions,
\begin{equation}
\int dt (L + L_{{\rm source}}) = \int dt L_{{\rm cl}}
+ S_{I,\theta} + S_{I,\phi} + S_{B,\theta} + S_{B,\phi} \,,
\end{equation}
and the partition function $Z(\omega)$ is a corresponding product,
\begin{equation}
Z(\omega) = e^{iTL_{{\rm cl}}(\omega)} Z_I(\omega) Z_B(\omega) \,.
\label{partition I B factor}
\end{equation}

The interior partition function is
\begin{equation}
Z_I(\omega) = \int \scrD\theta_I e^{iS_{I,\theta}} \int \scrD\phi_I e^{iS_{I,\phi}}
\Det\left[\frac{r^2}{\sqrt{-g}}\right] \,,
\label{Z_I def}
\end{equation}
where $\Det[r^2 / \sqrt{-g}]$ is replaced by its classical value.
The boundary partition function is
\begin{equation}
Z_B(\omega) = \frac{1}{Z_b} \int \prod_{i=1}^2
\left[\scrD\theta_{b,i} \scrD\phi_{b,i} \scrD r_i \Det[\bar R_i^2]\right]
e^{iS_{B,\theta} + iS_{B,\phi}} \,.
\label{Z_B def}
\end{equation}
We will evaluate these two parts separately in the next two sections.

\section{Evaluation of the Interior Partition Function $Z_I(\omega)$}

\label{interior integral section}

In this section, we evaluate $Z_I(\omega)$ and derive \eqnref{L fluc pre renorm}
for $L^{{\rm string}}_{{\rm fluc}}$, generalizing the results of~\cite{Baker+Steinke2}
to the case where the quark masses are
unequal. The interior action depends exclusively on
the functions $\tilde\theta_I(r,\nu)$ and $\tilde\phi_I(r,\nu)$,
which vanish at $r = (-1)^i \bar R_i$. We simplify
$S_{I,\theta}$ and $S_{I,\phi}$ by changing coordinates from $r$ to
\begin{equation}
x = \frac{1}{\omega} \arcsin \omega r \,.
\label{x coord def}
\end{equation}
We also change our integration variables $\tilde\theta_I$ and
$\tilde\phi_I$ to differently normalized functions,
\begin{eqnarray}
\tilde\theta_I(r, \nu) &=& \frac{1}{r} k(x,t) \,,
\nonumber \\
\tilde\phi_I(r, \nu) &=& \frac{1}{\gamma r} f(x,t) \,.
\label{f and k def}
\end{eqnarray}
The components of the action become
\begin{eqnarray}
S_{I,\theta} = -\frac{\sigma}{2} \int \frac{d\nu}{2\pi} \int_{-X_1}^{X_2} dx
k^* \left[ - \frac{\partial^2}{\partial x^2} - \nu^2 \right] k \,,
\nonumber \\
S_{I,\phi} = -\frac{\sigma}{2} \int \frac{d\nu}{2\pi} \int_{-X_1}^{X_2} dx
f^* \left[ - \frac{\partial^2}{\partial x^2} + 2 \omega^2 \sec^2 \omega x
 - \nu^2 \right] f \,,
\label{interior X}
\end{eqnarray}
where the limits of integration are
\begin{eqnarray}
X_i &=& \frac{1}{\omega} \arcsin \omega \bar R_i \,.
\label{X_i def}
\end{eqnarray}

In terms of the new variables \eqnlessref{f and k def}, the
interior partition function \eqnlessref{Z_I def} is
\begin{equation}
Z_I(\omega) = \int \scrD k \scrD f e^{iS_{I,\theta} + iS_{I,\phi}} \,.
\end{equation}
Doing the integrals over $f$ and $k$ gives
\begin{equation}
Z_I(\omega) = \Det^{-1/2}\left[-\nabla^2\right] \Det^{-1/2}\left[-\nabla^2
+ 2\omega^2 \sec^2 \omega x\right] \,,
\label{Z_I determinants}
\end{equation}
where $-\nabla^2$ is the Laplacian in the $x, t$ coordinate system.
This coordinate system is conformally flat, i.e. $g_{xx} = - g_{tt}$,
$g_{xt} = 0$, so we can use the result of
L\"uscher~\cite{Luscher1}, that for a static string in the large time limit,
\begin{equation}
\Det^{-1}\left[-\nabla^2\right] = e^{-\frac{\pi}{12} \frac{T}{R}}
\kern 1 in \hbox{: static quark background,}
\label{Luscher determinant}
\end{equation}
where $R$ is the length of the string and $T$ is the time elapsed.
In \eqnref{interior X}, the string length $X_1 + X_2$ obtained
from \eqnlessref{X_i def} is $R_p$, given by \eqnlessref{proper length},
which is the ``proper length'' of a relativistic rotating
string. Making the replacement $R = R_p$ in \eqnlessref{Luscher determinant}
gives
\begin{equation}
\Det^{-1}\left[-\nabla^2\right] = e^{-\frac{\pi}{12} \frac{T}{R_p}}
\kern 1 in \hbox{: rotating quark background.}
\end{equation}
We therefore see that
\begin{equation}
Z_I(\omega) = e^{-\frac{\pi}{12} \frac{T}{R_p}}
\Det^{-1/2} \left[\frac{-\nabla^2 + 2\omega^2 \sec^2 \omega x}{-\nabla^2}\right]
\equiv e^{-T L^{{\rm string}}_{{\rm fluc}}(\omega)} \,,
\label{Z_I after Luscher}
\end{equation}
so that
\begin{equation}
L^{{\rm string}}_{{\rm fluc}}(\omega) = \frac{\pi}{12R_p} - \frac{i}{T} \Tr\log\left[
\frac{-\nabla^2 + 2\omega^2 \sec^2 \omega x}{-\nabla^2}\right] \,,
\label{L_I tr log}
\end{equation}
is the contribution of the interior degrees of freedom to $L_{{\rm fluc}}$.

We can express the trace in \eqnlessref{L_I tr log} in terms of the
eigenvalues $\mu_n$ and $\lambda_n$ determined by the spatial boundary
problems
\begin{eqnarray}
\left(-\frac{\partial^2}{\partial x^2} + 2\omega^2 \sec^2 \omega x\right)
f_n(x) &=& \mu_n f_n(x) \,,
\nonumber \\
-\frac{\partial^2}{\partial x^2} k_n(x) &=& \lambda_n k_n(x) \,,
\label{trace eigenvalue difeqs}
\end{eqnarray}
where $-X_1 < x < X_2$, and $f_n(x)$ and $k_n(x)$ vanish at the
boundaries. The differential equations \eqnlessref{trace eigenvalue difeqs}
are identical to Eqs.~\eqnlessref{phi difeq} and \eqnlessref{theta difeq},
and the eigenvalues $\mu_n$ and $\lambda_n$, as well as the corresponding
eigenfunctions, are given by Eqs. \eqnlessref{theta eigenfunctions},
\eqnlessref{appendix theta eigenvalues}, \eqnlessref{phi eigenfunctions},
and \eqnlessref{phi eigenvalue def}. The eigenvalues are obtained from the
equations
\begin{eqnarray}
\tan\left(\sqrt{\mu_n}R_p\right) &=& \sqrt{\mu_n} \omega
\frac{v_1 \bar\gamma_1 + v_2 \bar\gamma_2}
{\mu_n - \omega^2 v_1 \bar\gamma_1 v_2 \bar\gamma_2} \,,
\nonumber \\
\sqrt{\lambda_n} &=& \frac{\pi n}{R_p} \,.
\end{eqnarray}
The solution $\mu_n = \omega^2$ to the first of these two equations does
not produce a valid eigenvalue, as the corresponding eigenfunction
vanishes everywhere.

Taking a Fourier transform in time and performing a Wick rotation on
\eqnlessref{L_I tr log} gives
\begin{eqnarray}
L^{{\rm string}}_{{\rm fluc}}(\omega) &=& \frac{\pi}{12 R_p}
+ \int \frac{d\nu}{2\pi} \sum_{n=1}^{\Lambda R_p}{\pi}
\ln\left[\frac{\nu^2 + \mu_n}{\nu^2 + \lambda_n}\right]
\nonumber \\
&=& \frac{\pi}{12 R_p} +  \sum_{n=1}^{\Lambda R_p}{\pi} \left[ \sqrt{\mu_n}
- \frac{\pi n}{R_p} \right] \,.
\label{L_I sum}
\end{eqnarray}
The sum over $n$ is logarithmically divergent, so we have imposed a cutoff,
restricting ourselves to spatial eigenvalues less than $\Lambda$.
Since our evaluation of the determinants took place in the $x$
coordinate system, this is a cutoff in the $x$ coordinate
space. It is related to the cutoff $M$ in the $r$ coordinate space
by the equation
\begin{equation}
\Lambda = M \frac{\partial r}{\partial x} = M \gamma^{-1} \,.
\label{cutoff coordinate change}
\end{equation}

Using the methods of~\cite{Baker+Steinke2} to
evaluate the sum \eqnlessref{L_I sum} gives the
result \eqnlessref{L fluc pre renorm} for
$L^{{\rm string}}_{{\rm fluc}}(\omega)$.

\section{Evaluating the Boundary Partition Function $Z_B(\omega)$}

\label{boundary integral section}

We begin our evaluation of $Z_B(\omega)$ \eqnlessref{Z_B def} by writing
down the explicit form of the partition function.
Inserting the expressions \eqnlessref{theta boundary action}
and \eqnlessref{phi boundary action} gives $Z_B$ the form
\begin{eqnarray}
Z_B(\omega) &=& \frac{1}{Z_b} \int \prod_{i=1}^2
\left[\scrD\tilde\theta_{b,i} \scrD\tilde\phi_{b,i} \scrD \tilde r_i
\Det[\bar R_i^2]\right]
\nonumber \\
& & \times \exp\Bigg\{ i \int \frac{d\nu}{2\pi} \Bigg[
- \frac{\sigma}{2} \sum_{i,j=1}^2 \left(
\tilde\theta^*_{b,i}(\nu) G_\theta^{ij}(\nu) \tilde\theta_{b,j}(\nu)
+ \tilde\phi^*_{b,i}(\nu) G_\phi^{ij}(\nu) \tilde\phi_{b,j}(\nu) \right)
\nonumber \\
& & + \sum_{i=1}^2 m_i \bar\gamma_i
\Bigg( \frac{1}{2} \left(\nu^2 + \left( 2\bar\gamma_i^2 - 1 \right)
\omega^2 \right) \left|\tilde r_i(\nu)\right|^2
- 2 \bar R_i \omega \bar\gamma_i^2 \nu
\Im\left(\tilde r_i^*(\nu) \tilde\phi_{b,i}(\nu)\right)
\nonumber \\
& & + \frac{1}{2} \bar R_i^2 \left( \nu^2 - \omega^2 \right)
\left|\tilde\theta_{b,i}(\nu)\right|^2
+ \frac{1}{2} \bar R_i^2 \bar\gamma_i^2 \nu^2
\left|\tilde\phi_{b,i}(\nu)\right|^2 \Bigg)
\nonumber \\
& & + \sum_{i=1}^2 (-1)^i \Bigg[
\hbox{$\tilde\rho^i_\theta$}^{\negthinspace*}(\nu)
\bar R_i \tilde\theta_{b,i}(\nu)
+ \frac{1}{2} \hbox{$\tilde\rho^i_\phi$}^{\negthinspace*}(\nu + \omega)
\left( \bar R_i \tilde\phi_{b,i}(\nu) - i \tilde r_i(\nu) \right)
\nonumber \\
& & + \frac{1}{2} \hbox{$\tilde\rho^i_\phi$}^{\negthinspace*}(\nu - \omega)
\left( \bar R_i \tilde\phi_{b,i}(\nu) + i \tilde r_i(\nu) \right)
\Bigg] \Bigg\} \,.
\label{partition quadratic}
\end{eqnarray}

Doing the integral over the $\tilde r_i$ gives
\begin{eqnarray}
& & \int \prod_{i=1}^2 \scrD \tilde r_i \exp\Bigg\{ i \int \frac{d\nu}{2\pi}
\sum_{i=1}^2 m_i \bar\gamma_i \Bigg[
\frac{1}{2} \left(\nu^2 + \left( 2\bar\gamma_i^2 - 1 \right)
\omega^2 \right) \left|\tilde r_i(\nu)\right|^2
\nonumber \\
& & - 2 \bar R_i \omega \bar\gamma_i^2 \nu 
\Im\left(\tilde r_i^*(\nu) \tilde\phi_{b,i}(\nu)\right)
+ \frac{(-1)^i}{m_i \bar\gamma_i} \left(
-i \hbox{$\tilde\rho^i_\theta$}^{\negthinspace*}(\nu + \omega)
+i \hbox{$\tilde\rho^i_\theta$}^{\negthinspace*}(\nu - \omega)
\right) \tilde r_i(\nu) \Bigg] \Bigg\}
\nonumber \\
&=& \Det^{-1/2}\left[\prod_{i=1}^2 m_i \bar\gamma_i \left( \nu^2
+ \left( 2\bar\gamma_i^2 - 1 \right) \omega^2 \right) \right]
\nonumber \\
& & \times \exp\Bigg\{i \int \frac{d\nu}{2\pi} \sum_{i=1}^2 \Bigg[
- 2 m_i \bar R_i^2 \bar\gamma_i^5 \frac{\nu^2\omega^2}
{\nu^2 + (2\bar\gamma_i^2 - 1) \omega^2}
\left|\tilde\phi_{b,i}(\nu)\right|^2
\nonumber \\
& & + \sum_{i=1}^2 \frac{(-1)^i}{2} \bar R_i \Re \Bigg(
\hbox{$\tilde\rho^i_\phi$}^{\negthinspace*}(\nu + \omega)
\frac{- 2\bar\gamma_i^2 \omega \nu}
{\nu^2 + (2\bar\gamma_i^2 - 1)\omega^2}
+ \hbox{$\tilde\rho_\phi^i$}^{\negthinspace*}(\nu - \omega)
\frac{2\bar\gamma_i^2 \omega \nu}
{\nu^2 + (2\bar\gamma_i^2 - 1)\omega^2}
\Bigg) \tilde\phi_{b,i}(\nu)
\nonumber \\
& & - \frac{1}{8} \sum_{i=1}^2
\frac{1}{m_i \bar\gamma_i \bar R_i^2}
\left(\nu^2 + \left(2\bar\gamma_i^2 - 1\right) \omega^2\right)^{-1}
\left| \tilde\rho_\phi^i(\nu + \omega)
- \tilde\rho_\phi^i(\nu - \omega) \right|^2 \Bigg] \Bigg\} \,.
\label{r_i integral done}
\end{eqnarray}

Inserting \eqnlessref{r_i integral done} into \eqnlessref{partition quadratic},
and using the fact that $Z_b^{-1}$ is equal to $\Det^3[\nu^2]$, up to an
overall constant, gives the following expression for $Z_B(\omega)$,
\begin{eqnarray}
Z_B(\omega) &=& \left( \hbox{const.} \right) \int \prod_{i=1}^2
\left[\scrD\tilde\theta_{b,i} \scrD\tilde\phi_{b,i} \right]
\Det^{-1/2}\left[\prod_{i=1}^2 \left( \nu^2
+ \left( 2\bar\gamma_i^2 - 1 \right) \omega^2 \right) \right]
\Det^3\left[\nu^2\right]
\nonumber \\
& & \times \exp\Bigg\{ i \int \frac{d\nu}{2\pi} \Bigg[
- \frac{1}{2} \sum_{i,j=1}^2 \left(
\tilde\theta^*_{b,i}(\nu) {\Gamma_\theta^{ij}}^{-1}(\nu) \tilde\theta_{b,j}(\nu)
+ \tilde\phi^*_{b,i}(\nu) {\Gamma_\phi^{ij}}^{-1}(\nu) \tilde\phi_{b,j}(\nu) \right)
\nonumber \\
& & + \sum_{i=1}^2 (-1)^i \bar R_i \Re\Bigg(
\frac{1}{2} \Bigg(\hbox{$\tilde\rho^i_\phi$}^{\negthinspace*}(\nu + \omega)
\frac{\nu^2 - \omega^2 - 2\bar\gamma_i^2 \omega (\nu - \omega)}
{\nu^2 + (2\bar\gamma_i^2 - 1)\omega^2}
\nonumber \\
& & + \hbox{$\tilde\rho_\phi^i$}^{\negthinspace*}(\nu - \omega)
\frac{\nu^2 - \omega^2 + 2\bar\gamma_i^2 \omega (\nu + \omega)}
{\nu^2 + (2\bar\gamma_i^2 - 1)\omega^2} \Bigg)
\tilde\phi_{b,i}(\nu)
+ \left(\hbox{$\tilde\rho^i_\theta$}^{\negthinspace*}(\nu)
\tilde\theta_{b,i}(\nu)\right) \Bigg)
\nonumber \\
& & - \frac{1}{8} \sum_{i=1}^2
\frac{1}{m_i \bar\gamma_i \bar R_i^2}
\left(\nu^2 + \left(2\bar\gamma_i^2 - 1\right) \omega^2\right)^{-1}
\left| \tilde\rho_\phi^i(\nu + \omega)
- \tilde\rho_\phi^i(\nu - \omega) \right|^2 \Bigg] \Bigg\} \,,
\label{Z after r int}
\end{eqnarray}
where
\begin{equation}
{\Gamma^{ij}_\theta}^{-1}(\nu) = \delta_{ij} m_i \bar\gamma_i \bar R_i^2
\left(\nu^2 - \bar\omega^2\right) - \sigma G^{ij}_\theta(\nu) \,,
\label{Gamma theta def}
\end{equation}
and
\begin{equation}
{\Gamma^{ij}_\phi}^{-1}(\nu) = \delta_{ij} m_i \bar\gamma_i^3
\bar R_i^2 \nu^2 \frac{\nu^2 - \left(2\bar\gamma_i^2 + 1\right) \omega^2}
{\nu^2 + \left(2\bar\gamma_i^2 - 1\right) \omega^2} - \sigma G^{ij}_\phi(\nu) \,.
\label{Gamma phi def}
\end{equation}

The quantities ${\Gamma^{ij}}^{-1}$ are the coefficients of the
quadratic terms in $\tilde\theta_{b,i}$
and $\tilde\phi_{b,i}$. They determine the contribution
of the boundary fluctuations to the string energy, and are
closely related to the physical propagator for driven
oscillations of the string modes.
Inserting the explicit forms \eqnlessref{G^ij_theta} and
\eqnlessref{G^ij_phi} of $G^{ij}_\theta$ and $G^{ij}_\phi$,
and using \eqnlessref{classical eqn} to replace $m_i$, we find
\begin{equation}
{\Gamma^{ij}_\theta}^{-1}(\nu) = \frac{\nu\sigma v_i v_j}{\omega^2}
\left( \delta_{ij} \left(\frac{\nu}{\omega v_i \bar\gamma_i} -
\cot\left(\nu R_p\right)\right)
+ \left(1 - \delta_{ij}\right) \csc\left(\nu R_p\right) \right) \,,
\label{Gamma theta inv expand}
\end{equation}
and
\begin{eqnarray}
{\Gamma^{ij}_\phi}^{-1}(\nu) &=& \delta_{ij} \frac{2\sigma \nu^2 v_i \bar\gamma_i
\left(\nu^2 - \omega^2\right)}
{\omega^3 \left(\nu^2 + (2\bar\gamma_i^2 - 1) \omega^2\right)}
- \frac{\sigma\nu}{\omega^3} (\nu^2 - \omega^2)
\nonumber \\
& & \times \frac{ \delta_{ij} \left(\nu v_i \bar\gamma_i \sin(\nu R_p)
- \omega v_1 \bar\gamma_1 v_2 \bar\gamma_2 \cos(\nu R_p)\right)
+ (1-\delta_{ij}) \omega v_i \bar\gamma_1 v_2 \bar\gamma_2}
{\left(\nu^2 - \omega^2 v_1 \bar\gamma_1 v_2 \bar\gamma_2\right)
\sin(\nu R_p) - \nu \omega \left(v_1\bar\gamma_1 + v_2\bar\gamma_2\right)
\cos(\nu R_p)} \,.
\label{Gamma phi inv expand}
\end{eqnarray}

Doing the $\tilde\theta_{b,i}$ and $\tilde\phi_{b,i}$ integrals
in \eqnlessref{Z after r int} gives
\begin{equation}
Z_B(\omega) = e^{iS_{{\rm boundary}} + iS_{{\rm sources}}} \,,
\label{Z_B eval}
\end{equation}
where
\begin{eqnarray}
e^{iS_{{\rm boundary}}} &=&
\Det^{-1/2}\left[\prod_{i=1}^2 \left( \nu^2
+ \left( 2\bar\gamma_i^2 - 1 \right) \omega^2 \right) \right]
\frac{\Det^{-1/2}\left[{\Gamma^{ij}_\theta}^{-1}\right]}
{\Det^{-1/2}\left[\delta_{ij} m_i \bar R_i^2 \bar\gamma_i\right]}
\nonumber \\
& & \times \frac{\Det^{-1/2}\left[{\Gamma^{ij}_\phi}^{-1}\right]}
{\Det^{-1/2}\left[\delta_{ij} m_i \bar R_i^2 \bar\gamma_i^3\right]}
\Det^3\left[\nu^2\right] \,,
\end{eqnarray}
defines the normalized boundary action. In the limit of large
elapsed time $T$, $S_{{\rm boundary}}$ is proportional to $T$,
so we define the effective boundary Lagrangian
\begin{equation}
L_{{\rm boundary}}(\omega) = \lim_{T\to\infty} \frac{-i}{T}
S_{{\rm boundary}}(\omega) \,.
\label{L boundary def}
\end{equation}
We evaluate $L_{{\rm boundary}}(\omega)$ in Appendix~\ref{boundary appendix}.

The source terms in \eqnlessref{Z_B eval} are,
\begin{eqnarray}
\negspace
S_{{\rm sources}} &=& \sum_{i,j=1}^2 \int \frac{d\nu}{2\pi}
(-1)^{i+j} \bar R_i \bar R_j \Bigg[ \frac{1}{2}
\hbox{$\tilde\rho_\theta^i$}^{\negthinspace*}(\nu)
\Gamma^{ij}_\theta(\nu) \tilde\rho_\theta^j(\nu)
+ \frac{1}{8} \Gamma^{ij}_\phi(\nu)
\nonumber \\
& & \times \left(\hbox{$\tilde\rho^i_\phi$}^{\negthinspace*}(\nu + \omega)
\frac{\nu^2 - \omega^2 - 2\bar\gamma_j^2 \omega (\nu - \omega)}
{\nu^2 + (2\bar\gamma_j^2 - 1)\omega^2}
+ \hbox{$\tilde\rho_\phi^i$}^{\negthinspace*}(\nu - \omega)
\frac{\nu^2 - \omega^2 + 2\bar\gamma_j^2 \omega (\nu + \omega)}
{\nu^2 + (2\bar\gamma_j^2 - 1)\omega^2} \right)
\negspace
\nonumber \\
& & \times \left(\tilde\rho_\phi^j(\nu + \omega)
\frac{\nu^2 - \omega^2 - 2\bar\gamma_i^2 \omega (\nu - \omega)}
{\nu^2 + (2\bar\gamma_i^2 - 1)\omega^2}
+ \tilde\rho_\phi^j(\nu - \omega)
\frac{\nu^2 - \omega^2 + 2\bar\gamma_i^2 \omega (\nu + \omega)}
{\nu^2 + (2\bar\gamma_i^2 - 1)\omega^2} \right)
\negspace
\nonumber \\
& & + \frac{\delta_{ij}}{m_i \bar\gamma_i \bar R_i^2}
\left(\nu^2 + \left(2\bar\gamma_i^2 - 1\right) \omega^2\right)^{-1}
\frac{1}{8} \left| \tilde\rho_\phi^i(\nu + \omega)
- \tilde\rho_\phi^i(\nu - \omega) \right|^2
\Bigg] \,.
\label{S source def}
\end{eqnarray}
The first two terms in \eqnlessref{S source def} were produced by the
$\tilde\theta_{b,i}$ and $\tilde\phi_{b,i}$ integrals. The third
was produced by the $\tilde r_i$ integral, and does not
couple the two ends of the string to each other.

\section{Explicit Evaluations of $L_{{\rm \lowercase{fluc}}}$
	and $S_{{\rm \lowercase{sources}}}$}

\label{quark mass limit section}

Combining Eqs. \eqnlessref{partition I B factor}, \eqnlessref{Z_I after Luscher},
and \eqnlessref{Z_B eval} gives
\begin{equation}
Z(\omega) = e^{iTL_{{\rm cl}}(\omega) + L_{{\rm fluc}}(\omega)
+ iS_{{\rm sources}}} \,,
\end{equation}
where $L_{{\rm cl}}(\omega)$ is defined by
\eqnlessref{L_cl def formal}, $S_{{\rm sources}}$ is defined by
\eqnlessref{S source def}, and $L_{{\rm fluc}}$ is
\begin{equation}
L_{{\rm fluc}}(\omega) = L^{{\rm string}}_{{\rm fluc}}(\omega)
+ L_{{\rm boundary}}(\omega) \,.
\label{L fluc expand}
\end{equation}
The terms on the right hand side of \eqnlessref{L fluc expand} are defined by
equations \eqnlessref{L fluc renorm} and \eqnlessref{L boundary def}.

We now evaluate the partition function for appropriate physical
limits of the quark masses. As we previously discussed
in Section~\ref{geodesic curvature section}, due to the renormalization of the
geodesic curvature our results are only valid when the
quark masses are either very large or exactly zero.
The case where both masses are large is 
relevant only to the evaluation of the potential.
We therefore have two physical limits: (1) the light--light
case, where $m_1 = m_2 = 0$ ($\bar\gamma_1,\bar\gamma_2 \to \infty$), and
(2) the heavy--light case, where $m_1 \to\infty$ ($v_1 \to 0$)
and $m_2 = 0$ ($\bar\gamma_2 \to\infty$). We now evaluate
$L_{{\rm fluc}}(\omega)$ and $S_{{\rm sources}}$ in these limits.
This will give us the zero point energy and the excitation energies of the
fluctuations.

We begin with $L_{{\rm fluc}}$, and its two
parts $L^{{\rm string}}_{{\rm fluc}}$ and $L_{{\rm boundary}}$.
In the light--light limit $L^{{\rm string}}_{{\rm fluc}}$ is
$(7/12) \omega$, and in the heavy--light limit it is $(5/12) \omega$
(see Eqs. \eqnlessref{L fluc ll} and \eqnlessref{L fluc hl}).
We evaluate $L_{{\rm boundary}}$ in Appendix~\ref{boundary appendix},
and find in the light--light limit
\begin{equation}
L_{{\rm boundary}} = 0 \,,
\end{equation}
and in the heavy--light limit
\begin{equation}
L_{{\rm boundary}} = -\frac{1}{4} \omega \,.
\end{equation}
Thus, in the light--light limit,
\begin{equation}
L_{{\rm fluc}}(\omega) = \frac{7}{12} \omega \,,
\label{L fluc ll num}
\end{equation}
and in the heavy--light limit,
\begin{equation}
L_{{\rm fluc}}(\omega) = \frac{1}{6} \omega \,.
\label{L fluc hl num}
\end{equation}

We next evaluate $S_{{\rm sources}}$ in these limits. Consider
first the excitations in the $\theta$ sector. We set $\rho_\phi^i$
equal to zero in \eqnlessref{S source def} to obtain
the the $\theta$ sector propagator,
\begin{equation}
K^{ij}_\theta(\nu) \equiv (-1)^{i+j} \bar R_i \bar R_j \Gamma_\theta^{ij}(\nu) \,.
\label{theta propagator def}
\end{equation}
The propagator $K^{ij}_\theta(\nu)$ has a simple pole wherever $\nu$
is equal to the energy of one of the excited modes. There is also
a double pole at $\nu = 0$ due to the invariance of the Lagrangian under
a translation of the string in the direction perpendicular to the plane
of rotation. This translation mode does not correspond to an excited
state of the string.

A general excited state will include multiple excitations of each mode,
so its energy will be a sum of multiples of the energy of each mode. Only
single excitations will appear in the propagator $K^{ij}_\theta(\nu)$,
because of the harmonic oscillator selection rules. We obtain an explicit form for
$K_\theta^{ij}(\nu)$ by inverting ${\Gamma_\theta^{ij}(\nu)}^{-1}$
(in the sense of inverting a two by two matrix).
In the light--light case, \eqnlessref{Gamma theta inv expand} becomes
\begin{equation}
{\Gamma^{ij}_\theta}^{-1}(\nu) = \frac{\nu\sigma}{\omega^2}
\left(-\delta_{ij} \cot\left(\pi\frac{\nu}{\omega}\right)
+ \left(1 - \delta_{ij}\right) \csc\left(\pi\frac{\nu}{\omega}\right) \right) \,,
\end{equation}
so the $\theta$ propagator is
\begin{equation}
K^{ij}_\theta(\nu) = \frac{1}{\nu\sigma}
\left(\delta_{ij} \cot\left(\pi\frac{\nu}{\omega}\right)
- \left(1 - \delta_{ij}\right) \csc\left(\pi\frac{\nu}{\omega}\right) \right) \,.
\end{equation}
This has a double pole at $\nu = 0$ due to the translation
mode in the direction perpendicular to the plane of rotation,
which does not correspond to an excited string state.

The single poles at $\nu = k\omega$ for $\nu \ne 0$ are due
to the excited states of the string. These are the same poles
as appear in $G^{ij}_\theta(\nu)$ \eqnlessref{G^ij_theta}, and consequently in
${K^{ij}_\theta}^{-1}(\nu)$, in the limit of massless quarks.
The locations of these poles are the same in $K^{ij}_\theta(\nu)$
and ${K^{ij}_\theta}^{-1}(\nu)$ because $\det K^{ij}_\theta(\nu)
= -1/\nu^2\sigma^2$ in the massless quark limit.

In the heavy--light
case, the components of ${\Gamma^{ij}_\theta}^{-1}$ are
\begin{eqnarray}
{\Gamma^{11}_\theta}^{-1}(\nu) &=& \frac{\nu^2\sigma}{\omega^3} v_1 + O(v_1^2) \,,
\nonumber \\
{\Gamma^{22}_\theta}^{-1}(\nu) &=& -\frac{\nu\sigma}{\omega^2}
\cot\left(\frac{\pi\nu}{2\omega}\right) + O(v_1) \,,
\nonumber \\
{\Gamma^{12}_\theta}^{-1}(\nu) &=& \frac{\nu\sigma}{\omega^2} v_1
\csc\left(\frac{\pi\nu}{2\omega}\right) + O(v_1^2) \,.
\label{Gamma phi inv hl}
\end{eqnarray}
Inverting the two by two matrix \eqnlessref{Gamma phi inv hl}
gives, up to an overall normalization, the components of the $\theta$ propagator,
\begin{eqnarray}
K^{11}_\theta(\nu) &=& \frac{\omega}{\nu^2\sigma} v_1 + O(v_1^2) \,,
\nonumber \\
K^{22}_\theta(\nu) &=& -\frac{1}{\nu\sigma}
\tan\left(\frac{\pi\nu}{2\omega}\right) + O(v_1) \,,
\nonumber \\
K^{12}_\theta(\nu) &=& \frac{\omega}{\nu^2\sigma}
\sec\left(\frac{\pi\nu}{2\omega}\right) v_1 + O(v_1^2) \,.
\end{eqnarray}

We see that, in the limit $m_1 \to \infty$ ($v_1 \to 0$),
only the light--light component $K^{22}_\theta(\nu)$ is nonvanishing.
It has poles at
\begin{equation}
\nu = (2k+1) \omega \,, \kern 1 in k = 0, 1, 2,... \,,
\end{equation}
corresponding to the normal modes of a string with one end fixed.
The remaining components of the propagator are proportional to the
heavy quark velocity. The heavy--heavy component $K^{11}_\theta(\nu)$
contains only a double pole at $\nu = 0$, corresponding to the translation
mode, and the heavy--light component $K^{12}_\theta(\nu)$ has poles
corresponding both to excited vibrational states and to the translation mode.

We next obtain the $\phi$ sector propagator by examining \eqnlessref{S source def}.
With some rearranging of terms, \eqnlessref{S source def} is
\begin{eqnarray}
S_{{\rm sources}} &=& \sum_{i,j=1}^2 \int \frac{d\nu}{2\pi}
\Bigg\{ \frac{1}{2} \hbox{$\tilde\rho_\theta^i$}^{\negthinspace*}(\nu)
K^{ij}_\theta(\nu) \tilde\rho_\theta^j(\nu)
+ \frac{1}{2} \hbox{$\tilde\rho^i_\phi$}^{\negthinspace*}(\nu)
K^{ij}_\phi(\nu) \tilde\rho^j_\phi(\nu)
\nonumber \\
& & + \frac{1}{4} \Re \Bigg[ \hbox{$\tilde\rho^i_\phi$}^{\negthinspace*}(\nu + \omega)
\Bigg( (-1)^{i+j} \bar R_i \bar R_j
\frac{\nu^2 - \omega^2 - 2\bar\gamma_i^2\omega(\nu - \omega)}
{\nu^2 + (2\bar\gamma_i^2 - 1)\omega^2}
\Gamma^{ij}_\phi(\nu)
\nonumber \\
& & \times \frac{\nu^2 - \omega^2 + 2\bar\gamma_j^2\omega(\nu + \omega)}
{\nu^2 + (2\bar\gamma_j^2 - 1)\omega^2}
- \frac{\delta_{ij}}{m_i \bar\gamma_i}
\frac{1}{\nu^2 + (2\bar\gamma_i^2 - 1) \omega^2}
\Bigg)
\tilde\rho^j_\phi(\nu - \omega) \Bigg] \Bigg\} \,,
\label{S source translated}
\end{eqnarray}
where $K^{ij}_\phi(\nu)$ is,
\begin{eqnarray}
K^{ij}_\phi(\nu) &\equiv&
\frac{\delta_{ij}}{4 m_i \bar\gamma_i} \left(
\frac{1}{\nu^2 - 2\nu\omega + 2\bar\gamma_i^2 \omega^2}
+ \frac{1}{\nu^2 + 2\nu\omega + 2\bar\gamma_i^2 \omega^2} \right)
+ \frac{(-1)^{i+j}}{4} \bar R_i \bar R_j
\nonumber \\
& & \times \Bigg[
\frac{\nu^2 - 2 \nu\omega - 2\bar\gamma_i^2\omega(\nu - 2\omega)}
{\nu^2 - 2\nu\omega + 2\bar\gamma_i^2 \omega^2}
\frac{\nu^2 - 2 \nu\omega - 2\bar\gamma_j^2\omega(\nu - 2\omega)}
{\nu^2 - 2\nu\omega + 2\bar\gamma_j^2 \omega^2}
\Gamma^{ij}_\phi(\nu - \omega)
\nonumber \\
& & + \frac{\nu^2 + 2 \nu\omega + 2\bar\gamma_i^2\omega(\nu + 2\omega)}
{\nu^2 + 2\nu\omega + 2\bar\gamma_i^2 \omega^2}
\frac{\nu^2 + 2 \nu\omega + 2\bar\gamma_j^2\omega(\nu + 2\omega)}
{\nu^2 + 2\nu\omega + 2\bar\gamma_j^2 \omega^2}
\Gamma^{ij}_\phi(\nu + \omega)
\Bigg] \,.
\label{phi propagator def}
\end{eqnarray}
In some of the terms in \eqnlessref{S source translated} we have translated
the integration variable $\nu$ by $\pm\omega$ to make the
argument of $\tilde\rho_\phi^i$ be $\nu$ instead of $\nu\pm\omega$.
There are three terms in \eqnlessref{S source translated}:
\begin{enumerate}
\item The $K^{ij}_\theta(\nu)$ term. This is the $\theta$ sector
propagator which we examined earlier.
\item The $K^{ij}_\phi(\nu)$ term, where the arguments of the two
$\tilde\rho_\phi^i$ factors are equal. This is the
$\phi$ sector propagator.
\item The $\hbox{$\tilde\rho_\phi^i$}^{\negthinspace*}(\nu + \omega)
\tilde\rho_\phi^j(\nu - \omega)$ term, where the arguments of the two
$\tilde\rho_\phi^i$ factors differ by $2\omega$. Since the natural
frequency of the classical rotating string is $\omega$, driving
it at a frequency $\nu$ produces sidebands at $\nu \pm 2\omega$
(to leading semiclassical order). The third term in
\eqnlessref{S source translated} is a manifestation of this effect.
\end{enumerate}

The poles in $K^{ij}_\phi(\nu)$ give the energies of the excited
modes of the string. We now consider how the poles in
$K^{ij}_\phi(\nu)$ relate to the poles in $\Gamma^{ij}_\phi(\nu)$.
Since $\Gamma^{ij}_\phi$ appears in \eqnlessref{phi propagator def}
with the argument $\nu\pm\omega$, the double pole
in $K^{ij}_\phi$ at $\nu = 0$ due to the translation modes
of the string will appear in $\Gamma^{ij}_\phi$ as double
poles at $\nu = \pm\omega$. Likewise, the single poles
in $K^{ij}_\phi$ giving the energies of singly excited states of the
string will be shifted by $\omega$ when they appear
in $\Gamma^{ij}_\phi$.

In the light--light limit, ${\Gamma^{ij}_\phi}^{-1}$,
\eqnlessref{Gamma phi inv expand}, is
\begin{equation}
{\Gamma^{ij}_\phi}^{-1}(\nu) = \frac{\sigma\nu(\nu^2 - \omega^2)}{\omega^4}
\left( -\delta_{ij} \cot\left(\pi\frac{\nu}{\omega}\right)
+ (1 - \delta_{ij}) \csc\left(\pi\frac{\nu}{\omega}\right) \right) \,,
\end{equation}
so $\Gamma^{ij}_\phi$ is
\begin{equation}
\Gamma^{ij}_\phi(\nu) = \frac{\omega^4}{\sigma\nu(\nu^2 - \omega^2)}
\left( \delta_{ij} \cot\left(\pi\frac{\nu}{\omega}\right)
+ (1 - \delta_{ij}) \csc\left(\pi\frac{\nu}{\omega}\right) \right) \,.
\label{Gamma phi massless}
\end{equation}
The factors $\left[\nu(\nu^2 - \omega^2)\right]^{-1}$, combined with
the trigonometric functions in \eqnlessref{Gamma phi massless}, produce
double poles in $\Gamma^{ij}_\phi(\nu)$ at $\nu = 0, \pm\omega$,
and single poles at $\nu = k\omega$ for $k = \pm 2, \pm 3, ...$.

The $\phi$ sector propagator, written in terms of $\Gamma^{ij}_\phi$, is
\begin{equation}
K^{ij}_\phi(\nu) = \frac{(-1)^{i+j}}{4}
\left(
\frac{(\nu - 2\omega)^2}{\omega^4}
\Gamma^{ij}_\phi(\nu - \omega)
+ \frac{(\nu + 2\omega)^2}{\omega^4}
\Gamma^{ij}_\phi(\nu + \omega)
\right) \,.
\label{phi propagator ll}
\end{equation}
The double poles in $\Gamma^{ij}_\phi(\nu)$ at $\nu = \pm\omega$
do not produce poles in the propagator $K^{ij}_\phi(\nu)$ due to
the prefactors in \eqnlessref{phi propagator ll}. Instead, they
produce a double pole at $\nu = 0$. This double pole is the
effect of the two translation modes in the directions parallel to
the plane of string rotation, which are present in
the propagator for all values of the quark mass.

The double pole in $\Gamma^{ij}_\phi(\nu)$ at $\nu = 0$
produces double poles in $K^{ij}_\phi(\nu)$ at $\nu = \pm\omega$.
These poles arise for the following reason: the condition
\eqnlessref{phi linear term} that the average of the fluctuations
of the angular momentum of the string vanishes means that the zero
frequency component of the motion of the string is constrained.
This constraint removes from the motion the zero frequency component
of a global rotation. This global rotation is contained in the
$\tilde\phi_{b,i}(\nu)$ boundary degrees of freedom. However, we
have coupled all components of $\tilde\phi_{b,i}(\nu)$ to the
sources $\tilde\rho_\phi^i(\nu)$. Elimination of the coupling
to the zero frequency component of the global rotational mode
will remove the double poles of $K^{ij}_\phi(\nu)$ at $\nu = \pm\omega$.

The single poles of $\Gamma^{ij}_\phi(\nu)$ at $\nu = \pm k\omega$
for $k \ge 2$ produce single poles in the propagator $K^{ij}_\phi(\nu)$
at
\begin{equation}
\nu = \nu_k = k\omega \,, \hbox{ for } k = 1,2,... \,.
\end{equation}
These are the true vibrational frequencies of the motion of the string
in the plane of rotation, and give the energies of the excited states of the
mesons corresponding to the excitation of a single quanta of frequency
$\nu_k$. The spectrum of excited states in the $\phi$ sector is then the
same as in the $\theta$ sector, for a meson composed of zero mass quarks.

In the heavy--light case, \eqnlessref{Gamma phi inv expand} is
\begin{eqnarray}
{\Gamma^{11}_\phi}^{-1}(\nu) &=& \frac{\sigma}{\omega^3} \frac{(\nu^2 - \omega^2)^2}
{\nu^2 + \omega^2} v_1 + O(v_1^2) \,,
\nonumber \\
{\Gamma^{22}_\phi}^{-1}(\nu) &=& \frac{\sigma\nu(\nu^2-\omega^2)}{\omega^4}
\tan\left(\frac{\pi\nu}{2\omega}\right) + O(v_1) \,,
\nonumber \\
{\Gamma^{12}_\phi}^{-1}(\nu) &=& -\frac{\sigma}{\omega^3}
\left(\nu^2 - \omega^2\right)^2 v_1 \sec\left(\frac{\pi\nu}{2\omega}\right)
+ O(v_1^2) \,,
\end{eqnarray}
so the components of $\Gamma^{ij}_\phi$ are
\begin{eqnarray}
\Gamma^{11}_\phi(\nu) &=& \frac{\omega^3}{\sigma} \frac{\nu^2 + \omega^2}
{(\nu^2 - \omega^2)^2} v_1^{-1} + O(1) \,,
\nonumber \\
\Gamma^{22}_\phi(\nu) &=& \frac{\omega^4}{\sigma\nu(\nu^2-\omega^2)}
\cot\left(\frac{\pi\nu}{2\omega}\right) + O(v_1) \,,
\nonumber \\
\Gamma^{12}_\phi(\nu) &=& - \frac{\omega^4}{\sigma\nu}
\frac{\nu^2 + \omega^2}{\nu^2 - \omega^2}
\csc\left(\frac{\pi\nu}{2\omega}\right) + O(v_1) \,.
\end{eqnarray}
The components of the $\phi$ sector propagator are
\begin{eqnarray}
K^{11}_\phi(\nu) &=& \frac{\omega v_1}{4\sigma} \left(
\frac{1}{\nu^2 - 2\nu\omega + 2\omega^2} + \frac{1}{\nu^2 + 2\nu\omega + 2\omega^2}
\right)
\nonumber \\
& & + \frac{(\nu - 2\omega)^4 v_1^2}
{4 \omega^2 (\nu^2 - 2\nu\omega + 2\omega^2)^2}
 \Gamma^{11}_\phi(\nu - \omega)
+ \frac{(\nu + 2\omega)^4 v_1^2}
{4 \omega^2 (\nu^2 + 2\nu\omega + 2\omega^2)^2}
 \Gamma^{11}_\phi(\nu + \omega)
+ O(v_1^2)
\nonumber \\
K^{22}_\phi(\nu) &=& 
\frac{(\nu - 2\omega)^2}{4 \omega^4}
 \Gamma^{22}_\phi(\nu - \omega)
+ \frac{(\nu + 2\omega)^2}{4 \omega^4}
 \Gamma^{22}_\phi(\nu + \omega)
+ O(v_1)
\nonumber \\
K^{12}_\phi(\nu) &=&
\frac{(\nu - 2\omega)^3 v_1}
{4 \omega^3 (\nu^2 - 2\nu\omega + 2\omega^2)}
 \Gamma^{12}_\phi(\nu - \omega)
- \frac{(\nu + 2\omega)^3 v_1}
{4 \omega^3 (\nu^2 + 2\nu\omega + 2\omega^2)}
 \Gamma^{12}_\phi(\nu + \omega)
+ O(v_1^2) \,.
\nonumber \\
\end{eqnarray}
Just as in the $\theta$ sector, the heavy end propagator
$K^{11}_\phi(\nu)$ only couples to the translation
mode as $m_1 \to \infty$. In this limit, the only
poles in $\Gamma^{11}_\phi$ are double poles at $\nu = \pm\omega$,
so the only pole in $K^{11}_\phi$ is a double pole at
$\nu = 0$. At the light quark end, $\Gamma^{22}_\phi$
does not have poles at $\nu = \pm\omega$, as the cotangent vanishes
there. Just as in the $\theta$ sector, in the heavy--light limit
the light end propagator $K^{22}_\phi$ does not couple to
the translation modes of the string. The simple poles
in $\Gamma^{22}_\phi$ at even multiples of $\omega$
produce poles in $K^{22}_\phi$ at $\nu = \pm (2n + 1) \omega$.

The spectrum of singly excited string modes in the $\phi$
sector is then the same as in the $\theta$ sector,
and the degeneracy of the light--light
excitations is repeated in the heavy--light excitations.
The remaining pole in $\Gamma^{22}_\phi$, a double
pole at $\nu = 0$, is present for reasons already discussed
in considering the light--light case. The heavy--light
component of the propagator $K^{12}_\phi(\nu)$ couples to
both light end and heavy end modes, just as it did in
the $\theta$ sector.

\section{The Meson Spectrum}

\label{meson spectrum section}

In this section, we use the results \eqnlessref{L fluc ll num} and \eqnlessref{L fluc hl num},
as well as the energies of the string excited states,
to derive the meson Regge trajectories
to leading semiclassical order. We begin by calculating classical
Regge trajectories from the classical
Lagrangian \eqnlessref{L_cl def formal}.
The angular momentum of the meson and its energy $E_{{\rm cl}}(\omega)$
are given by \eqnlessref{J class} and \eqnlessref{E class},
\begin{eqnarray}
J = \frac{\partial L_{{\rm cl}}}{\partial\omega}
&=& \sum_i \left( \sigma \frac{R_i^2}{2v_i} \left( \frac{\arcsin v_i}{v_i}
- \gamma_i^{-1} \right) + m_i R_i v_i \gamma_i \right) \,,
\nonumber \\
E = \omega \frac{\partial L_{cl}}{\partial\omega} - L_{cl}
&=& \sum_i \left( \sigma R_i \frac{\arcsin v_i}{v_i} + m_i \gamma_i \right) \,.
\label{J and E equations}
\end{eqnarray}
From the classical equation of motion we derived
\eqnlessref{classical eqn}, which shows that $R_i$ is proportional
to $\gamma^2_i$ for large $\gamma_i$. Evaluating \eqnlessref{L_cl def formal}
and \eqnlessref{J and E equations} in the limit of massless
quarks, where
the quark velocity $v_i$ goes to one, yields the classical results,
\begin{equation}
L_{{\rm cl}}(\omega) = - \frac{\pi\sigma}{2\omega} \,, \kern 0.5in
J = \frac{\pi\sigma}{2\omega^2} \,, \kern 0.5in
E_{{\rm cl}} = \frac{\pi\sigma}{\omega} \,, \kern 0.5in
J = \frac{E_{{\rm cl}}^2}{2\pi\sigma} \,.
\label{classical Regge eqn ll}
\end{equation}
In the heavy--light case $v_1$ goes to zero and $v_2$ goes to
one, so
\begin{equation}
J = \frac{\left(E_{{\rm cl}} - m_1\right)^2}{\pi\sigma} \,.
\label{classical Regge eqn hl}
\end{equation}

We now include the correction \eqnlessref{E of J} to the energy due
to fluctuations. In the light--light case Eqs. \eqnlessref{E of J},
\eqnlessref{L fluc ll num}, and \eqnlessref{classical Regge eqn ll} give
\begin{equation}
E(J) = E_{cl}(\omega) - L_{{\rm fluc}}(\omega)
= \frac{\pi\sigma}{\omega} - \frac{7}{12} \omega \,,
\label{E of omega ll}
\end{equation}
for the case of two light quarks. The value of
$\omega$ is given as a function of $J$ through the
classical relation $\omega = \sqrt{\pi\sigma/2J}$. Squaring
both sides of \eqnlessref{E of omega ll} and dropping the term
quadratic in $L_{{\rm fluc}}$ yields
\begin{eqnarray}
J &=& \frac{E^2}{2\pi\sigma} + \frac{7}{12}
+ O\left(\frac{\sigma}{E^2}\right) \,.
\label{J of E ll}
\end{eqnarray}
Using the WKB quantization condition $J = l + 1/2$ in \eqnlessref{J of E ll}
gives the leading Regge trajectory, relating the angular momentum quantum
number $l$ to the meson energy $E$,
\begin{equation}
l = \frac{E^2}{2\pi\sigma} + \frac{1}{12} + O\left(\frac{\sigma}{E^2}\right) \,.
\label{l of E ll}
\end{equation}
In the heavy--light case, Eqs. \eqnlessref{L fluc hl num} and
\eqnlessref{classical Regge eqn hl} give the Regge trajectory
\begin{equation}
l = \frac{(E-m_1)^2}{\pi\sigma} + \frac{1}{6} - \frac{1}{2}
= \frac{(E-m_1)^2}{\pi\sigma} - \frac{1}{3} \,.
\label{l of E hl}
\end{equation}

The energies of the excited states of the light mesons are
obtained by adding the excitation energies $n\omega$ to
\eqnlessref{E of omega ll}
\begin{equation}
E_n(\omega) = \frac{\pi\sigma}{\omega} - \frac{7}{12} \omega + n \omega \,.
\label{hybrid energy ll}
\end{equation}
Since there are many combinations of string normal modes
which give the same $n$ (e.g., a doubly excited $k=1$ mode
and a singly excited $k=2$ mode each give $n=2$), the spectrum
is highly degenerate. There are two $n=1$ trajectories,
each corresponding to a single excitation of one of
the $k=1$ normal modes. Higher values of $n$ have higher
degeneracies.

From \eqnref{hybrid energy ll} for the $n$th excited hybrid energy
level we derive the ``daughter'' Regge trajectory,
\begin{equation}
l = \frac{E^2}{2\pi\sigma} + \frac{1}{12} - n
+ O\left(\frac{\sigma}{E^2}\right) \,.
\label{hybrid l of E ll}
\end{equation}
In the heavy--light case,
the normal modes with frequencies of $(2k+1) \omega$
can combine to form states with excitation energies
$n\omega$ for any $n$, though the degeneracies are
different. The ``daughter'' Regge trajectories of
\eqnlessref{l of E hl} are then
\begin{equation}
l = \frac{(E-m_1)^2}{\pi\sigma} - \frac{1}{3} - n
+ O\left(\frac{\sigma}{(E-m_1)^2}\right) \,.
\label{hybrid l of E hl}
\end{equation}

Eqs. \eqnlessref{hybrid l of E ll} and \eqnlessref{hybrid l of E hl}
give the leading semiclassical correction
to the classical Regge formulae \eqnlessref{classical Regge eqn ll}
and \eqnlessref{classical Regge eqn hl}. To compute the
$O(\sigma/E^2)$ corrections to this result, it would be necessary to compute
the contribution of two loop vacuum diagrams in the two dimensional
field theory to find the energy of the lowest lying trajectory, and
to compute one loop corrections to the propagators to find the energies
of the excited states.

\section{Generalization to $D \ne 4$ Dimensions}

\label{D dimen section}

It is interesting to compare the results \eqnlessref{l of E ll}
with the corresponding result from classical bosonic string
theory. To do this, we need to generalize \eqnlessref{l of E ll}
to $D$ dimensions. This dependence comes from the
dependence of $L_{{\rm eff}}(\omega)$ on $D$, which
in turn comes from $L^{{\rm string}}_{{\rm fluc}}(\omega)$
and $L_{{\rm boundary}}(\omega)$,
since the classical string energy is independent of $D$.

Our calculation of $L_{{\rm boundary}}(\omega)$
separated the boundary fluctuations perpendicular to
the plane of rotation of the string and the fluctuations
in that plane. Neither of these contributed to $L_{{\rm boundary}}(\omega)$
(see Appendix~\ref{boundary appendix}). Going from four
dimensions to $D$ dimensions only adds $D-4$ additional
directions for perpendicular fluctuations. These fluctuations
will each give the same contribution to $L_{{\rm boundary}}(\omega)$
that the fluctuations perpendicular to the plane of rotation
did in the four dimensional case, namely zero. The
function $L_{{\rm boundary}}(\omega)$ is therefore zero,
independent of $D$.

The function $L^{{\rm string}}_{{\rm fluc}}(\omega)$
was derived from $Z_I(\omega)$, which was
expressed in \eqnlessref{Z_I determinants} as the product of two determinants,
one due to string modes perpendicular to the plane of rotation,
and one due to string modes in the plane of rotation.
Just as in the case of $L_{{\rm boundary}}(\omega)$,
adding more dimensions adds additional string modes
perpendicular to the plane of rotation,
so the generalization of \eqnlessref{Z_I determinants} to $D$ dimensions is
\begin{eqnarray}
Z_I(\omega) &=& \Det^{-(D-3)/2}\left[-\nabla^2\right] \Det^{-1/2}\left[-\nabla^2
+ 2\omega^2 \sec^2 \omega x\right]
\nonumber \\
&=& \Det^{-(D-2)/2}\left[-\nabla^2\right] \Det^{-1/2}\left[\frac{-\nabla^2
+ 2\omega^2 \sec^2 \omega x}{-\nabla^2}\right] \,,
\end{eqnarray}
The first of the determinants produces a term $(D-2)\pi/24R_p$
in $L^{{\rm string}}_{{\rm fluc}}(\omega)$,
equal to the L\"uscher term in $D$ dimensions, with the length $R$
of the string replaced with its proper length $R_p$ (see \eqnref{proper length}).
We have already evaluated the second determinant, which, after renormalization,
gave the second and third terms in \eqnlessref{L fluc renorm}. For massless quarks,
$R_p = \pi/\omega$, and the contribution of the second determinant is $\omega/2$,
so that the generalization of \eqnlessref{l of E ll} to $D$ dimensions is
\begin{equation}
E_n(\omega) = \frac{\pi\sigma}{\omega} - \frac{D-2}{24} \omega
- \frac{\omega}{2} + n \omega \,.
\end{equation}
This can be rewritten to get $E^2$ as a function of $l$,
\begin{equation}
E^2 = 2\pi\sigma \left( l - \frac{D-2}{24} + n
+ O\left(\frac{1}{l}\right) \right) \,.
\end{equation}
In 26 dimensions, the equation for the energy is
\begin{equation}
E^2 = 2\pi\sigma \left( l - 1 + n + O\left(\frac{1}{l}\right) \right) \,.
\label{our energy 26 dim}
\end{equation}
The spectrum \eqnlessref{our energy 26 dim} coincides with the spectrum
of open strings in classical bosonic string theory. However, in our
approach \eqnlessref{our energy 26 dim} is valid only in the leading
semiclassical approximation, so it cannot be used for $l=0$, where
it would yield the scalar tachyon of the open bosonic string.

%% file: results.tex
\def\myfiguresize{3.5in}
\def\mycolumnlabelheight{1.2in}

\section{Comparison with Meson Masses}

\begin {figure}[Ht]
    \begin{center}
	\begin{tabular}{rc}
	    \vbox{\hbox{$l+1$ \hskip 0in \null} \vskip \mycolumnlabelheight} &
	    \epsfxsize = \myfiguresize
	    \epsfbox{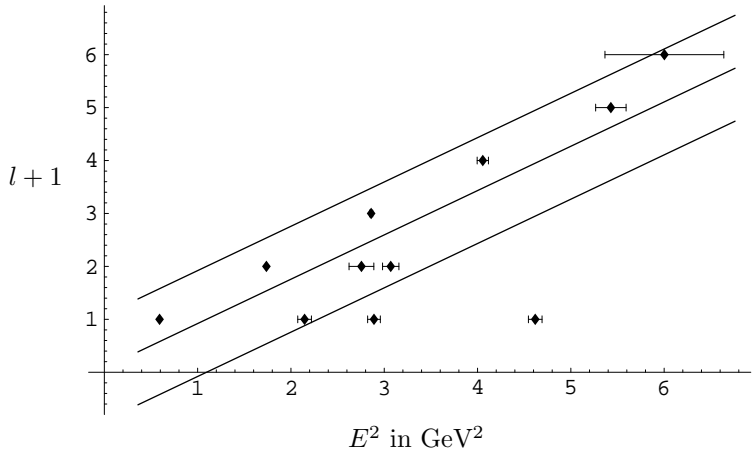} \\
	    &
	    \hbox{$E^2$ in GeV$^2$} \\
	\end{tabular}
    \end{center}
    \caption[$\rho$--$a_2$ Regge plot]
	{Regge trajectories \eqnlessref{hybrid l of E ll}
	with $n=0,1,2$, and meson masses in the $\rho$--$a_2$ sector}
    \label{rho plot}
\end {figure}

In Fig.~\ref{rho plot}, we plot the leading trajectory and first
two daughters (\eqnref{l of E ll} for $n = 0,1,2$) using the
string tension $\sigma = (0.436)^2$ GeV$^{-2}$, corresponding to
a value $\alpha' = 0.89$ GeV$^{-2}$ for the slope of the
$\rho$ trajectory.
The plotted points are meson masses on the
$\rho$ -- $a_2$ trajectory, and on possible daughters of this
trajectory. We have added one to the orbital
angular momentum to account for the spin of the quarks
($J = l + s = l + 1$). The plotted points lie to the right of
the leading trajectory, so the predicted masses are too low. This may
be due to the fact that we are using scalar instead of fermionic quarks.
In any case, the semiclassical correction is small, and the leading
trajectory lies close to the classical one.

We therefore compare the differences in the squared masses between the
lowest lying meson for each $l$ and higher energy states with the
predictions of the semiclassical formula \eqnlessref{l of E ll},
\begin{equation}
\frac{m_{{\rm excited}}^2 - m_{{\rm lowest}}^2}{2\pi\sigma} = n \,, \kern 1in
n = 1\,, 2\,,\,... \,.
\label{delta m^2}
\end{equation}
\begin{table}[Ht]
	\begin{center}
	\begin{tabular}{||@{\kern 0.15in}c@{\kern 0.15in}*{6}{|c@{\kern 0.15in}}||}
			\hline
			l & lowest state & excited state & $m_{{\rm lowest}}$ &
				$m_{{\rm excited}}$ & $\Delta m^2$ &
				$\Delta m^2 / 2\pi\sigma$ \\
			\hline \hline
			1 & $\rho$ & $\rho(1450)$ & 0.769 GeV & 1.465 GeV &
				1.555 $\hbox{GeV}^2$ & 1.30 \\
			1 & $\rho$ & $\rho(1700)$ & 0.769 GeV & 1.700 GeV &
				2.299 $\hbox{GeV}^2$ & 1.92 \\
			1 & $\rho$ & $\rho(2150)$ & 0.769 GeV & 2.149 GeV &
				4.027 $\hbox{GeV}^2$ & 3.37 \\
			2 & $a_2(1320)$ & $a_2(1660)$ & 1.318 GeV & 1.660 GeV &
				1.019 $\hbox{GeV}^2$ & 0.85 \\
			2 & $a_2(1320)$ & $a_2(1750)$ & 1.318 GeV & 1.752 GeV &
				1.332 $\hbox{GeV}^2$ & 1.12 \\
			\hline
		\end{tabular}
	\end{center}
	\caption{Squared mass differences for the excited states of the
		$\rho$ trajectory}
	\label{rho table}
\end{table}
This energy difference is entirely due to the excited states of the
string, and therefore may not be so sensitive to
the kind of quarks used in the model. The values of the mass difference
\eqnlessref{delta m^2} are shown in Table~\ref{rho table} for the excited
states of the $\rho$ ($l = 0$) and $a_2$ ($l = 1$). For $n=1$, the
semiclassical string theory predicts 2 degenerate states. This double
degeneracy will be broken by higher order corrections, and we expect
the predicted $n=1$ mass to lie halfway between the physical masses
of the two $n=1$ particles. This works very well for the $a_2$ meson,
where averaging the masses of the two excited states gives
$\Delta m^2/2\pi\sigma = 0.98$, compared to the predicted value of
unity. For the $\rho$ mesons, with $l = 0$, the semiclassical
theory is not applicable, and the excited states in table~\ref{rho table}
are not readily identified with the predictions of \eqnlessref{delta m^2}.

\section{Summary and Conclusions}

\begin{enumerate}
\item Beginning with an effective string theory of vortices which
describes long distance QCD, we have calculated, in the semiclassical
approximation, the effect of string fluctuations on Regge trajectories,
both for mesons containing light (zero mass) quarks, and for mesons
containing one heavy and one light quark.
The semiclassical correction to the leading Regge
trajectory for light quarks adds a constant $(D-2)/24$ to the classical Regge
formula.
The small size of this semiclassical correction for $D = 4$
could explain why Regge trajectories are linear at values
of $l$ of order one.
\item These results depended on two extensions of our previous
work:
	\begin{enumerate}
	\item The renormalization of the geodesic curvature in the
	semiclassical expansion about a rotating string solution,
	needed to take the zero quark mass limit.
	\item The decoupling of the boundary and interior degrees
	of freedom of the string to obtain the back reaction of
	the interior degrees of freedom on the boundary.
	\end{enumerate}
\item The spectrum of the energies of the excited states
formally coincides with the spectrum of the open
string of Bosonic string theory in its critical dimension
$D = 26$. Here, we obtained this spectrum for any $D$ from
the semiclassical expansion of an effective string theory. The
functional determinant $\Delta_{FP}$ determining the measure
for the path integral \eqnlessref{Wilson loop def} made the theory conformally
invariant in the limit of zero mass quarks. Perhaps this quantization
of effective string theory might prove useful towards efforts in
quantizing fundamental string theories in non-critical dimensions.
\item We treated the light quarks as massless scalar
particles. This is appropriate at best for determining the
energies of excited states of the string, where the
dependence on the boundary of the string is small.
The effect of chiral symmetry breaking, generating a constituent quark mass,
must play a dominant role in determining the masses of mesons
which are ground states of quark--antiquark systems.
However, this constituent mass should approximately cancel
in the mass differences between mesons on the leading
and first daughter trajectories. The effective string theory
should then describe the excitation energies of the mesons.
\end{enumerate}

\section*{Acknowledgements}

We would like to thank D. Gromes and A. Kaidalov for very helpful conversations.

%% file: string_quant.tex
\section{Quantizing the Angular Momentum}

\label{DHN appendix}

In this appendix, we quantize the angular momentum of the
string semiclassically, using the semiclassical methods of
Dashen, Hasslacher, and Neveu~\cite{DHN} (DHN) for obtaining
the energies of periodic orbits. DHN find the energies
of these states by looking at the trace of the propagator
\begin{equation}
G(E) = i \tr \int_0^\infty dT e^{i(E-H) T} \,,
\label{DHN propagator}
\end{equation}
where $H$ is the Hamiltonian. The operator on the right
hand side is defined in terms of a partition function
with periodic boundary conditions. In our case,
it is
\begin{eqnarray}
e^{-iHT} \equiv Z^{{\rm periodic}} &=& \frac{1}{Z_b}
\int \scrD f^1(\xi) \scrD f^2(\xi) \scrD {\bf\vec x}_1(t)
\scrD {\bf\vec x}_2(t) \Delta_{FP}
\nonumber \\
& & e^{-i\sigma \int d^2\xi \sqrt{-g}
- i \sum_{i=1}^2 m_i \int_{-T/2}^{T/2} dt \sqrt{1 - \dot{\bf\vec x}_i^2(t)}} \,,
\label{string partition periodic}
\end{eqnarray}
where the variables $f^i$ and ${\bf\vec x}_i$ are required to satisfy
the boundary conditions
\begin{equation}
f^i \Big|_{T/2} = f^i \Big|_{-T/2} \,, \kern 1 in
{\bf\vec x}_i \Big|_{T/2} = {\bf\vec x}_i \Big|_{-T/2} \,.
\end{equation}

\subsection{Euler Angles}

The first step in quantizing the angular
momentum is to introduce collective coordinates for the
rotational degrees of freedom of the string.
We parameterize the rigid body
rotations of a straight string using the Euler angles $\alpha$,
$\beta$, and $\gamma$, defined by the rotation matrix $M$,
\begin{equation}
M = \left(\matrix{
\cos\alpha\cos\beta\cos\gamma - \sin\alpha\sin\gamma &
\cos\alpha\cos\beta\sin\gamma + \sin\alpha\cos\gamma &
\cos\alpha\sin\beta \cr
-\sin\alpha\cos\beta\cos\gamma - \cos\alpha\sin\gamma &
- \sin\alpha\cos\beta\sin\gamma + \cos\alpha\cos\gamma &
-\sin\alpha\sin\beta \cr
- \sin\beta\cos\gamma &
-\sin\beta\sin\gamma &
\cos\beta
}\right) \,.
\end{equation}
The angles $\alpha$, $\beta$, and $\gamma$ are functions of the
time $t$. The rate of change of the matrix $M$ acting on a fixed vector
$\hat n$ defines the angular velocity $\vec\omega$,
\begin{equation}
\frac{d}{dt} (M \hat n) = \dot M \hat n \equiv \vec\omega \times (M \hat n)
\,.
\end{equation}
Since $\hat n$ is an arbitrary vector, we find
\begin{eqnarray}
\vec\omega[\gamma,\beta,\alpha] &\equiv& -\frac{1}{2} \ehat_i \epsilon^{ijk}
\left(\dot M M^{-1}\right)^{jk}
\nonumber \\
&=& - \dot\alpha \ehat_3 + \dot\beta
\left(\cos\alpha \ehat_2 + \sin\alpha \ehat_1\right) - \dot\gamma\left(
\cos\beta \ehat_3 - \sin\beta\sin\alpha \ehat_2 + \sin\beta\cos\alpha
\ehat_1\right) \,.
\nonumber \\
\end{eqnarray}

In the limit of small fluctuations about a straight
rotating string, the string only has two rotational
degrees of freedom. Classically,
the angular velocity about the axis of the string must be zero.
Any contribution to this component of the angular velocity
must be of quadratic order in small fluctuations about
the classical solution, since it takes one fluctuation to
give the string a moment of inertia about its own axis, and
another to give it rotation about that axis. Therefore,
to quadratic order,
\begin{equation}
\vec\omega \cdot \vec x_0 = 0 \,,
\end{equation}
where $\vec x_0$ is the classical position of the string.

The two physical angular degrees of freedom are the Euler
angles $\alpha(t)$ and $\beta(t)$ determining the 
orientation of the vector $\hat x_0$, chosen to be the $\ehat'_3$
axis in the body fixed frame,
\begin{equation}
\hat x_0 = \cos\alpha\sin\beta \ehat_1 - \sin\alpha\sin\beta \ehat_2
+ \cos\beta \ehat_3 \,.
\end{equation}
The condition $\hat x_0 \cdot \vec\omega = 0$, that there
are no rotations about the string axis, is
\begin{equation}
\dot\alpha\cos\beta + \dot\gamma = 0 \,.
\label{gamma def}
\end{equation}
This means that $\gamma$ is superfluous. Substituting for
$\gamma$ using \eqnref{gamma def} gives
\begin{equation}
\vec\omega[\beta,\alpha] = - \dot\alpha \sin\beta \left( \sin\beta \ehat_3
+ \sin\alpha\cos\beta \ehat_2 - \cos\alpha\cos\beta \ehat_1\right)
+ \dot\beta \left(\cos\alpha \ehat_2 + \sin\alpha \ehat_1\right) \,.
\label{omega of alpha and beta}
\end{equation}

To introduce the Euler angles into the functional integral
\eqnlessref{string partition periodic}, we must define
$\alpha$ and $\beta$ as functionals
of the string position $\tilde x^\mu$. Let the function
$\vec\Omega[\tilde x^\mu](t)$ be the angular velocity of
the string $\tilde x^\mu$ at time $t$.
The form of this function is not needed for our calculation.
We fix $\vec\omega[\alpha, \beta]$ at all times by inserting a factor of one into the
partition function,
\begin{equation}
1 = \int \scrD\beta \scrD\alpha
\delta^{(2)}\left[ \vec\omega - \vec\Omega[\tilde x^\mu]\right]
\Det\left[\epsilon^{ijk} \hat x_0^i \frac{\partial\omega^j}{\partial\alpha}
\frac{\partial\omega^k}{\partial\beta} \right] \,.
\label{omega 1}
\end{equation}
Because the argument of the $\delta$ function only has two nonzero
components, \eqnlessref{omega 1} only contains two $\delta$ functions.
Inserting the definition \eqnlessref{omega of alpha and beta} of $\vec\omega$ in the
determinant gives
\begin{eqnarray}
1 &=& \int \scrD(\cos\beta) \scrD\alpha
\delta^{(2)}\left[ \vec\omega - \vec\Omega[\tilde x^\mu]\right]
\Det\left[-\frac{d^2}{dt^2}\right] \,.
\label{omega 1 proper}
\end{eqnarray}

Inserting the factor \eqnlessref{omega 1 proper} into the partition
function allows us to write the center of mass partition
function in terms of rotational degrees of freedom $\alpha$ and $\beta$,
\begin{eqnarray}
Z^{{\rm periodic}} &=& \frac{1}{Z_b} \int \scrD(\cos\beta) \scrD\alpha
\scrD f_1 \scrD f_2 \scrD\vec x_1 \scrD\vec x_2 \Delta_{FP}
\nonumber \\
& & \times \delta^{(2)}\left[ \vec\omega - \vec\Omega[\tilde x^\mu]\right]
\Det\left[-\frac{d^2}{dt^2}\right] e^{i\int dt L[x^\mu]} \,.
\label{Z periodic Euler}
\end{eqnarray}
$\alpha$ is the angle between the $y$ axis and the normal to the
plane of rotation. $\beta$ is the (angular) position of the
end of the string in the plane of rotation.

\subsection{Extracting the Sum Over Classical Solutions}

We evaluate the partition function \eqnlessref{Z periodic Euler}
semiclassically. It will then
contain a sum over all classical solutions of period $T$.
We now explicitly extract this sum from the functional
integral.
The classical solutions in \eqnlessref{Z periodic Euler}
correspond to motion where the axis $\hat x_0(t)$ of the string
rotates with uniform angular velocity. We can parameterize
these solutions by the Euler angles
\begin{equation}
\alpha = \hbox{constant,} \kern 1 in \beta = \omega t \,.
\label{classical alpha beta}
\end{equation}
The constant $\omega$ satisfies the equation
\begin{equation}
\omega = \frac{2\pi n}{T}
\end{equation}
for some integer $n$.
We can always make a global rotation to ensure that the
classical solution for $\alpha$ and $\beta$ has the form 
\eqnlessref{classical alpha beta}, so we write
\begin{equation}
Z^{{\rm periodic}} = \int \scrD(\cos\beta) \scrD\alpha
\Det\left[-\frac{d^2}{dt^2}\right] e^{iS_{{\rm rotate}}[\alpha, \beta]} \,,
\end{equation}
where
\begin{eqnarray}
e^{iS_{{\rm rotate}}[\alpha, \beta]} &\equiv& \frac{1}{Z_b} \int dR^{ij} \int
\scrD f_1 \scrD f_2 \scrD\vec x_1 \scrD\vec x_2 \Delta_{FP}
\delta^{(2)}\left( R^{ij} \hat x_0^j - \ehat_3^i \right)
\nonumber \\
& & \times \delta^{(2)}\left[ \vec\omega^i
- R^{ij} \vec\Omega^j[\tilde x^\mu]\right] e^{i\int dt L[x^\mu]} \,.
\label{S rotate def}
\end{eqnarray}

To enable us to extract the sum over classical solutions,
we will explicitly divide $\alpha$ and $\beta$ into parts
which change the classical solution and parts which
perturb the fields away from the classical solution.
We first divide the field $\alpha(t)$ into fluctuations
which change the classical solution and fluctuations which
move the field away from its classical value. Classically, $\alpha(t)$ can be any
constant $\alpha_0$. The choice of this constant determines
which of the planes passing through $\beta = 0$ will
contain the string rotation.
We note that, since $\vec\omega^2 = \dot\alpha^2 \sin^2\beta + \dot\beta^2$,
$\alpha$ always appears in the action in terms
of the form $\dot \alpha \sin \beta$, which is classically zero. We define
a new variable
\begin{equation}
a = \dot \alpha \sin \beta \,.
\end{equation}
In terms of the function $a(t)$, $\alpha(t)$ is
\begin{equation}
\alpha(t) = \alpha_0 + \int_0^t dt' \frac{a(t')}{\sin\beta(t')} \,.
\end{equation}
Because $a$ only depends on $\dot \alpha$, its classical value
is independent of the choice of the constant $\alpha_0$.

\begin {figure}[Htbp]
    \epsfxsize = 3 in
    \leavevmode{\hfill \epsfbox{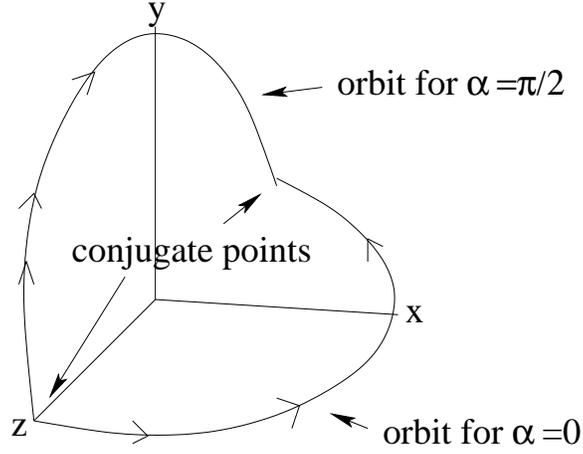} \hfill}
    \medskip
    \caption{Orbits of the end of the string for different values
	of $\alpha$}
    \label{conj points figure}
\end{figure}

The factor of $\sin \beta$ in the definition of $a$ produces
a complication. Comparing the inverse propagators of $\alpha$ and $a$,
we see that
\begin{equation}
\frac{\partial^2 S}{\partial \alpha(t) \partial \alpha(t')}
= - \frac{d}{dt} \sin \beta(t) \frac{\partial^2 S}{\partial a(t) \partial a(t')}
\sin \beta(t') \frac{d}{dt'} \,.
\end{equation}
When $\beta$ is a multiple of $\pi$, the inverse propagator of $\alpha$ vanishes,
but the same is not true of the inverse propagator of $a$.
At these points, the end of the string is at either the point
$\beta = 0$ or its antipode $\beta = \pi$, independent of the value
of $\alpha$ (see Fig.~\ref{conj points figure}).
These points are called conjugate points.
At these points classical trajectories with different values of
$\alpha_0$ meet. Due to the singularity in the
propagator of $\alpha$, the partition function picks
up a phase of $\pi/2$ at each of these points when
we do the $\alpha$ integral, analogous to
the phase shift at a WKB turning point~\cite{Gutzwiller}.
The $a$ propagator does not have this singularity, so
the partition function will not receive a phase shift
from the $a$ integral.
In changing variables from $\alpha$ to $a$, we must add this
phase shift to the partition function. The integration
measure for $\alpha$ is
\begin{equation}
\scrD \alpha = d\alpha_0 \scrD a \Det^{-1/2}[-\partial_t^2]
\Det^{-1}[\sin \beta] e^{i/2 \int_{-T/2}^{T/2} dt \dot \beta} \,.
\end{equation}
The argument of the exponential is equal to $\pi/2$ multiplied
by the number of times $\beta$ passes through
a multiple of $\pi$, which is the phase shift.
The integral over $\alpha_0$ is present because $a$ is
independent of the constant part of $\alpha$. $\alpha_0$
varies between $0$ and $2\pi$. In terms of these
new variables, the partition function is
\begin{equation}
Z^{{\rm periodic}} = \int d\alpha_0 \scrD\beta \scrD a
\Det^{1/2}\left[-\frac{d^2}{dt^2}\right] e^{iS_{{\rm rotate}}[\alpha, \beta]
 + i/2 \int_{-T/2}^{T/2} dt \dot \beta} \,.
\end{equation}

We next extract the sum over the classical frequencies
$\omega = 2\pi n/T$ from the integral over $\beta$.
Let $b(t) = \dot \beta(t) - \omega$, and $\beta_0 = \beta(t=0)$.
Then $\beta(t)$ is given by
\begin{equation}
\beta(t) = \beta_0 + \omega t + \int_0^t dt' b(t') \,.
\label{beta parameterization}
\end{equation}
Due to the boundary conditions
on $\beta$, $b(t)$ is subject to the restriction
\begin{equation}
\int_{-T/2}^{T/2} dt b = \beta(T/2) - \beta(-T/2) - \omega T = 0 \,.
\label{b condition}
\end{equation}
The function $b(t)$ is also independent of $\beta_0$.

The change of variables from $\beta$ to $b$ produces a change in
the functional integration measure,
\begin{equation}
\scrD \beta = \sum_{\omega = 2\pi n/T} \scrD b d\beta_0
\Det^{-1/2}[-\partial_t^2] 2\sqrt{\frac{\pi}{T}} \,.
\label{beta measure change}
\end{equation}
The factor of $2\sqrt{\pi/T}$
appears in \eqnlessref{beta measure change} because the integral has
periodic boundary conditions. The definition of
the determinant of $-\partial_t^2$ is~\cite{Cameron+Martin}
\begin{equation}
\Det^{-1/2}[-\partial_t^2] = \int ds_1 ... ds_j
\exp\left\{ -\frac{s_1^2}{t_1} - \frac{(s_2-s_1)^2}{t_2 - t_1}
- ... - \frac{(s_j - s_{j-1})^2}{t_j - t_{j-1}} \right\} \,.
\end{equation}
However, in the case of periodic boundary conditions, $s_j =
s_0 = 0$. Because of this, we need a Lagrange multiplier
to identify the values of $s$ at these two points,
\begin{equation}
\Det^{-1/2}_{{\rm periodic}}[-\partial_t^2]
 = \int d\lambda \int ds_1 ... ds_j
\exp\left\{ -\frac{s_1^2}{t_1} - \frac{(s_2-s_1)^2}{t_2 - t_1}
- ... - \frac{(s_j - s_{j-1})^2}{t_j - t_{j-1}} + i\lambda s_j \right\} \,.
\end{equation}
Due to the additional term in the action, we must translate each
of the $s_i$ to be able to do the integral. This translation is
\begin{equation}
s_i \to s_i + i\frac{\lambda}{2} t_i \,.
\end{equation}
The effect of this translation on each $s^2$ term is
\begin{equation}
-\frac{(s_i - s_{i-1})^2}{t_i - t_{i-i}}
\to -\frac{(s_i - s_{i-1})^2}{t_i - t_{i-i}}
- i\lambda(s_i - s_{i-1}) + \frac{\lambda^2}{4}(t_i - t_{i-i}) \,,
\end{equation}
while the effect on the $\lambda s_j$ term is
\begin{equation}
i\lambda s_j \to i\lambda s_j - \frac{\lambda^2}{2} t_j \,.
\end{equation}
The total effect of this transformation is that
\begin{eqnarray}
\Det^{-1/2}_{{\rm periodic}}[-\partial_t^2]
&=& \int d\lambda \int ds_1 ... ds_j \exp\left\{ -\frac{s_1^2}{t_1}
- \frac{(s_2-s_1)^2}{t_2 - t_1} - ...
- \frac{(s_j - s_{j-1})^2}{t_j - t_{j-1}} - \frac{\lambda^2}{4} t_j \right\}
\nonumber \\
&=& \Det^{-1/2}[-\partial_t^2] \int d\lambda \exp\left\{ -\frac{\lambda^2}{4}T
\right\}
\nonumber \\
&=& \Det^{-1/2}[-\partial_t^2] 2\sqrt{\frac{\pi}{T}} \,.
\end{eqnarray}
This gives the factor we have included in \eqnref{beta measure change}.
A more general version of this derivation, valid for all Gaussian
functional integrals, is done in \cite{DHN}. Making the change
of variables \eqnlessref{beta parameterization} and implementing
the restriction \eqnlessref{b condition}
gives the partition function the form
\begin{equation}
Z^{{\rm periodic}} = 2\sqrt{\frac{\pi}{T}} \sum_{\omega = 2\pi n/T}
\int d\alpha_0 d\beta_0 \scrD a \scrD b
\delta\left(\int_{-T/2}^{T/2} dt b(t)\right)
e^{iS_{{\rm rotate}}[\alpha, \beta] + iT\omega/2} \,.
\label{Z of a and b}
\end{equation}

\eqnref{Z of a and b} gives $Z^{{\rm periodic}}$ as a sum over
semiclassical integrals about classical solutions,
\begin{equation}
Z^{{\rm periodic}} = 2\sqrt{\frac{\pi}{T}} \sum_{\omega = 2\pi n/T}
e^{iT\omega/2} Z(\omega) \,,
\label{sum over Z of omega}
\end{equation}
where $Z(\omega)$ is
\begin{eqnarray}
Z(\omega) &\equiv& \frac{1}{Z_b}
\int d\alpha_0 d\beta_0 \scrD a \scrD b dR^{ij}
\scrD f_1 \scrD f_2 \scrD\vec x_1 \scrD\vec x_2 \Delta_{FP}
\delta^{(2)}\left( R^{ij} \hat x_0^j - \ehat_3^i \right)
\nonumber \\
& & \times \delta\left(\int_{-T/2}^{T/2} dt b(t)\right)
\delta^{(2)}\left[ \vec\omega^i - R^{ij} \vec\Omega^j[\tilde x^\mu]\right]
e^{i\int dt L[x^\mu]} \,.
\label{Z of omega def}
\end{eqnarray}
Doing the integrals over $a$, $b$, $\alpha_0$, and $\beta_0$ in
\eqnlessref{Z of omega def} gives
\begin{eqnarray}
Z(\omega) &=& \frac{1}{Z_b} \int
\scrD f_1 \scrD f_2 \scrD\vec x_1 \scrD\vec x_2 \Delta_{FP}
\delta\left(\omega - \left|\langle\vec\Omega[\tilde x^\mu]
\rangle\right|\right) e^{i\int dt L[x^\mu]} \,.
\label{Z with delta}
\end{eqnarray}
The constraints that have been placed on $\alpha$ and $\beta$
restrict $Z(\omega)$ to those string configurations with angular velocity $\omega$.
The expression \eqnlessref{Z with delta} is equivalent to the partition
function $Z(\omega)$ defined in \eqnref{partition param}, up to
the $\delta$ function
\begin{equation}
\delta\left(\omega - \left|\langle\vec\Omega[\tilde x^\mu]
\rangle\right|\right) \,.
\end{equation}
This $\delta$ function implements the boundary condition
\eqnlessref{phi linear term} as a constraint on $\tilde x^\mu$.
The constraint is due to the fact that, in obtaining the sum over
classical solutions, we have removed the zero frequency component of
one degree of freedom from the partition function.

\subsection{Summing over Classical Solutions}

We now insert the sum over classical solutions \eqnlessref{sum over Z of omega} into
\eqnlessref{DHN propagator} and use \eqnlessref{partition effective Lagrangian}.
This expresses the propagator $G(E)$ in terms of $L_{{\rm eff}}(\omega)$,
\begin{equation}
G(E) = i \int_0^\infty dT \, 2\sqrt{\frac{\pi}{T}} \sum_{\omega = 2\pi n/T}
e^{iT(E + \omega/2 + L_{{\rm eff}}(\omega))} \,.
\label{before T integral}
\end{equation}
We evaluate of the $T$ integral
by the method of stationary phase. The poles in the propagator $G(E)$
appear at those values of $E$ for which the sum over $n$ diverges.
Therefore, we must approximate the $T$ integral in a way which
is valid for large $n$. As $n$ becomes large, the classical solution
will consist of many orbits at some frequency $\omega$ determined
by the energy. Therefore, $T$ is large in the large $n$ limit,
and the phase in the exponential fluctuates wildly.
We do the $T$ integral by expanding about the stationary point
of this phase.
This point defines $T$ as a function of $E$,
\begin{equation}
\frac{d}{dT}\left[ T \left( L_{{\rm eff}}(\omega)
+ E + \frac{\omega}{2} \right) \right] = L_{{\rm eff}}(\omega)
+ E + \frac{\omega}{2} + T \frac{d\omega}{dT}
\left( \frac{dL_{{\rm eff}}}{d\omega} + \frac{1}{2} \right) = 0 \,.
\end{equation}
The definition $\omega = 2\pi n/T$ gives
\begin{equation}
\frac{d\omega}{dT} = - \frac{\omega}{T} \,,
\end{equation}
so the energy is
\begin{equation}
E = \omega \frac{dL_{{\rm eff}}}{d\omega}
- L_{{\rm eff}}(\omega) \,.
\label{E def}
\end{equation}
\eqnref{E def} implicitly defines $\omega$ as a function of $E$, so
$\omega$ is independent of $n$, while $T$ is proportional to $n$.
The integral also produces a factor of
\begin{equation}
\sqrt{\pi} \left(\frac{1}{2} \frac{d^2}{dT^2} \left[ T \left( L_{{\rm eff}}(\omega)
+ E + \frac{\omega}{2} \right) \right] \right)^{-1/2}
= \sqrt{2 \pi T} \left(
 \omega^2 \frac{dL_{{\rm eff}}}{d\omega^2} \right)^{-1/2} \,,
\end{equation}
which cancels the factor of $T^{-1/2}$ in \eqnlessref{before T integral}.
The propagator is
\begin{equation}
G(E) = 2i\pi\sqrt{2} \sum_{n=1}^\infty
\left( \omega^2 \frac{dL_{{\rm eff}}}{d\omega^2}
\right)^{-1/2} e^{iT(L_{{\rm eff}}(\omega) + E + \omega/2)} \,.
\end{equation}

To do the sum over $n$, we make the $n$ dependence
explicit by writing $T = 2\pi n/\omega$ everywhere,
\begin{eqnarray}
G(E) &=& 16i\pi^3\sqrt{2} \left(\omega^2
\frac{dL_{{\rm eff}}}{d\omega^2} \right)^{-1/2} \sum_{n=1}^\infty
\exp\left\{2\pi i n\left( \frac{L_{{\rm eff}}(\omega)}{\omega}
+ \frac{E}{\omega} + \frac{1}{2} \right) \right\} \,.
\end{eqnarray}
Doing the sum, and replacing $E$ using \eqnref{E def}, gives
\begin{equation}
\sum_{n=1}^\infty
\exp\left\{2\pi i n\left( \frac{L_{{\rm eff}}(\omega)}{\omega}
+ \frac{E}{\omega} + \frac{1}{2} \right) \right\}
= \frac{1}{1 - \exp\left\{2\pi i\left(\frac{dL_{{\rm eff}}}{d\omega}
+ \frac{1}{2}\right)\right\}} - 1 \,.
\end{equation}
Therefore, the propagator has poles whenever $\frac{dL_{{\rm eff}}}{d\omega}$
is an odd half integer,
\begin{equation}
\frac{dL_{{\rm eff}}}{d\omega} = l + \frac{1}{2} \,.
\label{omega quantization}
\end{equation}
This is the WKB quantization condition for angular momentum. It tells
us the angular velocities of the angular momentum
states. \eqnref{E def} then gives the energies of those states.
The poles in the propagator $G(E)$ come from the divergence
of the sum over orbits for large numbers of orbits (large $T$),
so $L_{{\rm eff}}(\omega)$ is defined by taking the
large $T$ limit,
\begin{equation}
L_{{\rm eff}}(\omega) \equiv \lim_{T\to\infty} \frac{-i}{T} \ln Z(\omega) \,.
\label{L straight def}
\end{equation}

%% file: Geval.tex
\section{Evaluation of $G^{\lowercase{ij}}(\nu)$}

\label{G^ij appendix}

We will now evaluate \eqnlessref{G^ij def},
\begin{equation}
G^{ij}_\psi(\nu) \equiv (-1)^{i+j} \Sigma((-1)^i \bar R_i)
\Sigma((-1)^j \bar R_j) \bar\gamma_i^{-2} \bar\gamma_j^{-2}
\frac{\partial^2}{\partial r \partial r'} G(r,r',\nu)
\Bigg|_{{r' = (-1)^i \bar R_i} \atop {r = (-1)^j \bar R_j}} \,.
\end{equation}
The Green's function $G(r,r',\nu)$ can be written in terms
of functions $l_i(r,\nu)$,
\begin{equation}
G(r,r',\nu) = - \frac{l_1(r_>) l_2(r_<)}{\Sigma(r) \gamma^{-2} W[l_1, l_2]} \,,
\end{equation}
and their Wronskian,
\begin{equation}
W[l_1, l_2] = \left( \frac{\partial l_1(r)}{\partial r}
l_2 (r) - \frac{\partial l_2(r)}{\partial r} l_1(r) \right) \,.
\end{equation}
The functions $l_i(r,\nu)$ satisfy the differential equation
\begin{equation}
\left( -\frac{\partial}{\partial r}
\gamma^{-2} \Sigma(r) \frac{\partial}{\partial r} 
- \Sigma(r) (\nu^2 - C) \right) l_i(r, \nu) = 0 \,,
\label{r diff}
\end{equation}
with the boundary conditions
\begin{equation}
l_i((-1)^j \bar R_j, \nu) = \delta_{ij} \,.
\label{r boundary}
\end{equation}
We will evaluate $G^{ij}_\psi$ in both the $\theta$ sector,
where $C = \omega^2$
and $\Sigma(r) = \gamma r^2$, and the $\phi$ sector, where
$C=0$ and $\Sigma(r) = \gamma^3 r^2$.

We begin with the $\theta$ sector. 
The first thing we do is change variables. We use the
`proper length' coordinate $x$, previously defined~\ref{x coord def} to be 
\begin{equation}
x = \frac{1}{\omega} \arcsin(\omega r) \,.
\end{equation}
We also change the normalization of the functions $l^\theta_i(r, \nu)$.
We define the functions $q^\theta_i(x,\nu)$,
\begin{equation}
q^\theta_i(x,\nu) = \frac{1}{r} l^\theta_i(r,\nu) \,.
\end{equation}
The $q^\theta_i$ satisfy the differential equation
\begin{equation}
\left(\nu^2 + \frac{\partial^2}{\partial x^2}\right)
q^\theta_i(x,\nu) = 0 \,,
\label{theta difeq}
\end{equation}
and the boundary conditions
\begin{equation}
q^\theta_i\left(\frac{(-1)^i}{\omega} \arcsin v_i, \nu\right)
= \delta_{ij} (-1)^i \bar R_i \,.
\end{equation}
Therefore, the $q^\theta_i$ are
\begin{equation}
q^\theta_i(x,\nu) = \bar R_i \frac{\sin\left(\nu x
+ (-1)^i \frac{\nu}{\omega} \arcsin v_{\hat \imath}\right)}
{\sin(\nu R_p)} \,,
\end{equation}
where $v_{\hat \imath}$ is the `other' velocity, i.e. $v_{\hat 2} = v_1$
and $v_{\hat 1} = v_2$, and $R_p$ is given by \eqnlessref{proper length}.

The functions $l_i^\theta(r,\nu)$ are therefore
\begin{equation}
l_i^\theta(r,\nu) = \frac{\bar R_i}{r} \frac{\sin\left(
\frac{\nu}{\omega} \left( \arcsin(\omega r)
+ (-1)^i \arcsin v_{\hat \imath} \right)\right)}{\sin(\nu R_p)} \,,
\end{equation}
and $G^{ij}_\theta(\nu)$ is
\begin{equation}
G^{ij}_\theta(\nu) = \delta_{ij} \left( - \frac{\bar R_i}{\bar\gamma_i}
+ \nu \bar R_i^2 \cot(\nu R_p) \right)
- (1-\delta_{ij}) \nu \bar R_i \bar R_j \csc(\nu R_p) \,.
\label{G^ij_theta}
\end{equation}

Next, consider the $\phi$ sector. We
define the functions $q_i^\phi(x,\nu)$,
\begin{equation}
l_i^\phi(r,\nu) = \frac{1}{\gamma r} q_i^\phi(x,\nu) \,.
\end{equation}
The $q_i^\phi$ satisfy the differential equation
\begin{equation}
\left(\nu^2 + \frac{\partial^2}{\partial x^2} - 2\omega^2
\sec^2(\omega x) \right) q_i^\phi(x,\nu) = 0 \,,
\label{phi difeq}
\end{equation}
with boundary conditions
\begin{equation}
q_i^\phi\left(\frac{(-1)^j}{\omega} \arcsin v_j, \nu \right)
= \delta_{ij} (-1)^i \bar R_i \bar\gamma_i \,.
\end{equation}
The $q_i^\phi$ are related to the eigenfunctions \eqnlessref{phi eigenfunctions},
and are given by
\begin{eqnarray}
q_i^\phi(x, \nu) &=& \bar R_i \bar\gamma_i \Bigg[
\left(\nu^2 - (-1)^i \omega^2 v_{\hat\imath}
\bar\gamma_{\hat\imath}\tan(\omega x) \right) (-1)^i
\sin\left(\nu x +(-1)^i \frac{\nu}{\omega} \arcsin v_{\hat i}\right)
\nonumber \\
& & - \nu \omega \left((-1)^i \tan(\omega x) + v_{\hat\imath}
\bar\gamma_{\hat\imath}\right) \cos\left(\nu x +(-1)^i
\frac{\nu}{\omega} \arcsin v_{\hat i}\right) \Bigg]
\nonumber \\
& & \times \left[\left(\nu^2 - \omega^2 v_1 \bar\gamma_1
v_2 \bar\gamma_2\right)
\sin(\nu R_p) - \nu \omega \left( v_1 \bar\gamma_1
+ v_2 \bar\gamma_2 \right) \cos(\nu R_p)\right]^{-1} \,.
\end{eqnarray}
This gives the value of $G^{ij}_\phi(\nu)$,
\begin{eqnarray}
G^{ij}_\phi(\nu) &=& -\delta_{ij} \bar\gamma_i \bar R_i
\frac{\nu^2}{\omega^2}
+ \frac{\nu}{\omega^3} (\nu^2 - \omega^2)
\nonumber \\
& & \times \frac{ \delta_{ij} \left(\nu v_i \bar\gamma_i \sin(\nu R_p)
- \omega v_1 \bar\gamma_1 v_2 \bar\gamma_2 \cos(\nu R_p)\right)
+ (1-\delta_{ij}) \omega v_i \bar\gamma_1 v_2 \bar\gamma_2}
{\left(\nu^2 - \omega^2 v_1 \bar\gamma_1 v_2 \bar\gamma_2\right)
\sin(\nu R_p) - \nu \omega \left(v_1\bar\gamma_1 + v_2\bar\gamma_2\right)
\cos(\nu R_p)} \,.
\label{G^ij_phi}
\end{eqnarray}

%% file: boundary_partition.tex
\section{Evaluation of $L_{\lowercase{{\rm boundary}}}$}

\label{boundary appendix}

We now evaluate the contribution $L_{{\rm boundary}}$ of
the boundary degrees of freedom to the effective Lagrangian,
\begin{eqnarray}
L_{{\rm boundary}} &=& \lim_{T \to \infty} \frac{-i}{T} \Bigg\{
-\frac{1}{2} \sum_{i=1}^2 \Tr\ln\left[\frac{\nu^2
+ \left( 2\bar\gamma_i^2 - 1 \right) \omega^2}{\nu^2} \right]
\nonumber \\
& & -\frac{1}{2} \Tr\ln\left[\frac{{\Gamma^{ij}_\theta}^{-1}}
{\delta_{ij} m_i \bar R_i^2 \bar\gamma_i \nu^2}\right]
- \frac{1}{2} \Tr\ln\left[\frac{{\Gamma^{ij}_\phi}^{-1}}
{\delta_{ij} m_i \bar R_i^2 \bar\gamma_i^3 \nu^2}\right] \Bigg\} \,.
\end{eqnarray}
Converting the traces to integrals, inserting the explicit
form \eqnlessref{Gamma theta def} of ${\Gamma^{ij}_\theta}^{-1}$
and \eqnlessref{Gamma phi def} of ${\Gamma^{ij}_\phi}^{-1}$, and
Wick rotating $\nu \to -i\nu$ gives $L_{{\rm boundary}}$ the form
\begin{eqnarray}
L_{{\rm boundary}} &=& - \frac{1}{2} \int \frac{d\nu}{2\pi} \Bigg\{
\tr\ln\left[\delta_{ij} \frac{\nu^2 - (2\bar\gamma_i^2 - 1)
\omega^2}{\nu^2}\right]
\nonumber \\
& & + \tr\ln\left[ \delta_{ij} \frac{\nu^2 + \omega^2}{\nu^2}
+ \frac{\omega^3}{\nu^2} \sqrt{\frac{\bar\gamma_i \bar\gamma_j}{v_i v_j}}
G^{ij}_\theta(-i\nu) \right]
\nonumber \\
& & + \tr\ln\left[ \delta_{ij} \frac{\nu^2 + (2\bar\gamma_i^2 + 1) \omega^2}
{\nu^2 - (2\bar\gamma_i^2 - 1) \omega^2}
+ \frac{\omega^3}{\nu^2} \frac{1}{\sqrt{v_i\bar\gamma_i v_j \bar\gamma_j}}
G^{ij}_\phi(-i\nu) \right] \Bigg\} \,.
\label{L boundary first}
\end{eqnarray}
The traces in \eqnlessref{L boundary first} are over the indices $i,j$.

The integral over the first trace in \eqnlessref{L boundary first}
is zero. The integrals over the second and third terms are logarithmically
divergent in the cutoff on the wavelengths of string modes. Three
of these modes are translation modes, so their contribution to
$L_{{\rm boundary}}$ should not be included in our calculation
of meson masses. Normally these modes would contribute nothing
to $L_{{\rm boundary}}$, since they appear at $\nu = 0$. However,
two of these modes are in the $\phi$ sector, and due to the
frequency shifting in that sector they appear as poles in
$\Gamma_\phi^{ij}$ at $\nu = \pm\omega$.
These modes contribute to $L_{{\rm boundary}}$ as harmonic
oscillators with frequency $\omega$, so the contribution
of the translation modes to $L_{{\rm boundary}}$ is
\begin{equation}
L_{{\rm boundary}}^{{\rm translation}} = 0 - \frac{\omega}{2} - \frac{\omega}{2}
= -\omega \,.
\end{equation}
Subtracting this contribution from $L_{{\rm boundary}}$ gives
\begin{eqnarray}
L_{{\rm boundary}} &=& -\frac{1}{2} \int \frac{d\nu}{2\pi} \Bigg\{
\tr\ln\left[ \delta_{ij} \frac{\nu^2 + \omega^2}{\nu^2}
+ \frac{\omega^3}{\nu^2} \sqrt{\frac{\bar\gamma_i \bar\gamma_j}{v_i v_j}}
G^{ij}_\theta(-i\nu) \right]
\nonumber \\
& & + \tr\ln\left[ \delta_{ij} \frac{\nu^2 + (2\bar\gamma_i^2 + 1) \omega^2}
{\nu^2 - (2\bar\gamma_i^2 - 1) \omega^2}
+ \frac{\omega^3}{\nu^2} \frac{1}{\sqrt{v_i\bar\gamma_i v_j \bar\gamma_j}}
G^{ij}_\phi(-i\nu) \right] \Bigg\} + \omega \,.
\label{L boundary no trans}
\end{eqnarray}

We see that there is a logarithmic divergence 
in $L_{{\rm boundary}}$ by noting that, without
a cutoff, $G^{ij}_\theta(-i\nu)$ and $G^{ij}_\phi(-i\nu)$ are
\begin{eqnarray}
G^{ij}_\theta(-i\nu) &=& \delta_{ij} \left( - \frac{\bar R_i}{\bar\gamma_i}
+ \nu \bar R_i^2 \coth(\nu R_p) \right)
- (1-\delta_{ij}) \nu \bar R_i \bar R_j \csch(\nu R_p) \,,
\nonumber \\
G^{ij}_\phi(-i\nu) &=& \delta_{ij} \bar\gamma_i \bar R_i
\frac{\nu^2}{\omega^2}
- \frac{\nu}{\omega^3} (\nu^2 + \omega^2)
\nonumber \\
& & \times \frac{ \delta_{ij} \left(\nu v_i \bar\gamma_i \sinh(\nu R_p)
+ \omega v_1 \bar\gamma_1 v_2 \bar\gamma_2 \cosh(\nu R_p)\right)
- (1-\delta_{ij}) \omega v_i \bar\gamma_1 v_2 \bar\gamma_2}
{\left(\nu^2 + \omega^2 v_1 \bar\gamma_1 v_2 \bar\gamma_2\right)
\sinh(\nu R_p) + \nu \omega \left(v_1\bar\gamma_1 + v_2\bar\gamma_2\right)
\cosh(\nu R_p)} \,.
\label{G^ij Wick rotate}
\end{eqnarray}
The functions \eqnlessref{G^ij Wick rotate} are proportional to
$\nu$ in the large $\nu$ limit. We pull this divergence outside
of the logarithm by adding and subtracting the trace of $G^{ij}$
from the integrals. This breaks \eqnlessref{L boundary no trans}
into four parts,
\begin{equation}
L_{{\rm boundary}} = L^{\hbox{\scriptsize log term}}_\theta
+ L^{\hbox{\scriptsize log term}}_\phi
+ L^{{\rm cutoff}}_\theta + L^{{\rm cutoff}}_\phi \,,
\end{equation}
where
\begin{eqnarray}
L^{\hbox{\scriptsize log term}}_\theta &=& - \frac{1}{2} \int \frac{d\nu}{2\pi} \Bigg\{
\tr\ln\left[ \delta_{ij} \left(1 + \frac{\omega v_i \bar\gamma_i}{\nu}
\coth\left(\nu R_p\right)\right)
- \left(1 - \delta_{ij}\right) \frac{\omega}{\nu} \sqrt{v_1 \bar\gamma_1
v_2 \bar\gamma_2} \csch\left(\nu R_p\right) \right]
\nonumber \\
& & - \sum_{i=1}^2 \frac{\omega^3}{\nu^2} v_i \bar\gamma_i
\left(\nu \bar R_i^2 \coth(\nu R_p)  - \frac{\bar R_i^2}{R_p} \right) \Bigg\} \,,
\nonumber \\
L^{\hbox{\scriptsize log term}}_\phi &=& - \frac{1}{2} \int \frac{d\nu}{2\pi} \Bigg\{
\tr\ln\Bigg[ 2 \delta_{ij} \frac{\nu^2 + \omega^2}
{\nu^2 - (2\bar\gamma_i^2 - 1) \omega^2} - (\nu^2 + \omega^2)
\nonumber \\
& & \times \frac{ \delta_{ij} \left(\sinh(\nu R_p)
+ \frac{\omega v_1 \bar\gamma_1 v_2 \bar\gamma_2}{\nu v_i \bar\gamma_i}
\cosh(\nu R_p)\right)
- (1-\delta_{ij}) \frac{\omega}{\nu} \sqrt{v_i \bar\gamma_1 v_2 \bar\gamma_2}}
{\left(\nu^2 + \omega^2 v_1 \bar\gamma_1 v_2 \bar\gamma_2\right)
\sinh(\nu R_p) + \nu \omega \left(v_1\bar\gamma_1 + v_2\bar\gamma_2\right)
\cosh(\nu R_p)} \Bigg]
\nonumber \\
& & - \sum_{i=1}^2 \frac{\omega^3}{\nu^2} \frac{1}{v_i \bar\gamma_i} \Bigg[
\bar\gamma_i \bar R_i \frac{\nu^2}{\omega^2}
+ \frac{v_1 \bar\gamma_1 v_2 \bar\gamma_2}{\omega (v_1 \bar\gamma_1
+ v_2 \bar\gamma_2) + \omega^2 R_p v_1 \bar\gamma_1 v_2 \bar\gamma_2}
- \frac{\nu}{\omega^3} (\nu^2 + \omega^2)
\nonumber \\
& & \times \frac{ \nu v_i \bar\gamma_i \sinh(\nu R_p)
+ \omega v_1 \bar\gamma_1 v_2 \bar\gamma_2 \cosh(\nu R_p)}
{\left(\nu^2 + \omega^2 v_1 \bar\gamma_1 v_2 \bar\gamma_2\right)
\sinh(\nu R_p) + \nu \omega \left(v_1\bar\gamma_1 + v_2\bar\gamma_2\right)
\cosh(\nu R_p)} \Bigg] \Bigg\} + \omega \,,
\end{eqnarray}
and where
\begin{eqnarray}
L^{{\rm cutoff}}_\theta &=& - \frac{1}{2} \int \frac{d\nu}{2\pi}
\sum_{i=1}^2 \frac{\omega^3}{\nu^2} \frac{\bar\gamma_i}{v_i}
\left( G^{ii}_\theta(-i\nu) - G^{ii}_\theta(0) \right) \,,
\nonumber \\
L^{{\rm cutoff}}_\phi &=& - \frac{1}{2} \int \frac{d\nu}{2\pi}
\sum_{i=1}^2 \frac{\omega^3}{\nu^2} \frac{1}{v_i \bar\gamma_i}
\left( G^{ii}_\phi(-i\nu) - G^{ii}_\phi(0) \right) \,.
\label{L boundary parts}
\end{eqnarray}
We have grouped the term $\omega$ with the $\phi$ sector,
since it was introduced to cancel the contribution of the
translation modes in the $\phi$ sector.
The presence of $G^{ii}_\theta(0)$ and $G^{ii}_\phi(0)$
in the cutoff terms removes
a divergence at $\nu = 0$. The logarithmic
divergence is now entirely in $L^{{\rm cutoff}}_\theta$
and $L^{{\rm cutoff}}_\phi$. Since $L^{\hbox{\scriptsize log term}}_\theta$
and $L^{\hbox{\scriptsize log term}}_\phi$ are cutoff independent,
we have inserted the explicit functions \eqnlessref{G^ij Wick rotate}
into their definitions \eqnlessref{L boundary parts}.

Naive insertion of the functions \eqnlessref{G^ij Wick rotate}
into $L^{{\rm cutoff}}_\theta$ and $L^{{\rm cutoff}}_\phi$
in \eqnlessref{L boundary no trans} would make these integrals divergent.
We must therefore include the dependence of the Green's function
$G(r,r',\nu)$ on the cutoff $\Lambda$ in the definitions of
$G^{ij}_\theta$ and $G^{ij}_\phi$.
We determine the cutoff dependence of $G(r,r',\nu)$
by writing it as a sum of functions $s_n(r)$,
\begin{equation}
G(r,r',\nu) = \sum_{n=1}^{n_{{\rm max}}} \frac{1}{\chi_n - \nu^2}
\frac{s_n(r) s_n(r')}{\int_{-\bar R_1}^{\bar R_2} dr'' \Sigma(r'') s_n^2(r'')} \,,
\label{Green's function eigen expand}
\end{equation}
which satisfy the eigenfunction equation
\begin{equation}
\left(- \frac{\partial}{\partial r} \Sigma(r) \gamma^{-2}
\frac{\partial}{\partial r} + \Sigma(r) (\chi_n + C) \right) s_n(r) = 0 \,,
\end{equation}
with eigenvalue $\chi_n$ and boundary conditions
\begin{equation}
s_n(-\bar R_1) = s_n(\bar R_2) = 0 \,.
\end{equation}
The upper limit of the sum in \eqnlessref{Green's function eigen expand}
is defined by the equation
\begin{equation}
\chi_{n_{{\rm max}}} \le \Lambda^2 < \chi_{n_{{\rm max}} + 1} \,.
\end{equation}
Inserting \eqnlessref{Green's function eigen expand} in the definition
\eqnlessref{G^ij def} for the $G^{ij}$ gives
\begin{equation}
G^{ij}_\psi(\nu) \equiv (-1)^{i+j} \Sigma((-1)^i \bar R_i)
\Sigma((-1)^j \bar R_j) \bar\gamma_i^{-2} \bar\gamma_j^{-2}
\sum_{n=1}^{n_{{\rm max}}} \frac{1}{\chi_n - \nu^2}
\frac{s_n'((-1)^i \bar R_i) s_n'((-1)^i \bar R_j)}
{\int_{-\bar R_1}^{\bar R_2} dr'' \Sigma(r'') s_n^2(r'')} \,.
\label{G^ij cutoff}
\end{equation}

The cutoff dependent part of $L_{{\rm boundary}}$ has the form
\begin{equation}
L^{{\rm cutoff}}_\psi = - \frac{1}{2} \int \frac{d\nu}{2\pi}
\sum_{i=1}^2 \frac{\omega^3}{\nu^2} \frac{v_i \bar\gamma_i^2}
{\omega^2 \Sigma((-1)^i \bar R_i)}
\left( G^{ii}_\psi(-i\nu) - G^{ii}_\psi(0) \right) \,.
\label{L cutoff gen def}
\end{equation}
Inserting \eqnlessref{G^ij cutoff} into \eqnlessref{L cutoff gen def} gives
\begin{eqnarray}
L^{{\rm cutoff}}_\psi &=& - \frac{1}{2} \int \frac{d\nu}{2\pi}
\sum_{i=1}^2 \frac{\omega}{\nu^2}
v_i \bar\gamma_i^{-2} \Sigma((-1)^i \bar R_i)
\sum_{n=1}^{n_{{\rm max}}}\frac{\left(s_n'((-1)^i \bar R_i)\right)^2}
{\int_{-\bar R_1}^{\bar R_2} dr'' \Sigma(r'') s_n^2(r'')}
\left(\frac{1}{\chi_n + \nu^2} - \frac{1}{\chi_n}\right)
\nonumber \\
&=& - \frac{\omega}{2} \sum_{i=1}^2
v_i \bar\gamma_i^{-2} \Sigma((-1)^i \bar R_i)
\sum_{n=1}^{n_{{\rm max}}} \chi_n^{-3/2}
\frac{\left(s_n'((-1)^i \bar R_i)\right)^2}
{\int_{-\bar R_1}^{\bar R_2} dr'' \Sigma(r'') s_n^2(r'')} \,.
\label{L cutoff formula}
\end{eqnarray}

For the $\theta$ sector, the eigenfunctions $s_n(r)$ are
\begin{equation}
s_n(r) = \frac{1}{r} k_n\left(\frac{\arcsin \omega r}{\omega}\right) \,,
\label{theta eigenfunctions for G}
\end{equation}
where $k_n$ is defined by
\begin{equation}
k_n(x) = \sin\left(\frac{\pi n}{R_p} (x + X_1)\right) \,,
\label{theta eigenfunctions}
\end{equation}
and the eigenvalues $\chi_n$ are
\begin{equation}
\chi_n = \left( \frac{\pi n}{R_p} \right)^2 \,.
\label{appendix theta eigenvalues}
\end{equation}
Replacing the $s_n(r)$ with \eqnlessref{theta eigenfunctions for G}
gives
\begin{equation}
L^{{\rm cutoff}}_\theta = - \frac{1}{2} \sum_{i=1}^2
\bar\gamma_i v_i \omega \sum_{n=1}^{\Lambda R_p/\pi} \frac{1}{\pi n} \,.
\end{equation}
Replacing the sum over $n$ with a contour integral with poles
at $z = \pi n/R_p$ gives
\begin{equation}
L^{{\rm cutoff}}_\theta = - \frac{1}{4\pi i} \sum_{i=1}^2
\bar\gamma_i v_i \omega \int dz \frac{1}{z}
\cot(R_p z) \,.
\label{L cutoff theta contour integral}
\end{equation}
The contour runs along the line $\Re\, z = \pi/(2R_p)$ and along a semicircle
where $|z| = \Lambda$, the cutoff, and the real part of $z$ is positive.

We divide \eqnlessref{L cutoff theta contour integral}
into two integrals over the parts of the contour to get
\begin{eqnarray}
L^{{\rm cutoff}}_\theta &=& - \frac{1}{4\pi} \sum_{i=1}^2
\bar\gamma_i v_i \omega \Bigg[
\int_{-\sqrt{\Lambda^2 - \pi^2/4R_p^2}}^{\sqrt{\Lambda^2 - \pi^2/4R_p^2}}
dy \frac{1}{y - i \frac{\pi}{2R_p}} \tanh(R_p y)
\nonumber \\
& & + \int_{-\arccos(\pi/2R_p \Lambda)}^{\arccos(\pi/2R_p \Lambda)}
d\theta \cot\left(\Lambda R_p e^{i\theta}\right) \Bigg] \,.
\end{eqnarray}
The cotangent in the $\theta$ integral is proportional to
the sign of $\theta$ for large $\Lambda$, so the $\theta$ integral
vanishes. The integral over $y$ is real, because the imaginary
part of the integrand changes sign when $y\to -y$,
\begin{equation}
L^{{\rm cutoff}}_\theta = - \frac{1}{2\pi} \sum_i
\bar\gamma_i v_i \omega \int_0^{\sqrt{\Lambda^2 - \pi^2/4R_p^2}}
dy \frac{y}{y^2 + \frac{\pi^2}{4R_p^2}} \tanh(R_p y) \,.
\end{equation}
We extract the cutoff dependence from the integral to get
\begin{equation}
L^{{\rm cutoff}}_\theta = - \sum_i \frac{\omega v_i \bar\gamma_i}{2\pi}
\left[\ln\left(\frac{2 M R_p}{\pi \bar\gamma_i}\right)
+ \int_0^\infty du \frac{u}{u^2 + \frac{\pi^2}{4}}
\left( \tanh u - 1 \right) \right] \,.
\label{L cutoff theta with M}
\end{equation}
We have replaced $\Lambda$, the cutoff for the coordinate $x$,
by $M/\bar\gamma_i$, where $M$ is the cutoff for the physical
coordinate $r$, as we did in \eqnref{cutoff coordinate change}.
We have also changed integration variables to $u = y R_p$.

The logarithmic dependence of \eqnlessref{L cutoff theta with M}
on $M$ is removed by renormalization of the coefficient $\kappa$
of the geodesic curvature term
defined in \eqnlessref{boundary action}. Since the geodesic curvature diverges in
the small quark mass limit, our choice of renormalization
point determines the coefficient of the leading term in the
effective Lagrangian in that limit.
In the final result, the renormalized geodesic curvature term
will cancel the divergence of the semiclassical corrections
in the small quark mass limit.
We therefore choose the renormalization
point for which there is no small quark mass divergence in
the semiclassical corrections. We must evaluate both
$L^{\hbox{\scriptsize log term}}_\theta$ and
$L^{{\rm cutoff}}_\theta$,
since both contribute to the small mass limit divergence.

The term $L^{\hbox{\scriptsize log term}}_\theta$
in $L_{{\rm boundary}}$ is
\begin{eqnarray}
L^{\hbox{\scriptsize log term}}_\theta &=& - \frac{1}{2} \int \frac{d\nu}{2\pi} \Bigg\{
\ln\left[ 1 + \frac{\omega}{\nu} \left(v_1 \bar\gamma_1
+ v_2 \bar\gamma_2\right) \coth(\nu R_p) + \frac{\omega^2}{\nu^2}
v_1 \bar\gamma_1 v_2 \bar\gamma_2 \right]
\nonumber \\
& & - \frac{\omega v_i \bar\gamma_i}{\nu}
\sum_i \left[ \coth(\nu R_p) - \frac{1}{\nu R_p} \right] \Bigg\} \,.
\end{eqnarray}
We can simplify this integral by rewriting the integrand,
\begin{eqnarray}
L^{\hbox{\scriptsize log term}}_\theta &=& - \int_0^\infty \frac{d\nu}{2\pi}
\ln\left[ 1 + \frac{\omega \nu \left( v_1 \bar\gamma_1
+ v_2 \bar\gamma_2 \right)}{\nu^2 + \nu \omega \left( v_1 \bar\gamma_1
+ v_2 \bar\gamma_2 \right) + \omega^2 v_1 \bar\gamma_1
v_2 \bar\gamma_2} \left( \coth(\nu R_p) - 1 \right) \right]
\nonumber \\
& & + \int_0^\infty \frac{d\nu}{2\pi} \left\{
\ln\left[ \left( 1 + \frac{\omega}{\nu} v_1 \bar\gamma_1 \right)
\left( 1 + \frac{\omega}{\nu} v_2 \bar\gamma_2 \right) \right]
- \sum_i \frac{\omega v_i \bar\gamma_i}{\nu}
\left[ \coth(\nu R_p) - \frac{1}{\nu R_p} \right] \right\} \,.
\nonumber \\
\label{theta log term separated}
\end{eqnarray}
Integrating by parts in the second integral gives
\begin{eqnarray}
L^{\hbox{\scriptsize log term}}_\theta &=& - \int_0^\infty \frac{d\nu}{2\pi}
\ln\left[ 1 + \frac{\omega \nu \left( v_1 \bar\gamma_1
+ v_2 \bar\gamma_2 \right)}{\nu^2 + \nu \omega \left( v_1 \bar\gamma_1
+ v_2 \bar\gamma_2 \right) + \omega^2 v_1 \bar\gamma_1
v_2 \bar\gamma_2} \left( \coth(\nu R_p) - 1 \right) \right]
\nonumber \\
& & + \sum_i \frac{\omega v_i \bar\gamma_i}{2\pi} \Bigg\{
\left[ \frac{\nu + \omega v_i \bar\gamma_i}{\omega v_i \bar\gamma_i}
\ln\left( \nu + \omega v_i \bar\gamma_i \right)
- \frac{\nu \ln\nu}{\omega v_i \bar\gamma_i}
- \ln(\nu R_p)\left( \coth(\nu R_p) - \frac{1}{\nu R_p} \right) \right]
\Bigg|_0^\infty \negspace
\nonumber \\
& & - R_p \int_0^\infty d\nu \ln(\nu R_p)\left[ \csch^2(\nu R_p)
- \frac{1}{\nu^2 R_p^2} \right] \Bigg\} \,.
\end{eqnarray}
Evaluating this expression and making the change of variables
$t = \nu R_p$ gives
\begin{eqnarray}
L^{\hbox{\scriptsize log term}}_\theta &=& - \int_0^\infty \frac{d\nu}{2\pi}
\ln\left[ 1 + \frac{\omega \nu \left( v_1 \bar\gamma_1
+ v_2 \bar\gamma_2 \right)}{\nu^2 + \nu \omega \left( v_1 \bar\gamma_1
+ v_2 \bar\gamma_2 \right) + \omega^2 v_1 \bar\gamma_1
v_2 \bar\gamma_2} \left( \coth(\nu R_p) - 1 \right) \right]
\nonumber \\
& & + \sum_i \frac{\omega v_i \bar\gamma_i}{2\pi} \Bigg\{
-\ln\left(R_p \omega v_i \bar\gamma_i\right) + 1
- \int_0^\infty dt \ln t \left[ \csch^2(t) - \frac{1}{t^2} \right] \Bigg\} \,.
\end{eqnarray}

The sum of $L^{{\rm cutoff}}_\theta$ and $L^{\hbox{\scriptsize log term}}_\theta$
is
\begin{eqnarray}
L^{{\rm cutoff}}_\theta + L^{\hbox{\scriptsize log term}}_\theta &=&
- \int_0^\infty \frac{d\nu}{2\pi}
\ln\left[ 1 + \frac{\omega \nu \left( v_1 \bar\gamma_1
+ v_2 \bar\gamma_2 \right)}{\nu^2 + \nu \omega \left( v_1 \bar\gamma_1
+ v_2 \bar\gamma_2 \right) + \omega^2 v_1 \bar\gamma_1
v_2 \bar\gamma_2} \left( \coth(\nu R_p) - 1 \right) \right]
\nonumber \\
& & + \sum_i \frac{\omega v_i \bar\gamma_i}{2\pi} \Bigg\{
\ln\left(\frac{2M m_i}{\pi\sigma}\right) + 1
+ \int_0^\infty du \frac{u}{u^2 + \frac{\pi^2}{4}} \left( \tanh u - 1 \right)
\nonumber \\
& & - \int_0^\infty dt \ln t \left[ \csch^2(t) - \frac{1}{t^2} \right] \Bigg\} \,,
\end{eqnarray}
where we have used the classical equation of motion \eqnlessref{classical eqn}
to simplify the argument of the logarithm. The terms in the sum
are renormalizations of the geodesic curvature, so, after renormalization,
\begin{equation}
L^{{\rm cutoff}}_\theta + L^{\hbox{\scriptsize log term}}_\theta =
- \int_0^\infty \frac{d\nu}{2\pi}
\ln\left[ 1 + \frac{\omega \nu \left( v_1 \bar\gamma_1
+ v_2 \bar\gamma_2 \right)}{\nu^2 + \nu \omega \left( v_1 \bar\gamma_1
+ v_2 \bar\gamma_2 \right) + \omega^2 v_1 \bar\gamma_1
v_2 \bar\gamma_2} \left( \coth(\nu R_p) - 1 \right) \right] \,.
\label{L theta renorm}
\end{equation}
For two massless quarks the integral \eqnlessref{L theta renorm} is zero,
and for one massless and one heavy
quark it has the value $\omega/8 + O(v_{{\rm heavy}})$.

We next evaluate $L^{{\rm cutoff}}_\phi$, using the formula
\eqnlessref{L cutoff formula} as we did for $L^{{\rm cutoff}}_\theta$.
For the $\phi$ sector, the eigenfunctions $s_n(r)$ are
\begin{equation}
s_n(r) = \frac{1}{\gamma r} f_n\left(\frac{\arcsin \omega r}{\omega}\right) \,,
\label{phi eigenfunctions for G}
\end{equation}
where $f_n$ is defined by
\begin{equation}
f_n(x) = \sqrt{\chi_n} \cos\left(\sqrt{\chi_n} x + \delta_n\right)
+ \omega \tan(\omega x) \sin\left(\sqrt{\chi_n} x + \delta_n\right) \,,
\label{phi eigenfunctions}
\end{equation}
and the eigenvalues $\chi_n$ satisfy the equation
\begin{equation}
\tan(\sqrt{\chi_n} R_p) = \sqrt{\chi_n} \omega
\frac{v_1 \bar\gamma_1 + v_2 \bar\gamma_2}
{\chi_n - \omega^2 v_1\bar\gamma_1 v_2\bar\gamma_2} \,.
\label{phi eigenvalue def}
\end{equation}
The solution $\chi_n = \omega^2$ to \eqnlessref{phi eigenvalue def}
is not a valid eigenvalue, as it causes \eqnlessref{phi eigenfunctions}
to vanish everywhere. The phases $\delta_n$ are
\begin{eqnarray}
\delta_n &=& \frac{\sqrt{\chi_n}}{\omega} \arcsin v_1
+ \arctan\left(\frac{\sqrt{\chi_n}}{\omega v_1 \bar\gamma_1}\right)
\nonumber \\
&=& -\frac{\sqrt{\chi_n}}{\omega} \arcsin v_2
- \arctan\left(\frac{\sqrt{\chi_n}}{\omega v_2 \bar\gamma_2}\right) \,.
\end{eqnarray}
The definition \eqnlessref{phi eigenvalue def} of $\chi_n$ makes the
two definitions for $\delta_n$ equivalent.

Replacing the $s_n(r)$ with \eqnlessref{phi eigenfunctions for G}
gives
\begin{eqnarray}
L^{{\rm cutoff}}_\phi &=& - \frac{1}{2} \sum_i \bar\gamma_i v_i
\omega \sum_n \frac{1}{\sqrt{\nu_n}}
\frac{(\nu_n - \omega^2)}
{\left(\nu_n + \omega^2 v_i^2 \bar\gamma_i^2\right)}
\left[ R_p + \frac{\omega v_1 \bar\gamma_1}{\nu_n
+ \omega^2 v_1^2 \bar\gamma_1^2}
+ \frac{\omega v_2 \bar\gamma_2}{\nu_n
+ \omega^2 v_2^2 \bar\gamma_2^2} \right]^{-1}
\nonumber \\
&=& \frac{1}{2} \sum_n (\nu_n - \omega^2) \frac{1}
{\sqrt{\nu_n}} \left[ 1 - \frac{R_p}{R_p
+ \frac{\omega v_1 \bar\gamma_1}{\nu_n
+ \omega^2 v_1^2 \bar\gamma_1^2}
+ \frac{\omega v_2 \bar\gamma_2}{\nu_n
+ \omega^2 v_2^2 \bar\gamma_2^2}} \right] \,.
\label{L cutoff phi sum}
\end{eqnarray}
The function
\begin{equation}
F(z) = \frac{d}{dz} \ln\left[\frac{\left(z^2
- \omega^2 v_1 \bar\gamma_1 v_2 \bar\gamma_2\right) \sin\left(R_p z\right)
- \omega z \left(v_1\bar\gamma_1 + v_2\bar\gamma_2\right)
\cos\left(R_p z\right)}{z^2 - \omega^2}\right] \,,
\end{equation}
has poles of residue one at $z = \pm \sqrt{\nu_n}$.
We rewrite the sum \eqnlessref{L cutoff phi sum}
as a contour integral
\begin{equation}
L^{{\rm cutoff}}_\phi = - \frac{1}{4\pi i}
\int dz (z^2-\omega^2) \frac{1}{z}
\left[ 1 - \frac{R_p}{R_p
+ \frac{\omega v_1 \bar\gamma_1}{z^2
+ \omega^2 v_1^2 \bar\gamma_1^2}
+ \frac{\omega v_2 \bar\gamma_2}{z^2
+ \omega^2 v_2^2 \bar\gamma_2^2}} \right]
F(z) \,.
\label{L cutoff phi contour integral}
\end{equation}
The contour is the same as for \eqnref{L cutoff theta contour integral}.
The poles in the term in square brackets in
\eqnlessref{L cutoff phi contour integral} all
lie on the imaginary axis, which lies outside
of the integration contour. Evaluating $F(z)$ gives
\begin{eqnarray}
L^{{\rm cutoff}}_\phi &=& - \frac{1}{4\pi i}
\int dz (z^2 - \omega^2) \frac{1}{z}
\left[ 1 - \frac{R_p}{R_p
+ \frac{\omega v_1 \bar\gamma_1}{z^2
+ \omega^2 v_1^2 \bar\gamma_1^2}
+ \frac{\omega v_2 \bar\gamma_2}{z^2
+ \omega^2 v_2^2 \bar\gamma_2^2}} \right]
\nonumber \\
& & \times \Bigg\{ \Big[
\left(2 + R_p \omega \left(v_1\bar\gamma_1
+ v_2\bar\gamma_2\right)\right) z \sin\left(R_p z\right)
\nonumber \\
& & + \left(z^2 R_p - \omega \left(v_1\bar\gamma_1 + v_2\bar\gamma_2\right)
- \omega^2 R_p v_1 \bar\gamma_1 v_2 \bar\gamma_2\right)
\cos\left(R_p z\right)\Big]
\nonumber \\
& & \times \left[\left(z^2 - \omega^2 v_1 \bar\gamma_1
v_2 \bar\gamma_2\right) \sin\left(R_p z\right)
- \omega z \left(v_1\bar\gamma_1 + v_2\bar\gamma_2\right)
\cos\left(R_p z\right)\right]^{-1} - \frac{2z}{z^2 - \omega^2} \Bigg\} \,.
\nonumber \\
\end{eqnarray}
The term $2z/(z^2-\omega^2)$ does not contribute to the integral,
since its poles are canceled by the factor
of $z^2 - \omega^2$ in the integrand.

We divide the integral over $z$ into an integral over the line
at $\Re\, z = \pi/(2R_p)$ and an integral over the semicircle
at $|z| = \Lambda$. The integral over the semicircle vanishes,
and the integral over the line is
\begin{eqnarray}
L^{{\rm cutoff}}_\phi &=& - \frac{1}{4\pi}
\int_{-\sqrt{\Lambda^2 - \pi^2/4R_p^2}}^{\sqrt{\Lambda^2 - \pi^2/4R_p^2}}
dy \left(\left( y - i \frac{\pi}{2R_p}\right)^2 + \omega^2 \right)
\frac{1}{y - i \frac{\pi}{2R_p}}
\nonumber \\
& & \times \left[ R_p \left(R_p
- \frac{\omega v_1 \bar\gamma_1}{\left( y - i \frac{\pi}{2R_p}\right)^2
- \omega^2 v_1^2 \bar\gamma_1^2}
- \frac{\omega v_2 \bar\gamma_2}{\left( y - i \frac{\pi}{2R_p}\right)^2
- \omega^2 v_2^2 \bar\gamma_2^2}\right)^{-1} - 1 \right]
\nonumber \\
& & \times \Bigg[\left( \left( y - i \frac{\pi}{2R_p}\right)^2 R_p
+ \omega \left(v_1\bar\gamma_1 + v_2\bar\gamma_2\right)
+ \omega^2 R_p v_1 \bar\gamma_1 v_2 \bar\gamma_2\right)
\sinh\left(R_p y\right)
\nonumber \\
& & + \left(2 + R_p \omega \left(v_1\bar\gamma_1
+ v_2\bar\gamma_2\right)\right) \left( y - i \frac{\pi}{2R_p}\right)
 \cosh\left(R_p y\right)\Bigg]
\nonumber \\
& & \times \Bigg[\left(\left( y - i \frac{\pi}{2R_p}\right)^2
+ \omega^2 v_1 \bar\gamma_1 v_2 \bar\gamma_2\right)
\cosh\left(R_p y\right)
\nonumber \\
& & + \omega \left( y - i \frac{\pi}{2R_p}\right)
\left(v_1\bar\gamma_1 + v_2\bar\gamma_2\right)
\sinh\left(R_p y\right)\Bigg]^{-1} \,.
\end{eqnarray}
Extracting the divergent part gives
\begin{eqnarray}
L^{{\rm cutoff}}_\phi &=& - \sum_i \frac{v_i^2 \bar\gamma_i}{2\pi \bar R_i}
\ln\left(\frac{2 M R_p}{\pi \bar\gamma_i}\right)
\nonumber \\
& & + \frac{1}{2\pi} \int_0^\infty dy
\Re\Bigg\{ \left(\left( y - i \frac{\pi}{2R_p}\right)^2 + \omega^2 \right)
\frac{1}{y - i \frac{\pi}{2R_p}}
\nonumber \\
& & \times \frac{\frac{\omega v_1 \bar\gamma_1}{\left( y - i \pi/2R_p\right)^2
- \omega^2 v_1^2 \bar\gamma_1^2}
+ \frac{\omega v_2 \bar\gamma_2}{\left( y - i \pi/2R_p\right)^2
- \omega^2 v_2^2 \bar\gamma_2^2}}
{R_p
- \frac{\omega v_1 \bar\gamma_1}{\left( y - i \pi/2R_p\right)^2
- \omega^2 v_1^2 \bar\gamma_1^2}
- \frac{\omega v_2 \bar\gamma_2}{\left( y - i \pi/2R_p\right)^2
- \omega^2 v_2^2 \bar\gamma_2^2}}
\nonumber \\
& & \times \Bigg[\left( \left( y - i \frac{\pi}{2R_p}\right)^2 R_p
+ \omega \left(v_1\bar\gamma_1 + v_2\bar\gamma_2\right)
+ \omega^2 R_p v_1 \bar\gamma_1 v_2 \bar\gamma_2\right)
\sinh\left(R_p y\right)
\nonumber \\
& & + \left(2 + R_p \omega \left(v_1\bar\gamma_1
+ v_2\bar\gamma_2\right)\right) \left( y - i \frac{\pi}{2R_p}\right)
 \cosh\left(R_p y\right)\Bigg]
\nonumber \\
& & \times \Bigg[\left(\left( y - i \frac{\pi}{2R_p}\right)^2
+ \omega^2 v_1 \bar\gamma_1 v_2 \bar\gamma_2\right)
\cosh\left(R_p y\right)
\nonumber \\
& & + \omega \left( y - i \frac{\pi}{2R_p}\right)
\left(v_1\bar\gamma_1 + v_2\bar\gamma_2\right)
\sinh\left(R_p y\right)\Bigg]^{-1}
- \sum_i \frac{\omega v_i \bar\gamma_i}{y - i \frac{\pi}{2R_p}}
\Bigg\} \,,
\label{L phi cutoff final}
\end{eqnarray}
where we have replaced $\Lambda$ with $M/\bar\gamma_i$ just
as we did with $L^{{\rm cutoff}}_\theta$.

We are interested in the value of $L^{{\rm cutoff}}_\theta$
in two limits, one where $\bar\gamma_1, \bar\gamma_2 \to \infty$,
and one where $v_1 \ll 1$, $\bar\gamma_2 \to \infty$.
In the first limit, the integral
\eqnlessref{L phi cutoff final} is dominated by the
region where $y$ is large. If $y$ is of order $\omega$,
the term in the integrand which contains the
hyperbolic functions is of order $\bar\gamma_i^{-1}$,
so we can make the approximation
\begin{equation}
\sinh\left(R_p y\right) \simeq \cosh\left(R_p y\right)
\simeq \frac{1}{2} e^{R_p y}
\end{equation}
to simplify the integrand,
\begin{eqnarray}
L^{{\rm cutoff}}_\phi &=& - \sum_i \frac{\omega v_i \bar\gamma_i}{2\pi}
\ln\left(\frac{2MR_p}{\pi \bar\gamma_i}\right)
+ \frac{1}{2\pi} \Re \int_0^\infty \frac{dy}{y - i \frac{\pi}{2R_p}}
\Bigg\{ \left( \left(y - i \frac{\pi}{2R_p}\right)^2 + \omega^2 \right)
\nonumber \\
& & \times \left( \sum_i \frac{\omega v_i \bar\gamma_i}
{\left(y - i \frac{\pi}{2R_p}\right)^2 - \omega^2 v_i^2 \bar\gamma_i^2}
\right) \frac{R_p + \sum_i \frac{1}{y - i \frac{\pi}{2R_p}
+ \omega v_i \bar\gamma_i}}{R_p - \sum_i \frac{\omega v_i \bar\gamma_i}
{\left(y - i \frac{\pi}{2R_p}\right)^2 - \omega^2 v_i^2 \bar\gamma_i^2}}
- \sum_i \omega v_i \bar\gamma_i \Bigg\} + O(\bar\gamma_i^{-1}) \,.
\nonumber \\
\end{eqnarray}
The terms in the integrand can be rearranged to extract the most
important part of the integrand for large $y$,
\begin{eqnarray}
L^{{\rm cutoff}}_\phi &=& - \sum_i
\frac{\omega v_i \bar\gamma_i}{2\pi} \left[
\ln\left(\frac{2MR_p}{\pi \bar\gamma_i}\right)
+ \Re \int_0^\infty \frac{dy}{y - i \frac{\pi}{2R_p}}
\left( \frac{\left(y - i \frac{\pi}{2R_p}\right)^2 + \omega^2}
{\left(y - i \frac{\pi}{2R_p}\right)^2 - \omega^2 v_i^2 \bar\gamma_i^2}
- 1 \right) \right]
\nonumber \\
& & + \frac{1}{2\pi} \Re \int_0^\infty dy
\left( \left(y - i \frac{\pi}{2R_p}\right)^2 + \omega^2 \right)
\left( \sum_i \frac{\omega v_i \bar\gamma_i}
{\left(y - i \frac{\pi}{2R_p}\right)^2 - \omega^2 v_i^2 \bar\gamma_i^2}
\right)
\nonumber \\
& & \times \left( \sum_i \frac{1} {\left(y - i \frac{\pi}{2R_p}\right)^2
- \omega^2 v_i^2 \bar\gamma_i^2} \right)
\left( R_p - \sum_i \frac{\omega v_i \bar\gamma_i}
{\left(y - i \frac{\pi}{2R_p}\right)^2 - \omega^2 v_i^2 \bar\gamma_i^2}
\right)^{-1} + O(\bar\gamma_i^{-1}) \,.
\nonumber \\
\label{split L phi cutoff}
\end{eqnarray}
The second integral in \eqnlessref{split L phi cutoff}
is zero. We can rewrite it as an integral from $-\infty$ to $\infty$,
since the real part of the integrand is symmetric when $y \to -y$.
We can then convert it into an integral over a closed contour
by adding a semicircle which passes through $y = -i\infty$.
The integrand has no poles inside this contour, so the integral is
zero. The first integral in \eqnlessref{split L phi cutoff}
can be done exactly, giving
\begin{equation}
L^{{\rm cutoff}}_\phi = - \sum_i \frac{\omega v_i \bar\gamma_i}{2\pi}
\ln\left(\frac{M}{\omega v_i \bar\gamma_i^2}\right)
+ O(\bar\gamma_1^{-1},\bar\gamma_2^{-1}) \,,
\label{L phi asymptote}
\end{equation}
in the limit $\bar\gamma_1, \bar\gamma_2 \to \infty$.

In the limit $v_1 \ll 1$, $\bar\gamma_2 \to \infty$,
$L^{{\rm cutoff}}_\phi$ splits into two parts. One part,
dominated by $y = \omega v_2 \bar\gamma_2 + O(1)$, is evaluated the
same way as in the light--light case. The other part is dominated
by small $y$ and is handled differently. Using the result
\eqnlessref{L phi asymptote} to evaluate the first part
and taking the limit $\bar\gamma_2 \to \infty$ in the
second part gives
\begin{eqnarray}
L^{{\rm cutoff}}_\phi &=& - \frac{\omega v_1 \bar\gamma_1}{2\pi}
\ln\left(\frac{2 M R_p}{\pi \bar\gamma_1}\right)
- \frac{\omega v_2 \bar\gamma_2}{2\pi}
\ln\left(\frac{M}{\omega v_2 \bar\gamma_2^2}\right)
\nonumber \\
& & + \frac{\omega v_1 \bar\gamma_1}{2\pi} \int_0^\infty \frac{dy}
{y - i \frac{\pi}{2R_p}}
\Re\Bigg\{ \frac{\left( y - i \frac{\pi}{2R_p}\right)^2 + \omega^2}
{R_p \left(\left( y - i \frac{\pi}{2R_p}\right)^2
- \omega^2 v_1^2 \bar\gamma_1^2\right) - \omega v_1 \bar\gamma_1}
\nonumber \\
& & \times \frac{\left( \omega + \omega^2 R_p v_1 \bar\gamma_1\right)
\sinh\left(R_p y\right)
+ R_p \omega \left( y - i \frac{\pi}{2R_p}\right)
 \cosh\left(R_p y\right)}
{\omega^2 v_1 \bar\gamma_1 \cosh\left(R_p y\right)
+ \omega \left( y - i \frac{\pi}{2R_p}\right)
\sinh\left(R_p y\right)} - 1 \Bigg\} + O(\bar\gamma_2^{-1}) \,.
\label{L cutoff gamma 2 limit}
\end{eqnarray}
Since $y - i\pi/2R_p$ is always at least of order $1$,
the integrand in \eqnlessref{L cutoff gamma 2 limit}
is of order $1$, and the term containing the integral is of order
$v_1$. The difference between the first two terms in
\eqnlessref{L cutoff gamma 2 limit} and the terms
in \eqnlessref{L phi asymptote} is also of order $v_1$.
Therefore, in the heavy--light limit,
\begin{equation}
L^{{\rm cutoff}}_\phi = - \sum_i \frac{\omega v_i \bar\gamma_i}{2\pi}
\ln\left(\frac{M}{\omega v_i \bar\gamma_i^2}\right) + O(v_1, \bar\gamma_2^{-1}) \,.
\end{equation}

We next evaluate $L^{\hbox{\scriptsize log term}}_\phi$,
We can simplify the integral
by separating it into two parts which are small
for $\nu \gg \omega$ and one part which does
not contain hyperbolic functions.
We break the $\nu$ integral into three pieces,
\begin{equation}
L^{\hbox{\scriptsize log term}}_\phi = - I_1 - I_2 - I_3 \,,
\end{equation}
where
\begin{eqnarray}
I_1 &=& \sum_i \int_0^\infty \frac{d\nu}{2\pi} \left[
\ln\left(\frac{(\nu^2 + \omega^2) (\nu^2 + 2 \nu \omega v_i \bar\gamma_i
+ (2\bar\gamma_i^2 - 1) \omega^2)}{(\nu^2 - (2\bar\gamma_i^2 - 1) \omega^2)
\nu (\nu + \omega v_i \bar\gamma_i)}\right)
- \frac{v_i \bar\gamma_i \omega}{\nu + v_i \bar\gamma_i \omega} \right] - \omega \,,
\nonumber \\
I_2 &=& \int_0^\infty \frac{d\nu}{2\pi} \Bigg\{
\tr\ln\Bigg[ 2 \delta_{ij} \frac{\nu^2 + \omega^2}
{\nu^2 - (2\bar\gamma_i^2 - 1) \omega^2} - (\nu^2 + \omega^2)
\nonumber \\
& & \times \frac{ \delta_{ij} \left(\sinh(\nu R_p)
+ \frac{\omega v_1 \bar\gamma_1 v_2 \bar\gamma_2}{\nu v_i \bar\gamma_i}
\cosh(\nu R_p)\right)
- (1-\delta_{ij}) \frac{\omega}{\nu} \sqrt{v_i \bar\gamma_1 v_2 \bar\gamma_2}}
{\left(\nu^2 + \omega^2 v_1 \bar\gamma_1 v_2 \bar\gamma_2\right)
\sinh(\nu R_p) + \nu \omega \left(v_1\bar\gamma_1 + v_2\bar\gamma_2\right)
\cosh(\nu R_p)} \Bigg]
\nonumber \\
& & - \sum_i
\ln\left(\frac{(\nu^2 + \omega^2) (\nu^2 + 2 \nu \omega v_i \bar\gamma_i
+ (2\bar\gamma_i^2 - 1) \omega^2)}{(\nu^2 - (2\bar\gamma_i^2 - 1) \omega^2)
\nu (\nu + \omega v_i \bar\gamma_i)}\right) \Bigg\} \,,
\nonumber \\
I_3 &=& - \sum_i \int_0^\infty \frac{d\nu}{2\pi} \Bigg\{
\frac{\omega^3}{\nu^2} \frac{1}{v_i \bar\gamma_i} \Bigg[
\bar\gamma_i \bar R_i \frac{\nu^2}{\omega^2}
+ \frac{v_1 \bar\gamma_1 v_2 \bar\gamma_2}{\omega (v_1 \bar\gamma_1
+ v_2 \bar\gamma_2) + \omega^2 R_p v_1 \bar\gamma_1 v_2 \bar\gamma_2}
- \frac{\nu}{\omega^3} (\nu^2 + \omega^2)
\nonumber \\
& & \times \frac{ \nu v_i \bar\gamma_i \sinh(\nu R_p)
+ \omega v_1 \bar\gamma_1 v_2 \bar\gamma_2 \cosh(\nu R_p)}
{\left(\nu^2 + \omega^2 v_1 \bar\gamma_1 v_2 \bar\gamma_2\right)
\sinh(\nu R_p) + \nu \omega \left(v_1\bar\gamma_1 + v_2\bar\gamma_2\right)
\cosh(\nu R_p)} \Bigg]
- \frac{\omega v_i \bar\gamma_i}
{\nu + v_i \bar\gamma_i \omega} \Bigg\} \,.
\label{three integrals def}
\end{eqnarray}
The second and third integrals are dominated by the region
where $\nu$ is small.
We can evaluate $I_1$ exactly, since the argument of
the log factorizes. We find
\begin{equation}
I_1 = \sum_i \left[ \frac{\omega v_i \bar\gamma_i}{2\pi}
\left( 1 - \ln\left( \frac{2\bar\gamma_i^2-1}{\bar\gamma_i^2-1} \right)
+ \frac{\pi}{v_i} + \frac{2}{v_i} \arctan(v_i) \right) \right] \,.
\end{equation}

In both the light--light and heavy--light limit we take the limit
$\bar\gamma_2 \to \infty$. In this limit, the three integrals
\eqnlessref{three integrals def} are
\begin{eqnarray}
I_1 &=& \frac{\omega v_1 \bar\gamma_1}{2\pi}
\left( 1 - \ln\left( \frac{2\bar\gamma_1^2-1}{\bar\gamma_1^2-1} \right)
+ \frac{\pi}{v_1} + \frac{2}{v_1} \arctan(v_1) \right)
+ \frac{\omega v_2 \bar\gamma_2}{2\pi} \left( 1 - \ln 2 + \frac{3}{2}\pi \right)
+ O(\bar\gamma_2^{-1}) \,,
\nonumber \\
I_2 &=& \frac{\omega}{2\pi} \int_0^\infty ds \ln\left[
\frac{ (s + v_1 \bar\gamma_1) \left( (s^2 + 2\bar\gamma_1^2 - 1)
\sinh(s \omega R_p) + 2 s v_i \bar\gamma_i \cosh (s \omega R_p)\right)}
{\left(s \cosh(s \omega R_p) + v_1 \bar\gamma_1 \sinh(s \omega R_p)\right)
(s^2 + 2 s v_1 \bar\gamma_1 + 2\bar\gamma_1^2 - 1)}\right]
+ O(\bar\gamma_2^{-1}) \,,
\nonumber \\
I_3 &=& -\frac{\omega}{2\pi} \int_0^\infty ds \Bigg[
\frac{s}{s + v_1 \bar\gamma_1} - \frac{(s^2 + 1) \cosh(s\omega R_p)}
{s^2 \cosh(s\omega R_p) + s v_1 \bar\gamma_1 \sinh(s\omega R_p)}
\nonumber \\
& & + \frac{1}{s^2} \frac{1}{1 + \omega R_p v_1 \bar\gamma_1} \Bigg]
+ O(\bar\gamma_2^{-1}) \,,
\nonumber \\
\label{three ints large gamma}
\end{eqnarray}
where we have changed integration variables in $I_2$ and $I_3$ to $s = \nu/\omega$.

In the light--light limit, we take $\bar\gamma_1 \to \infty$ and find
\begin{eqnarray}
I_1 &=& \sum_i \frac{\omega v_i \bar\gamma_i}{2\pi}
\left( 1 - \ln 2 + \frac{3}{2}\pi \right)
+ O(\bar\gamma_1^{-1}, \bar\gamma_2^{-1}) \,,
\nonumber \\
I_2 &=& O(\bar\gamma_1^{-1}, \bar\gamma_2^{-1}) \,,
\nonumber \\
I_3 &=& O(\bar\gamma_1^{-1}, \bar\gamma_2^{-1}) \,,
\end{eqnarray}
so $L^{\hbox{\scriptsize log term}}_\phi$ is
\begin{equation}
L^{\hbox{\scriptsize log term}}_\phi = - \sum_i \frac{\omega v_i \bar\gamma_i}{2\pi}
\left( 1 - \ln 2 + \frac{3}{2}\pi \right) + O(\bar\gamma_1^{-1}, \bar\gamma_2^{-1})
\end{equation}
in this limit. The combination $L^{\hbox{\scriptsize log term}}_\phi
+ L^{{\rm cutoff}}_\phi$ is then
\begin{equation}
L^{\hbox{\scriptsize log term}}_\phi + L^{{\rm cutoff}}_\phi =
- \sum_i \frac{\omega v_i \bar\gamma_i}{2\pi} \left[
\ln\left(\frac{M}{\omega v_i \bar\gamma_i^2}\right)
+ 1 - \ln 2 + \frac{3}{2}\pi \right]
+ O(\bar\gamma_1^{-1}, \bar\gamma_2^{-1}) \,.
\end{equation}
After renormalizing the geodesic curvature, we obtain
\begin{equation}
L^{\hbox{\scriptsize log term}}_\phi + L^{{\rm cutoff}}_\phi = 0 \,,
\label{L phi light light renorm}
\end{equation}
in the limit of massless quarks.

In the heavy--light limit, we take $v_1 \to 0$ in
\eqnlessref{three ints large gamma} and find,
\begin{eqnarray}
I_1 &=& \frac{\omega v_2 \bar\gamma_2}{2\pi} \left( 1 - \ln 2
+ \frac{3}{2}\pi \right) + \frac{1}{2} \omega + O(v_1 \ln v_1, \bar\gamma_2^{-1}) \,,
\nonumber \\
I_2 &=& \frac{\omega}{2\pi} \int_0^\infty ds \ln\tanh\left(\frac{\pi}{2} s\right)
+ O(v_1, \bar\gamma_2^{-1}) = -\frac{1}{8} \omega + O(v_1, \bar\gamma_2^{-1}) \,,
\nonumber \\
I_3 &=& O(v_1, \bar\gamma_2^{-1}) \,.
\end{eqnarray}
Thus, in this limit, after renormalizing the geodesic curvature,
\begin{equation}
L^{\hbox{\scriptsize log term}}_\phi + L^{{\rm cutoff}}_\phi
= -\frac{3}{8} \omega
+ O(v_1 \ln v_1, \bar\gamma_2^{-1}) \,.
\label{L phi heavy light renorm}
\end{equation}

Combining \eqnlessref{L theta renorm} with
\eqnlessref{L phi light light renorm} gives $L^{{\rm boundary}}$
in the light--light limit,
\begin{equation}
L_{{\rm boundary}} =  0 \,,
\end{equation}
and combining \eqnlessref{L theta renorm} with
\eqnlessref{L phi heavy light renorm} gives $L^{{\rm boundary}}$
in the heavy--light limit,
\begin{equation}
L_{{\rm boundary}} = -\frac{1}{4} \omega \,.
\end{equation}

%% file: excited_states.bbl
\begin{thebibliography}{10}

\bibitem{Abrikosov}
A.~A. Abrikosov.
\newblock On the magnetic properties of superconductors of the second group.
\newblock {\em Soviet Physics JETP}, 5:1174--1182, 1957.

\bibitem{ACPZ}
E.~T. Akhmedov, M.~N. Chernodub, M.~I. Polikarpov, and M.~A. Zubkov.
\newblock Quantum theory of strings in the {A}belian {H}iggs model.
\newblock {\em Phys. Rev. D}, 53:2087--2095, 1996.

\bibitem{Allen+Olsson+Veseli:1998}
T.~J. Allen, M.~G. Olsson, and S.~Veseli.
\newblock Excited glue and the vibrating flux tube.
\newblock {\em Phys. Lett.}, B434:110--114, 1998.

\bibitem{Alvarez:1983}
O.~Alvarez.
\newblock Theory of strings with boundaries: {F}luctuations, topology, and
  quantum geometry.
\newblock {\em Nuc. Phys.}, B216:125--184, 1983.

\bibitem{Baker+Ball+Brambilla+Prosperi+Zachariasen}
M.~Baker, J.~S. Ball, N.~Brambilla, G.~M. Prosperi, and F.~Zachariasen.
\newblock Confinement: Understanding the relation between the {W}ilson loop and
  dual theories of long distance {Y}ang--{M}ills theories.
\newblock {\em Phys. Rev. D}, 54:2829--2844, 1996.

\bibitem{Baker+Ball+Zachariasen:1990}
M.~Baker, J.~S. Ball, and F.~Zachariasen.
\newblock {QCD} flux tubes for {SU(3)}.
\newblock {\em Phys. Rev. D}, 41:2612--2618, 1990.

\bibitem{Baker+Ball+Zachariasen:1995}
M.~Baker, J.~S. Ball, and F.~Zachariasen.
\newblock Effective quark--antiquark potential for the constituent quark model.
\newblock {\em Phys. Rev. D}, 51:1968--1988, 1995.

\bibitem{Baker+Steinke2}
M.~Baker and R.~Steinke.
\newblock Effective string theory of vortices and {R}egge trajectories.
\newblock {\em Phys. Rev. D}, 69:094013, 2001.
\newblock hep-ph/0006069.

\bibitem{Cameron+Martin}
R~H. Cameron and W.~T. Martin.
\newblock Evaluation of various {W}iener integrals by use of certain
  {S}turm--{L}iouville differential equations.
\newblock {\em Bull. Am. Math. Soc.}, 51:73--90, 1945.

\bibitem{DHN}
R.~Dashen, B.~Hasslacher, and A.~Neveu.
\newblock Nonperturbative methods and extended-hadron models in field theory.
  {I.} {S}emiclassical functional methods.
\newblock {\em Phys. Rev. D}, 10:4114--4129, 1974.

\bibitem{Dubin+Kaidalov+Simonov2}
A.~Yu. Dubin, A.~B. Kaidalov, and Yu.~A. Simonov.
\newblock String with quarks. {I.} {S}pinless quarks.
\newblock {\em Phys. Atom. Nucl.}, 56:1745--1759, 1993.
\newblock hep-ph/9311344.

\bibitem{Dubin+Kaidalov+Simonov}
A.~Yu. Dubin, A.~B. Kaidalov, and Yu.~A. Simonov.
\newblock Dynamical regimes of the {QCD} string with quarks.
\newblock {\em Phys. Lett.}, 323B:41--45, 1994.

\bibitem{Forster}
D.~Forster.
\newblock Dynamics of relativistic vortex lines and their relation to dual
  theory.
\newblock {\em Nuc. Phys.}, B81:84--92, 1974.

\bibitem{Gervais+Sakita}
J.~L Gervais and B.~Sakita.
\newblock Quantized relativistic string as a strong coupling limit on the
  {H}iggs model.
\newblock {\em Nuc. Phys.}, B91:301--316, 1975.

\bibitem{Gutzwiller}
M.~C. Gutzwiller.
\newblock Phase--integral approximation in momentum space and the bound states
  of an atom. {II}.
\newblock {\em J. Math. Phys}, 10:1004--1020, 1969.

\bibitem{Isgur+Paton}
N.~Isgur and J.~Paton.
\newblock Flux-tube model for hadrons in {QCD}.
\newblock {\em Phys. Rev. D}, 31:2910--2929, 1985.

\bibitem{Kikkawa+Kotani+Sato}
K.~Kikkawa, T.~Kotani, M.~Sato, and M.~Kemmoku.
\newblock Semiclassical approach to the quark--string model and the hadron
  spectrum.
\newblock {\em Phys. Rev. D}, 18:2606--2622, 1978.

\bibitem{LaCourse+Olsson}
D.~LaCourse and M.~G. Olsson.
\newblock String potential model: {S}pinless quarks.
\newblock {\em Phys. Rev. D}, 39:2751--2757, 1989.

\bibitem{Landau+Lifshitz:Mechanics}
L.~D. Landau and E.~M. Lifshitz.
\newblock {\em Mechanics}.
\newblock Pergamon Press, 1976.
\newblock \S 40, pp. 131-133.

\bibitem{Luscher2}
M.~L{\"u}scher.
\newblock Symmetry-breaking aspects of the roughening transition in gauge
  theories.
\newblock {\em Nuc. Phys.}, B180:317--329, 1981.

\bibitem{Luscher1}
M.~L{\"u}scher, K.~Symanzik, and P.~Weisz.
\newblock Anomalies of the free loop wave equation in the {WKB} approximation.
\newblock {\em Nuc. Phys.}, B173:356--396, 1980.

\bibitem{Mandelstam}
S.~Mandelstam.
\newblock Vortices and quark confinement in non-{A}belain gauge theories.
\newblock {\em Phys. Rep.}, 23C:245--249, 1976.

\bibitem{Nambu}
Y.~Nambu.
\newblock Strings{,} monopoles{,} and gauge fields.
\newblock {\em Phys. Rev. D}, 10:4262--4268, 1974.

\bibitem{Nielsen+Olesen}
H.~B. Nielsen and P.~Olesen.
\newblock Vortex-line models for dual strings.
\newblock {\em Nuc. Phys.}, B61:45--61, 1973.

\bibitem{Pol+Strom}
J.~Polchinski and A.~Strominger.
\newblock Effective string theory.
\newblock {\em Phys. Rev. Lett.}, 67:1681--1684, 1991.

\bibitem{Polyakov:book}
A.~M. Polyakov.
\newblock {\em Gauge Fields and Strings}, pages 151--191.
\newblock Harwood Academic Publishers, Chur, Switzerland, 1987.

\bibitem{tHooft}
G.~`t~Hooft.
\newblock Gauge theories with unified weak, electromagnetic, and strong
  interactions.
\newblock In A.~Zichichi, editor, {\em High Energy Physics, Proceedings of the
  European Physical Society Conference, Palermo, 1975}, pages 1225--1249.
  Editrice Compositori, Bologna, 1976.

\end{thebibliography}
